\documentclass[twocolumn,preprintnumbers,amsmath,showpacs,amssymb]{revtex4}

\usepackage{graphicx}
\usepackage{dcolumn}
\usepackage{bm}
\usepackage{color}
\usepackage{bbm}

\begin{document}


\title{Quantum Dynamics of Kerr Optical Frequency Combs \\
       below and above Threshold: \\
       Spontaneous Four-Wave-Mixing, Entanglement and Squeezed States of Light}

\author{Yanne K. Chembo}
\thanks{Email address: yanne.chembo@femto-st.fr}
\affiliation{FEMTO-ST Institute, CNRS \& University Bourgogne Franche-Comt\'e, \\
             Optics Department, 15B Avenue des Montboucons, 25030 Besan\c con cedex, France.}
\date{\today}

\begin{abstract}
The dynamical behavior of Kerr optical frequency combs is very well understood today from the perspective of 
the semi-classical approximation.
These combs are obtained by pumping an ultra-high~$Q$ whispering-gallery mode resonator with a continuous-wave laser. 
The long-lifetime photons are trapped within the torus-like eigenmodes of the resonator, where they interact nonlinearly \textit{via} the Kerr effect.
In this article, we use quantum Langevin equations to provide a theoretical understanding of the non-classical behavior of these combs when pumped below and above threshold. 
In the configuration where the system is under threshold, the pump field is the unique oscillating mode inside the resonator, and triggers the phenomenon of spontaneous four-wave mixing, where two photons from the pump are symmetrically up- and down-converted in the Fourier domain. This phenomenon can only be understood and analyzed from a fully quantum perspective as a consequence of the coupling between the field of the central (pumped) mode and the vacuum fluctuations of the various sidemodes.
We analytically calculate the power spectra of the spontaneous emission noise, 
and we show that these spectra can be either single- or double peaked depending on the value of the laser frequency, chromatic dispersion, pump power, and spectral distance between the central mode and the sidemode of interest.
We also calculate as well the overall spontaneous noise power per sidemode, and propose simplified analytical expressions for some particular cases. 
In the configuration where the system is pumped above threshold, we investigate the
phenomena of quantum correlations and multimode squeezed states of light that can occur in the Kerr frequency combs originating from stimulated four-wave mixing. 
We show that for all stationary spatio-temporal patterns, the side-modes that are symmetrical relatively to the pumped mode in the frequency domain display quantum correlations that can lead to squeezed states of light under some optimal conditions that are analytically determined.
These quantum correlations can persist regardless the dynamical state of the system (rolls or solitons), 
regardless of the spectral extension of the comb (number sidemodes), and 
regardless of the dispersion regime (normal or anomalous). 
We also explicitly determine the phase quadratures leading to photon entanglement, and analytically calculate their quantum noise spectra. 
For both the below- and above-threshold cases, 
we study with particular emphasis the two principal architectures for Kerr comb generation, namely the add-through and add-drop configurations. 
It is found that regardless of the configuration, an essential parameter is the ratio between out-coupling and total losses, which plays a key role as it directly determines the efficiency of the detected spontaneous noise or squeezing spectra. 
We finally discuss the relevance of Kerr combs for quantum information systems at optical telecommunication wavelengths, below and above threshold.
\end{abstract}
\maketitle

\section{Introduction}
\label{intro}

Kerr optical frequency combs are sets of equidistant spectral lines that are generated after pumping a whispering gallery mode (WGM) or ring resonator with a continuous-wave (cw) laser~\cite{Vahala_PRL_2004,Maleki_PRL_LowThres,DelhayeKipp}.  
When the bulk resonator has both an ultra-high quality factor and a Kerr nonlinearity, it can at the same time 
trap the pump photons for a significantly long time in the torus-like eigenmodes of the resonator, and host the nonlinear interactions amongst them. 
When the pump power is sufficiently low, the intra-cavity photons remain in a single cavity-mode and their frequency essentially remains the same as the one of the pump laser.
However, above a certain threshold, these confined and long-lifetime pump photons are transferred through four-wave mixing (FWM) to neighboring cavity-modes, provided that phase-matching, energy and momentum conservation conditions are fulfilled. This process can be further cascaded and yield a frequency comb with all-to-all coupling, and involving up to several hundred modes over several THz.
In comparison to optical frequency comb generators based on femtosecond mode-locked lasers, 
Kerr comb generators are fairly simple, compact, robust and energy-efficient:
they are expected to be a core photonic systems for many applications, such as integrated photonics, metrology, aerospace and communication engineering~\cite{Lipson_NatPhot,Review_Kerr_combs_Science,Nature_Ferdous,PRL_Vahala_2012,PRL_NIST,Nat_Gaeta_rev,
PfeifleNatPhot,NIST_Optica,Self_injection_NIST,Kipp_Soliton,PRL_WDM,Nanophotonics}.

Beyond these potential applications, which have been a very powerful drive, Kerr comb are also actually an ideal 
test-bench systems for fundamental physics, and particularly, for quantum optics. 
In fact, understanding  Kerr comb generation is strikingly simple when one considers the photon picture and describes the process as the photonic interaction
$\hbar \omega_m + \hbar \omega_p \rightarrow \hbar \omega_n + \hbar \omega_q$, where two input photons labelled 
$m$ and $p$ interact coherently via the Kerr nonlinearity to yield two output photons $n$ and $q$. 
Without further analysis, this interpretation already suggests that purely quantum phenomena based on
the non-classical nature of light can eventually arise in Kerr combs.

From a theoretical point of view, it is well known today that in the semi-classical limit, 
Kerr combs can be described using either a set of coupled ordinary differential equations (one equation per mode~\cite{YanneNanPRL,YanneNanPRA,YanneNanOL}), 
or using a single partial differential equation (one equation for the sum of the modes~\cite{Matsko_OL_2,PRA_Yanne-Curtis,Coen}).
It is also well known that both formalisms are in fact perfectly equivalent~\cite{PRA_Yanne-Curtis}, with the first one emphasizing the {spectro-temporal} dynamics of the system, while the second emphasizes the {spatio-temporal} dynamics.
It is important to note here that these Kerr comb models are singularly accurate: the comparison between the numerical power spectra obtained using the models and the experimental ones is excellent across a dynamical range that can be as large as 
$80$~dB~\cite{YanneNanPRL,IEEE_PJ,OL_phaselocking,Chaos_paper}. 

The spatio-temporal formalism is generally known as the Lugiato-Lefever equation (LLE), and was introduced for the first time by Lugiato and Lefever in the context of ring resonators where the semi-classical cavity fields where subjected to Kerr nonlinearity and diffraction~\cite{LL}.
In the approximation of 1D-diffraction, some of the key dynamical properties of this optical system
had also been derived in the same article, such as for example the super- and sub-critical Turing instability leading to roll patterns.  
The LLE used to model Kerr combs has an essential dissimilarity with the one initially introduced by Lugiato and Lefever:
diffraction is replaced by dispersion. 
This difference is of no importance from the mathematical point of view. 
However, from the physical standpoint, the difference is significant.
On the one hand, Kerr comb generation is genuinely 1D, originates from a small bulk cavity (from $\mu$m- to mm-size), and  involves guided fields: the system is experimentally  compact, simple, low-power, versatile,  controllable, and its behavior can be described by the LLE with high accuracy as emphasized above despite its high dimensionality (from three to up to several hundred modes).
On the other hand, in the initial system, the approximation of 1D diffraction is rather poor (the 2D approximation is much better), the fields are propagating freely, and the cavity is set up with mirrors: the system is experimentally very complex and the LLE is a rather simplistic model, even though the number of interacting modes is always very limited
(rarely more that $10$). \\

In the scientific literature, several researchers have explored the quantum properties of optical resonators with Kerr nonlinearity when pumped under or above threshold.

In the case of a resonator pumped below threshold, the classical viewpoint assumes that the pump field is the unique oscillating mode inside the resonator, while all the sidemodes have zero power (hence, there is technically no comb in this case).
From a quantum standpoint, the pump field is actually at the origin of \textit{spontaneous four-wave mixing} where two pump photons are symmetrically up- and down-converted in the Fourier domain, thereby leading to the simultaneous and spontaneous generation of \textit{signal} and \textit{idler} photons, respectively. This phenomenology corresponds to the photonic interaction 
$2\hbar \omega_{\rm p} \rightarrow \hbar \omega_{\rm i} + \hbar \omega_{\rm s}$, where 
$\omega_{\rm p}$, $\omega_{\rm i}$ and $\omega_{\rm s}$ are the pump, idler and signal angular frequencies, respectively.
The phenomenon of spontaneous FWM (which is also sometimes referred to as \textit{parametric fluorescence}) can only be understood and analyzed from a fully quantum perspective, because it results from the coupling between the intracavity pump photons and the vacuum fluctuations of the various sidemodes.
This topic is the focus of a very large body of literature, particularly related to the generation of correlated pairs of entangled photons with chip-scale and integrated ring-resonators 
(see for example 
refs.~\cite{Sharping_OE,Clemmen_OE,Helt_OL,Chen_OE,Azzini_OE,Helt_scale_JOSAB,Azzini_OL,Camacho_OE,Takesue_SciRep,
Reimer_OE,Vernon_Arxiv,Engin_OE,Grassani_Optica,Qubit_entanglement,Wakabayashi} and references therein).

When the system is pumped above threshold, the photonic interaction $2\hbar \omega_{\rm p} \rightarrow \hbar \omega_{\rm i} + \hbar \omega_{\rm s}$ becomes steadily sustained:  from a classical perspective, the signal and idler sidemodes are correlated twin beams in the frequency domain, yielding a roll pattern in the spatial domain. By analogy to laser theory, it is considered that this phenomenon corresponds to \textit{stimulated four-wave mixing}~\cite{Agrawal}.    
In ref.~\cite{Lugiato_Castelli}, Lugiato and Castelli have pioneered investigations on the quantum properties of the paradigmatic system described in~\cite{LL} when pumped above threshold in the  approximation of 1D-diffraction.
In that work, they have demonstrated that the intensity difference exhibits fluctuations below the standard quantum noise limit~(QNL). This important result, which for the first time predicted squeezing in optical systems ruled by the LLE, was obtained in the three-mode approximation (central pumped mode and two sidemodes), and therefore, was only valid close to the threshold leading to the rolls in the super-critical case. Zambrini \textit{et~al.} numerically showed later on  that the squeezing behavior when certain additional degrees of freedom are accounted for is still consistent with the one of the reduced three-mode truncation~\cite{Zambrini_PRA}.
Further research on the quantum properties of optical systems ruled by the LLE was performed with the more realistic case of 2D-diffraction. However, in that case, the roll pattern is unstable and instead, the simplest  non-trivial solution is 
an hexagonal structure which emerges through a sub-critical bifurcation. As a consequence, the number of modes involved in the dynamics increases significantly because of the hexagonal structure itself (the smallest order truncation now involves $7$~modes, instead of $3$ for the roll pattern), and because of its sub-critical nature (the higher-order sidemodes can not be legitimately discarded anymore, even close to threshold, so that even the $7$-modes truncation is not very accurate). 
However, using that lowest-order truncation, Grynberg and Lugiato had shown very early that these hexagons 
can display four-fold mode squeezing in a lossless cavity close to threshold~\cite{Grynberg_Lugiato}, while Gatti and Mancini have extended the results and shown that squeezing and multimode entanglement persists even in the presence of losses, and even far above threshold as long as the $7$-mode truncation remains a good approximation~\cite{Gatti_Mancini}. 
In view of these preceding results, it could therefore be foreshadowed that Kerr combs, which can be described with great accuracy by the LLE in the semi-classical limit, can display a non-classical behavior as well. In this regard, 
an elegant demonstration of the theoretical prediction of Lugiato and Castelli has been achieved recently:
in the research work reported in ref.~\cite{Arxiv_squeezing_Cornell}, squeezing in a Kerr comb is experimentally demonstrated between the two-sidemodes of a $15$th-order roll pattern. 
Most important, this experiment is also the very first demonstration of squeezing in Kerr optical frequency combs, to the best of our knowledge. \\

From a purely technical point of view, other important parameters to consider are the central frequency of the comb, its spectral span, and the frequency separation between the comb lines.
Even though some works have shown that the combs can be obtained with a pump close to the lower and upper limits of the near-infrared range ($\sim 800$~nm~\cite{Nature_Selectable_freq}  and $\sim 2500$~nm~\cite{Nature_Mid_IR}), 
the overwhelming majority  of Kerr combs are generated today with laser pumps around $1550$~nm. 
Since this wavelength corresponds to the well known telecom spectral window,
there is a plethora of commercial off-the-shelf optical components (lasers, photodiodes, narrow filters, amplifiers, phase shifters, etc.) that are available for the manipulation of the photons around that wavelength, even at the single-photon level. 
It is also noteworthy that many nonlinear amorphous and crystalline materials have low dispersion and losses in that wavelength window, and these are two features that are of extreme importance in Kerr comb generation.

Moreover, Kerr combs originate from stimulated FWM which is an hyper-parametric process: hence, the frequency separation  between the spectral lines generally ranges from $\sim 1$~GHz  to $\sim 1$~THz for the Kerr combs of interest, instead of $\sim 100$~THz for parametric processes. Hence, in Kerr combs, the photo-detected signals fall into the microwave range where there is a very wide variety of technological solutions for the careful handling of low-noise signals. 

For the above reasons, Kerr combs have many singular advantages for quantum optics experiments, powered 
by the possibility to manipulate the photons in the optical frequency domain, and 
measure their slowly-varying attributes (amplitude and phase) in the microwave frequency domain. 
They also have the potential to play a major role in compact or integrated quantum-information systems
at optical telecommunication wavelengths~\cite{OL_QKD_Bloch,PRA_Merolla_2010,PRA_Merolla_2014}.\\

Despite the aforementioned theoretical works in the context of quantum phenomena of LLE-based systems, and despite 
the promising technological opportunities highlighted above, several critical problems remain wide open for the understanding of the quantum properties of spontaneous and stimulated FWM combs in WGM resonators.

The first topic of interest is the analysis of the spontaneous FWM comb spectra when the system is pumped below threshold. 
Many groups have investigated experimentally the main characteristics of this phenomenon, but a coherent theoretical basis explaining the influence of the various parameters of the system (dispersion, frequency detuning, etc.) on the output spectra is still lacking.  

A second challenge is that in the literature, the available research results to this date only consider minimally truncated expansions, whose validity is automatically restricted to a parameter range close to threshold.
However, Kerr combs are generally operated  far above threshold, and can be very large -- up to several hundreds of modes.
They can also correspond to different kinds of spatiotemporal patterns such as rolls (super- and sub-critical) or solitons (bright and dark), for example. It is therefore important to investigate in detail the quantum correlations in the case of Kerr combs where spectrum amplitude, size and span restrictions do not apply. 
 
A third  issue is  related to the sources of quantum noise in the system.
Previous theoretical works on LLE-based systems focused on gedanken experiments were the unique source of losses was the semi-reflecting mirror used to couple the light in and outside the cavity (the intrinsic losses were null). 
The corresponding equations therefore included only one vacuum fluctuation term. 
However, in the case or Kerr combs, the resonators are bulk, and then, necessarily lossy.
This introduces an extra term related to vacuum fluctuations induced by these intrinsic losses. 
Actually, the in- and out-coupling processes might also be distinct (like in the add-drop configuration, for example), so that overall, we might have up to three vacuum fluctuation terms, instead of just one.
In order to remain close to the experimental reality, it is therefore necessary to understand the effect of all these intrinsic and extrinsic vacuum fluctuations at the quantum level.

The fourth open point is the explicit determination of the quadratures that can potentially lead to multimode squeezing.
The conjugate variable of the  photon number operator is the phase operator~\cite{comment_phase_operators}, so that when the squeezing occurs for a linear combination of modal intensities, there exists is necessarily a corresponding linear combination of correlated phase quadratures in the system.
In Kerr combs, the large number of modes and the complexity of the all-to-all coupling amongst them allows for a large variety of phase-locking patterns in the semi-classical limit: the determination of the equivalent quantum correlations in terms of phase quadratures is therefore of particular relevance.

Our objective is to provide answers to the four open points highlighted above, and the article is therefore organized as follows.
In the next section, we present a brief overview of the mean-field models used to model the dynamics of Kerr combs in the semi-classical limit. Important physical considerations such as orders of magnitudes and system architecture will be discussed in detail.  
In Sec.~\ref{Quantummodel}, we build the quantum models for Kerr combs, using both the canonical quantization and the Hamiltonian formalism. Particular emphasis will be laid on the various sources of quantum noise that have to be accounted for depending on the in- and out-coupling configuration.
The dynamics of the system below threshold is investigated in Sec.~\ref{Spontaneous_FWM}, where the spontaneous FWM spectra are explicitly calculated as a fonction of the system's parameters.
Quantum correlations and squeezing for the photon numbers is investigated in 
Sec.~\ref{squeezzphotnumbers}, where we will explain why  the squeezing properties of the comb are degraded as the size of the comb increases. 
Section~\ref{squeezedquad} is devoted to the study of the quantum correlations and squeezing behavior in both the amplitude and phase quadratures, after the explicit derivation of the relevant quantum Langevin equations.
Particular emphasis is laid on the analysis of squeezing in  rolls and solitons (bright and dark), which are the most prevalent spatiotemporal patterns in Kerr comb generation, and their squeezing spectra will be investigated in Sec.~\ref{quadssqueezingrollssolitons}.
We sum up our main results in the last section, which concludes this article.

\section{Semi-classical models for Kerr optical frequency combs}
\label{semiclassicalmodels}

We provide here a brief overview of the semi-classical models for Kerr combs, which are useful to gain a deep understanding of the quantum models that will be developed in the next section, and which are also needed to introduce the key macroscopic parameters needed to describe the system.

\subsection{Modal expansion model}
\label{modalmodel}

WGM resonators, as well as ring-resonators, generally have several families of longitudinal (azimuthal) 
modes~\cite{SelTop_Matsko_I,SelTop_Matsko_II,Feron_Review}. 
Let us consider that only one family is involved in our case, and without loss of generality, we also consider that it is the fundamental family (torus-like modes).
In that case, the modes of interest, which are sometimes referred to as azimuthal, can be unambiguously defined by a single integer wavenumber $\ell$, which
characterizes each member's angular momentum. In the case of WGM resonators, this number $\ell$ can be considered as equal to
the total number of reflections that a photon undergoes during one round trip in the cavity (ray-optics interpretation).
Let us also consider that the eigennumber of the mode that is pumped by the external laser is $\ell_0$. 
In the spectral neighborhood of $\ell_0$, the eigenfrequencies of the resonator
can be expanded in a Taylor series, following 
\begin{eqnarray}
\omega_\ell = \omega_{\ell_0} +  \sum_{n=1}^{n_{\rm max}} \frac{\zeta_n}{n !} (\ell - \ell_0)^n   \, ,
\label{frequency_expansion}
\end{eqnarray}
where  $ \omega_{\ell_0}$ is the eigenfrequency at $\ell = \ell_0$ and $n_{\rm max}$ is the order of truncation for the expansion.

For a disk resonator with main radius $a$, the parameter $\zeta_1 = c/n_g a = \Delta \omega_{_{\rm FSR}}$ stands for the free-spectral range (FSR), with $c$ being the velocity of light and $n_g$ the group-velocity refraction index at
$\omega_{\ell_0}$.  This intermodal angular frequency is, of course, linked to the
round-trip period of a photon through the resonator as $T_{_{\rm FSR}} = 2 \pi/\zeta_1$.
The parameter $\zeta_2$ stands for the second-order group-velocity dispersion of the eigenmodes (normal GVD for $\zeta_2
<0$, and anomalous GVD when $\zeta_2 > 0$). 
We recall that $\zeta_2$ is generally the sum of two contributions, namely the geometrical dispersion (normal)
and the material dispersion (normal or anomalous). 
The parameters $\zeta_n$ for $n \ge 2$ stand for higher-dispersion terms and in this study, these terms will be considered as uniformly null. Note that perfect equidistance for the eigenfrequencies is achieved when $\zeta_n \equiv 0$ for all $n \ge 2$.
More details can be found in refs.~\cite{YanneNanPRA,YanneNanPRL,PRA_Yanne-Curtis,Jove}, for example.

The resonator is also characterized by its losses, which can be internal or external.
For each mode, the internal losses (bulk absorption, surface scattering, etc.) are quantified by the linewidth $\Delta \omega_{{\rm int},\ell}$.
On the other hand, the external losses $\Delta \omega_{{\rm ext},\ell}$ are here considered to be induced by both the in- or out-coupling processes of the optical fields. 
The total losses are just defined as the sum of the two aforementioned contributions following
$\Delta \omega_{{\rm tot},\ell} = \Delta \omega_{{\rm int},\ell}+\Delta \omega_{{\rm ext},\ell}$.
The loaded (or total) $Q$ factor for each mode can be defined as
$Q_{{\rm tot},\ell}^{-1} = Q_{{\rm int},\ell}^{-1}+Q_{{\rm ext},\ell}^{-1}= \Delta \omega_{{\rm tot},\ell}/\omega_\ell $, and the modal photon lifetime is $\tau_{{\rm ph}, \ell} = 1/  \Delta \omega_{{\rm tot},\ell}$.

The total electric field (in V/m) inside the cavity can be expanded as 
\begin{equation}
{\bf E}({\bf r}, t)= \sqrt{ 2 \, \frac{\hbar \omega_{_{\rm L}}}{\varepsilon_0 n_{_{\rm L}}^2} } \, \sum_\ell \frac{1}{2} \, {\cal A}_\ell (t) \, e^{i \omega_\ell t}  {\bf \Upsilon}_\ell ({\bf r})   + {\rm c.c.}  
\label{field_expansion}
\end{equation}
where $t$ is the time, 
${\cal A}_\ell (t)$ is the complex-valued slowly-varying amplitude of the $\ell$-th mode, 
${\bf \Upsilon}_\ell ({\bf r}) $ is the corresponding spatial mode profile (units of m$^{-\frac{3}{2}}$),
$\varepsilon_0$ is the permittivity of vacuum, 
$n_{_{\rm L}}$ is the refraction index at the laser pump wavelength,
and {\rm c.c.} stands for the ``complex conjugate'' of all the preceding terms~\cite{YanneNanPRA}.
It is important to note that in Eq.~(\ref{field_expansion}), and the fields have been normalized such that
$|\mathcal{A}_\ell|^2$ is equal to the number of photons in the $\ell$-th mode.

It has been shown in ref.~\cite{YanneNanPRA} that the slowly varying envelopes $\mathcal{A}_\ell$ of the
modes are governed by the following system of equations:
\begin{eqnarray}
\frac{d {\cal A}_\ell}{dt}  &=& -\frac{1}{2} \Delta \omega_{{\rm tot},\ell} \, {\cal A}_\ell  + \frac{1}{2}  \Delta \omega_{{\rm tot},\ell} \, {\cal F}_{\ell} \, e^{i\sigma t} \delta (\ell- \ell_0) \label{modal_equations} \\
&& -ig_0 \sum_{\ell_m,\ell_n,\ell_p}  {\cal A}_{\ell_m} {\cal A}_{\ell_n}^* {\cal A}_{\ell_p} e^{[i(\omega_{\ell_m} - \omega_{\ell_n} + \omega_{\ell_p} -\omega_{\ell} )t ] } \nonumber \\
&&  \,\,\,\,\,\,\,\,\, \,\,\,\,\,\,\,\,\,\,\,\,\,\,\,\,\,\,\,\,\,\,\, \times \Lambda_{\ell}^{\ell_m \ell_n \ell_p} \delta (\ell_m - \ell_n + \ell_p - \ell)  \nonumber \, ,
\end{eqnarray}
where $\delta (x)$ is the Kronecker delta-function that equals $1$ when
$x=0$ and equals zero otherwise.  
In the above equation, the Kronecker functions 
indicate that only the mode $\ell=\ell_0$ is pumped, and that the allowed four-wave mixing interactions will be those for which the total angular momentum of the interacting photons is conserved, following $\ell_m + \ell_p = \ell_n + \ell$. 

The four-wave mixing gain is $g_0 = n_2 c\hbar\omega_{\ell_0}^2/n_0^2  V_{\rm eff}$, where $\hbar$ is Planck's
constant, $n_2$ is the Kerr coefficient at
$\ell = \ell_0$, and $ V_{\rm eff} = [\int_V \| {\bf \Upsilon}_{\ell_0} ({\bf r_{\bot}}) \|^4 \, dV]^{-1}$ is the effective
mode volume of the pumped mode. 
The parameter $\Lambda_\ell^{\ell_m \ell_n
\ell_p}$ is an intermodal coupling tensor which weights the spatial overlap amongst the various modes. 
The laser pump field is characterized by the detuning $\sigma = \omega_{_{\rm L}} - \omega_\ell $ between its angular frequency
$\omega_{_{\rm L}} = 2 \pi c/\lambda_{_{\rm L}}$ and the resonance frequency $\omega_{\ell_0}$ of the pumped mode,
and by ${\cal F}_{\ell_0}  = [4 \Delta \omega_{{\rm ext},\ell_0} /\Delta \omega_{{\rm tot},\ell_0}^2 ]^{\frac{1}{2}} 
[P/\hbar \omega_{_{\rm L}}]^{\frac{1}{2}}$ which stands for the external pumping field, 
with $\Delta \omega_{{\rm ext}}$ representing in-coupling losses only.

Equation~(\ref{modal_equations}) can be further simplified and rewritten in a more convenient form, suitable for the canonical quantization.
The first step is to introduce the reduced eigennumber $l= \ell - \ell_0$, so that the pumped mode is now $l=0$, while the various sidemodes symmetrically expand as $l= \pm 1, \pm 2, \dots$, with ``$+$" and ``$-$" standing respectively for higher and lower frequency sidemodes. The modes $\ell_m$, $\ell_n$ and $\ell_p$ in the four-wave mixing sum will now be simply replaced by their reduced counterpart as $\{m,n,p \} = \ell_{\{m,n,p \}} - \ell_0$.
The second step is to consider that the spectral extension of the comb is narrow enough to consider that the modes are 
quasi-degenerate in space and frequency ($\Lambda_l^{m n p} \equiv 1$), and that the modal losses are quasi-degenerate as well, with $\Delta \omega_l \equiv \Delta \omega_{{\rm tot},0} = \Delta \omega_{\rm tot}$.
The last step is to replace the fields ${\cal A}_\ell \equiv {\cal A}_l $ in Eq.~(\ref{modal_equations}) by 
${\cal A}_l^* \exp [i (\sigma - \frac{1}{2} \zeta_2 l^2)t]$, so that explicit time dependence is removed in
Eq.~(\ref{modal_equations}). From a physical viewpoint, this latter transformation corresponds to setting the frequency reference at the laser frequency instead of the cold-cavity resonance of the pumped mode, and to express the modal frequencies with respect to the equidistant (FSR-spaced) frequency grid, instead of the dispersion-detuned eigenfrequency grid~\cite{PRA_Unified}.

After implementing these mathematical transformations, 
it can be shown that the new modal fields ${\cal A}_l$ obey the following set of autonomous, nonlinear and coupled ordinary differential equations:
\begin{eqnarray}
\dot{{\cal A}}_l  &=& -\frac{1}{2} \Delta \omega_{\rm tot} \, {\cal A}_l 
                      + i \left[ \sigma  - \frac{1}{2} \zeta_2 l^2  \right] \, {\cal A}_l \nonumber \\
                  &&  + \delta(l) \sqrt{\Delta \omega_{\rm ext}} \, {\cal A}_{\rm in}  \nonumber  \\ 
                  &&  + i g_0 \sum_{m,n,p} \delta(m-n+p-l)  \, {\cal A}_m {\cal A}_n^* {\cal A}_p  \, .
\label{modal_model_new}
\end{eqnarray}
where the overdot indicates the time derivative. 
Note that higher-order dispersion at arbitrary order can be accounted for by replacing 
$\zeta_2 l^2/2$ by $\sum_{n=2}^{n_{\rm max}} \zeta_n l^n / n!$ which is obtained from Eq.~(\ref{frequency_expansion}).
Without loss of generality, we can arbitrarily consider the phase of the external pump field as a reference and set it to zero, so that this field becomes real-valued and can be written as
\begin{eqnarray}
{\cal A}_{\rm in} \equiv {A}_{\rm in} =\sqrt{\frac{P}{\hbar \omega_{_{\rm L}}}} \, .
\label{def_Ain}
\end{eqnarray}
It is important to recall the normalization in the semi-classical Eqs.~(\ref{modal_model_new})
is such that $|{\cal A}_l|^2$ is a number of photons (cavity fields), 
while  $|{{A}}_{\rm in}|^2$ is a number of photons per second (propagating fields).
This normalization is  physically the most appropriate at the time to perform the canonical quantization.

\begin{figure}
\begin{center}
\includegraphics[width=7cm]{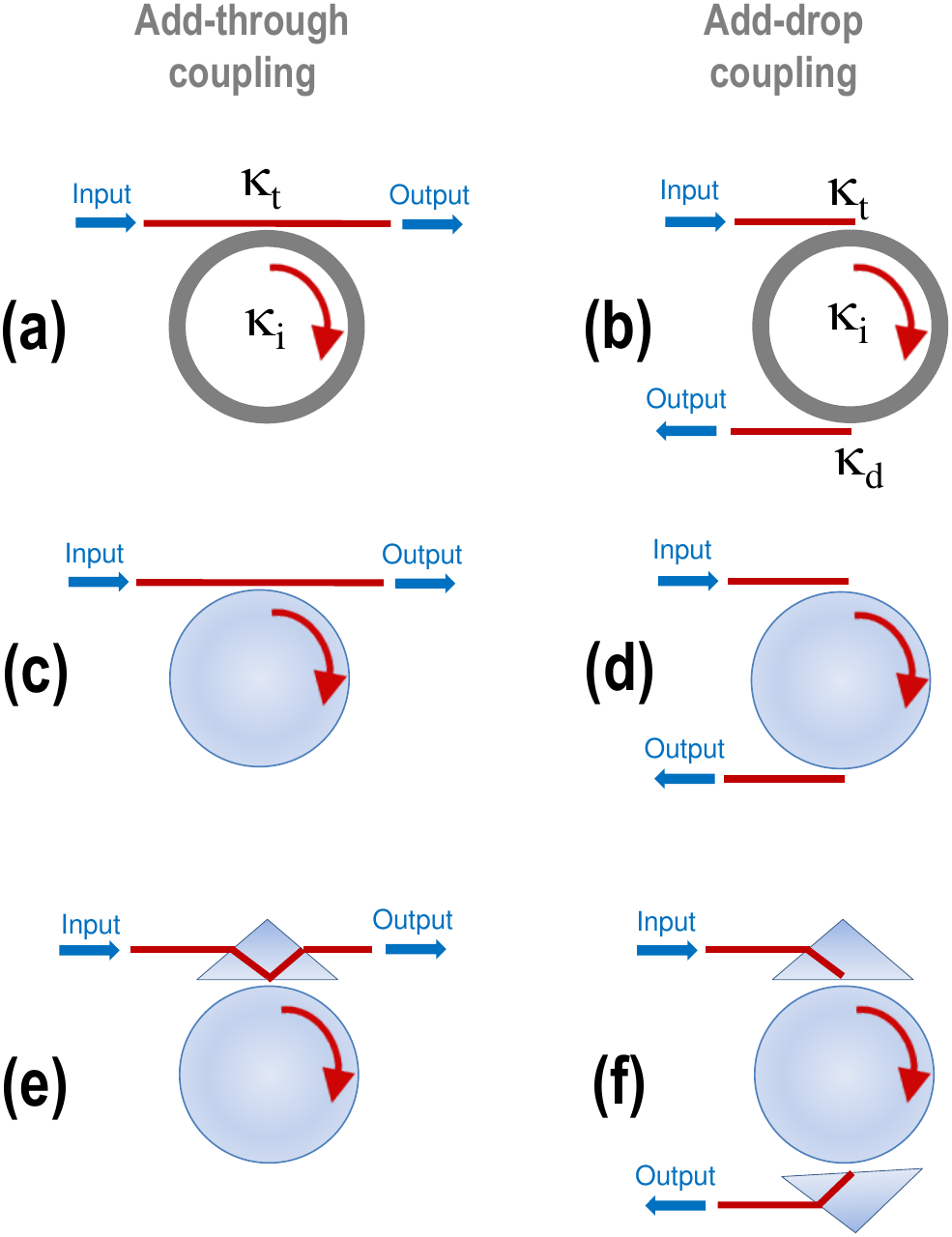}
\end{center}
\caption[Figure_Squeezing_Setup_AT_AD_resized_v2]
{\label{Figure_Squeezing_Setup_AT_AD_resized} 
(Color online) The two main configurations for Kerr comb generation with monolithic resonators, namely the add-through (left column) and add-drop (right column) configurations. 
Each architecture features a certain number of loss mechanisms (quantified by their half-linewidths $\kappa_{\rm t, i, d}$), which are associated with vacuum fluctuations. The related quantum noise contributions have to be accounted for when calculating the squeezing spectra. 
(a) and (b): waveguide coupling of integrated ring-resonators;
(c) and (d): Tapered-fiber coupling of WGM resonators;
(e) and (f): Prism coupling of WGM resonators.}
\end{figure}

\subsection{The two configurations under study}
\label{configunderstudy}

Two configurations are routinely used to generate Kerr optical frequency combs, as displayed in 
Fig.~\ref{Figure_Squeezing_Setup_AT_AD_resized}.
It therefore important to identify precisely all the loss terms as well as the out-coupled fields in each case, because as we will see later on, the vacuum quantum noise terms are closely related to these losses and out-coupling configurations. 

In the first architecture, that we call \textit{add-through}~\cite{comment_AT}, a single coupler is used to pump the cavity and to retrieve the comb signal, which is detected at the through port. This architecture allows for limited coupling losses (and therefore, low threshold power for Kerr comb generation). However, a disadvantage of this architecture is that the output signal is a superposition of the intra-cavity and a portion of the pump which is directly passing through the coupling
waveguide~\cite{IEEE_PJ}.
In this {add-through} configuration, the total and external linewidths in Eq.~(\ref{modal_model_new})
can be written as 
\begin{eqnarray}
\Delta \omega_{\rm tot} & \equiv & \Delta \omega_{\rm int} + \Delta \omega_{\rm ext,t} \label{d_omegatot_AT} \\ 
\Delta \omega_{\rm ext} & \equiv & \Delta \omega_{\rm ext,t} \, , \label{d_omegaext_AT}  
\end{eqnarray}
while the modal output fields obey 
\begin{eqnarray}
{{\cal A}}_{{\rm out},l} =\sqrt{\Delta \omega_{\rm ext,t}} \, {\cal A}_l - {A}_{\rm in} \delta(l) \, .
\label{Aout_addthrough}
\end{eqnarray}
with $\Delta \omega_{\rm ext,t}$ standing for the coupling losses in the through port.

In the second architecture, referred to as \textit{add-drop}, two different couplers are used to perform in- and out-coupling tasks.
The comb is therefore retrieved at the drop port. 
This double-coupling has the disadvantage to increase the overall losses (thereby increasing the threshold for Kerr comb generation), but however, at the opposite of the precedent case, 
the output signal is proportional to the intra-cavity field and provides an unambiguous representation
of the physical processes that are taking place inside the resonator.  
For the {add-drop} configuration, the linewidths in Eq.~(\ref{modal_model_new}) are explicitly defined as 
\begin{eqnarray}
\Delta \omega_{\rm tot} & \equiv & \Delta \omega_{\rm int} + \Delta \omega_{\rm ext,t} + \Delta \omega_{\rm ext,d} \label{d_omegatot_AD} \\ 
\Delta \omega_{\rm ext} & \equiv & \Delta \omega_{\rm ext,t} \, , \label{d_omegaext_AD}  
\end{eqnarray}
and the modal output fields simply obey 
\begin{eqnarray}
{{\cal A}}_{{\rm out},l} =\sqrt{\Delta \omega_{\rm ext,d}} \, {\cal A}_l \, .
\label{Aout_adddrop}
\end{eqnarray}
where $\Delta \omega_{\rm ext,d}$ stands for the coupling losses in the drop port~\cite{JSTQE_AurYanne}.

In all cases, the various linewidths are related to their corresponding quality factors by
$\Delta \omega_{{\rm int,ext,tot}} = \omega_{_{\rm L}} /Q_{{\rm int,ext,tot}}$.
A technique routinely used to determine the various quality factors at the experimental level
is the cavity-ring-down method~\cite{Feron_CRD}.

\subsection{Spatiotemporal formalism}
\label{spatiotemporal_formalism}

Several studies on the quantum properties of  self-organized dissipative optical structures are performed 
on systems that are ruled by the LLE. In the case of Kerr combs, it has be shown in ref.~\cite{PRA_Yanne-Curtis}  that the above modal expansion model is exactly equivalent to the following normalized LLE
\begin{eqnarray}
  \frac{\partial {\cal A}}{\partial t} &=&  -\frac{1}{2} \Delta \omega_{\rm tot} \, {\cal A}
                                          + i \sigma {\cal A} 
                                          + i g_0 |{\cal A}|^2 {\cal A} \nonumber \\
                                       &&   + i \frac{\zeta_2}{2} \frac{\partial^2 {\cal A}}{\partial \theta^2}
                                          + \sqrt{\Delta \omega_{\rm ext,t}} \, {A}_{\rm in} 
\label{eq:LLE_phot_nber}
\end{eqnarray}
where ${\cal A} (\theta, t) = \sum_l {\cal A}_l (t) e^{i l\theta} $  is the total intra-cavity field and $\theta \in [- \pi, \pi]$ is the azimuthal angle along the circumference of the resonator. 
Higher-order dispersion can be accounted for by replacing 
$({\zeta_2}/{2}) \partial^2 \! {\cal A}/\partial {\theta}^2 $ by
$v_g \sum_{k=2}^{k_{\rm max}} (i\Omega_{_{\rm FSR}})^k ({\beta_k}/{k !}) \partial^k \! {\cal A}/\partial {\theta}^k$
where the dispersion coefficients $\beta_k = -[v_g (-\Omega_{_{\rm FSR}})^k] \, \zeta_k$ exactly correspond to those used in fiber optics.   
The total number of intracavity photons is therefore $|{\cal A}|^2$, while the output field is ${{\cal A}}_{{\rm out}} =\sqrt{\Delta \omega_{\rm ext,t}} \, {\cal A} - {A}_{\rm in}$ in the add-through configuration, and ${{\cal A}}_{{\rm out}} =\sqrt{\Delta \omega_{\rm ext,d}} \, {\cal A}$ in the add-drop configuration.
In several theoretical studies, Eq.~(\ref{eq:LLE_phot_nber}) is further normalized to 
\begin{equation}
  \frac{\partial \psi}{\partial \tau} = -(1+i\alpha)\psi + i|\psi|^2\psi
  -i\frac{\beta}{2} \frac{\partial^2\psi}{\partial \theta^2} + F 
  \label{eq:LLE}
\end{equation}
where $\psi (\theta, \tau) = (2g_0/\Delta \omega_{\rm tot})^{1/2} {\cal A}$ is the dimensionless intra-cavity field, 
and $\tau = \Delta \omega_{\rm tot} t/2 = t /2 \tau_{\rm ph} $ is the dimensionless time.
The dimensionless parameters of this normalized equation are the frequency detuning
$\alpha = -2 \sigma/\Delta \omega_{\rm tot}$, the cavity second-order
dispersion $\beta = -2\zeta_2/\Delta \omega_{\rm tot}$, and the external excitation
$F = (8 g_0 \Delta \omega_{\rm ext,t} / \Delta \omega_{\rm tot}^3 )^{1/2}  \sqrt{P/\hbar \omega_{_{\rm L}}}$. 
In the context of Kerr comb generation, the LLE has been extensively investigated in several articles since the pioneering works of refs.~\cite{Matsko_OL_2,PRA_Yanne-Curtis,Coen}.

In ref.~\cite{PRA_Unified}, an exhaustive study of the various dynamical regimes of the LLE has been performed,
and the stability basin of the various solutions has been determined. 
In the anomalous dispersion regime, the stationary solutions are rolls (super- and sub-critical), bright solitons (isolated or coexisting), and soliton molecules (isolated or coexisting). 
In the case of normal dispersion, the stationary solutions can be rolls, dark solitons (isolated or coexisting), and non-smooth dark solitons (sometimes referred to as \textit{platicons}, see ref.~\cite{platicons}).
For all these stationary  solutions, the Kerr comb is perfectly symmetric in the semi-classical limit, and we will see in Sec.~\ref{squeezzphotnumbers} that this symmetry opens the way for multimode squeezing when quantum noise is accounted for.

\begin{figure*}
\begin{center}
\includegraphics[width=16cm]{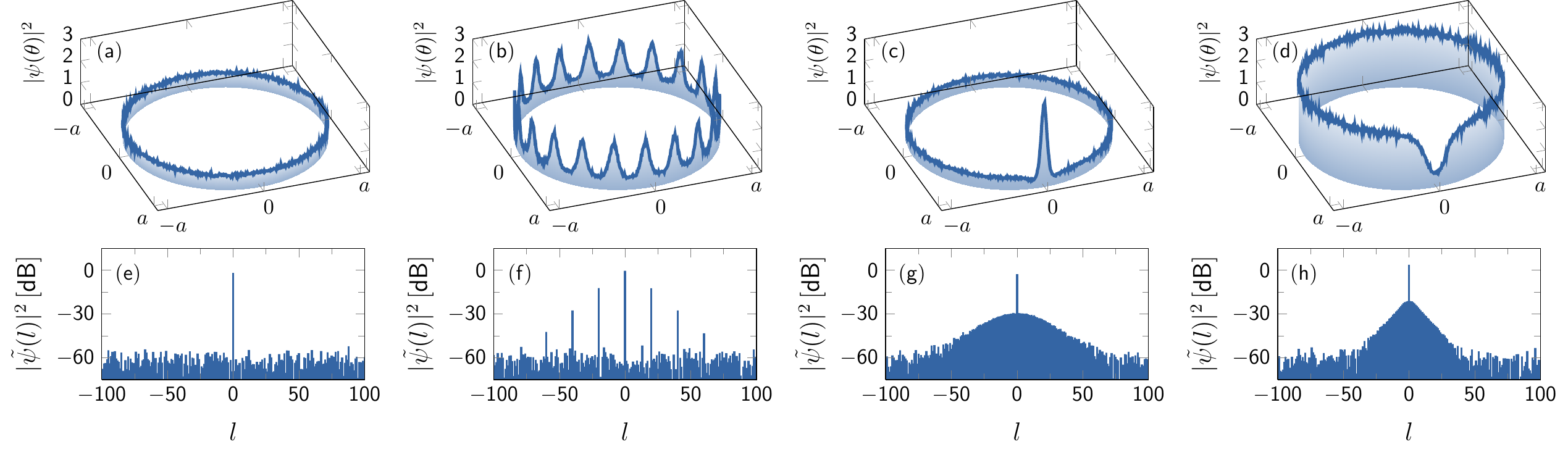}
\end{center}
\caption[spatiospectroplots]
{\label{spatiospectroplots} 
(Color online) Spatiotemporal (upper row) and spectro-temporal (lower row) representation  at a given time $t$ (snapshot) of some  stationary solutions for the normalized intra-cavity field  $\psi(\theta)= \sum_l \psi_l e^{il\theta}$ when quantum noise is accounted for.  
The spatio-temporal representation displays the intra-cavity field intensity 
$|\psi|^2$ (which is proportional to the total intra-cavity photon number) as a function of the azimuthal angle $\theta$ along the circumference of the resonator of radius $a$. 
The spectro-temporal representation displays the corresponding stem plot for the modal intensities $|\psi_l|^2 \equiv |\tilde{\psi}(l)|^2$ as a function of the reduced eigenumber $l$. 
The quantum noise has been added to the  deterministic stationary solutions (flat state, rolls, ans solitons) and in the figure, the noise intensity has been set at a much stronger intensity than realistic quantum noise for the sake of visual clarity. The parameters of the system are defined in Sec.~\ref{Corrsqueezspec}.
Note that the pumped mode is $l=0$, so that the sidemodes expand as $l = \pm 1, \pm 2, \dots$ 
(a) and (e): flat state ($P=1.5$~mW and $\sigma=-\kappa$). The system is here under threshold. The deterministic intracavity field is constant and in the spectral domain there is only one spectral line. The effect of quantum noise is to induce a random modulation of the flat state in the spatial domain, and to generate noisy sidemodes around the pump in the spectral domain.
(b) and (f): roll pattern of order $L=20$ ($P=2.5$~mW and $\sigma=-\kappa$). There are $20$~rolls in the spatial domain, and the deterministic oscillating  sidemodes in the spectral domain have a $20$-FSR spacing.  
(c) and (g): bright soliton  ($P=3.5$~mW and $\sigma=-2\kappa$);
(d) and (h): dark soliton  ($P=5.3$~mW and $\sigma=-2.5\, \kappa$).}
\end{figure*}

\subsection{Orders of magnitude in experimental systems}
\label{orderofmagnitude}

In order to facilitate comparisons between theory and experiments, 
it is important to link the normalized parameters and variables to their counterparts in SI units.
In particular, knowing the power levels involved provides key information at the time to choose the  low-noise, high sensitivity components needed to perform experiments with non-classical light~\cite{Bachor_Ralph_book}.
 
In our Eq.~(\ref{modal_model_new}), the dispersion parameter $\zeta_2$ is linked to the parameter $\beta_2$
used in fiber optics by $\beta_2= - \zeta_2/v_g \Delta \omega_{_{\rm FSR}}^2$ (in s$^2$m$^{-1}$),
where $v_g = c/n_g$ is the group velocity. 
The coefficient $g_0$ can be converted to the nonlinear coefficient
$\gamma  = \omega_{_{\rm L}} n_2/c A_{\rm eff} = g_0 T_{_{\rm FSR}} /v_g \hbar \omega_{_{\rm L}}$ (in W$^{-1}$m$^{-1}$) which is also well known in fiber optics, 
where   $A_{\rm eff} = V_{\rm eff} / 2 \pi a $ is the effective area, and 
$V_{\rm eff}$ is the effective volume. For a spherical resonator of radius $a$, an approximation of the effective volume of a WGM of azimuthal eigenumber $\ell$ and polar eigennumber $m$ is given in ref.~\cite{Braginski_PLA} as $V_{\rm eff} \simeq 3.4 \, \pi^{\frac{3}{2}} (\lambda_{_{\rm L}}/2 \pi n_g)^3 \ell^{\frac{11}{6}} \sqrt{\ell -m +1} $. 
Since $\ell \simeq m$ for the WGMs of interest, the effective area can therefore be approximated as 
$A_{\rm eff}  \sim (\lambda_{_{\rm L}} / n_g)^{\frac{7}{6}} a^{\frac{5}{6}}$ for a spherical WGM resonator, and 
this is generally a higher bound estimate for WGM disks or ring resonators.
Finally, the intra-cavity and output dimensionless intensities $|{{\cal A}}_{l}|^2$ and 
$|{{\cal A}}_{{\rm out},l}|^2$ can be converted in watts following
$|{{\cal E}}_{l}|^2 = \hbar \omega_{_{\rm L}} |{{\cal A}}_{l}|^2/T_{_{\rm FSR}}$ and 
$|{{\cal E}}_{{\rm out},l}|^2 = \hbar \omega_{_{\rm L}} |{{\cal A}}_{{\rm out},l}|^2$.  

The theory based on the stability analysis of the normalized LLE indicates that Kerr combs can scarcely  be generated when the normalized intra-cavity power $|\psi|^2$ and external pump power $F^2$ are inferior to $1$.
Therefore, the condition $F_{\rm min}^2 = 1$ leads the following absolute minimum pump power (in watts) to trigger Kerr comb generation
\begin{eqnarray}
P_{\rm min} = \frac{\hbar \omega_{_{\rm L}}}{8 g_0} \frac{\Delta \omega_{{\rm tot}}^3}{\Delta \omega_{{\rm ext,t}}}
            = 2 \pi a \, \frac{\omega_{_{\rm L}}^2}{8 \gamma v_g^2}  \frac{Q_{{\rm ext,t}}}{Q_{{\rm tot}}^3} \, ,                       
\label{minpowersinwattspump}
\end{eqnarray}
which correspond to an absolute minimum photon flux of $|{A}_{\rm in}|_{\rm min}^2  = {P_{\rm min}}/{\hbar \omega_{_{\rm L}}}$.
On the other hand, the condition $|\psi|_{\rm min}^2 = 1$ yields the following formula for the minimum intra-cavity power (in watts)
\begin{eqnarray}
|{{\cal E}}_{\rm min}|^2 = \frac{\hbar \omega_{_{\rm L}}}{2 g_0} \frac{\Delta \omega_{{\rm tot}}}{T_{_{\rm FSR}}}
                         = \frac{\omega_{_{\rm L}}}{ 2 \gamma v_g \, Q_{{\rm tot}}} \, ,                       
\label{minpowersinwattsitracav}
\end{eqnarray}
which corresponds to a minimal intra-cavity number of photon equal to 
$|{\cal A}_{0}|_{\rm min}^2  = \Delta \omega_{{\rm tot}} / 2 g_0$.
The above values are therefore absolute minima (necessary but not sufficient for comb generation), that can be reached when the laser is accurately detuned to
$\sigma = - \frac{1}{2} \Delta \omega_{\rm tot}$ in the anomalous dispersion regime 
(see refs.~\cite{YanneNanPRA,PRA_Unified}).
For any other detuning, and in both dispersion regimes,
the threshold pump power $P_{\rm th}$ for Kerr comb generation will necessarily be higher than  $P_{\rm min}$, 
up to a factor $100$. 
However, the threshold number of intra-cavity threshold number of photons $|{\cal A}_{\rm th}|^2$ will still be equal, or very close, to the minimal value $|{\cal A}_{0}|_{\rm min}^2$~\cite{PRA_Unified}.

Therefore, for mm-size crystalline resonator with $10$~GHz free-spectral range ($T_{_{\rm FSR}} = 100$~ps),
$\gamma \sim  1$~W$^{-1}$km$^{-1}$, $n_g \sim 1.4$, and $Q_{\rm int} =  Q_{\rm ext} \sim  10^9$ at $1550$~nm in the add-through configuration, the absolute minimum threshold power can be as low as $P_{\rm min} \sim 1$~mW.
Such low pumping power has already been demonstrated experimentally, like in ref.~\cite{Comb_2mW}
where a threshold power of $\sim 2$~mW was sufficient to trigger Kerr comb generation.
On the other hand, for an integrated silicon nitride resonator with $100$~GHz repetition rate,
$\gamma \sim  10$~W$^{-1}$km$^{-1}$, $n_g \sim 2$, and quality factors 
$Q_{\rm int} = Q_{\rm ext} \sim 3 \times 10^6$ at $1550$~nm in the add-through configuration,
the absolute minimum threshold pump power is rather $P_{\rm min} \sim 1$~W.

\begin{figure*}
\begin{center}
\includegraphics[width=16cm]{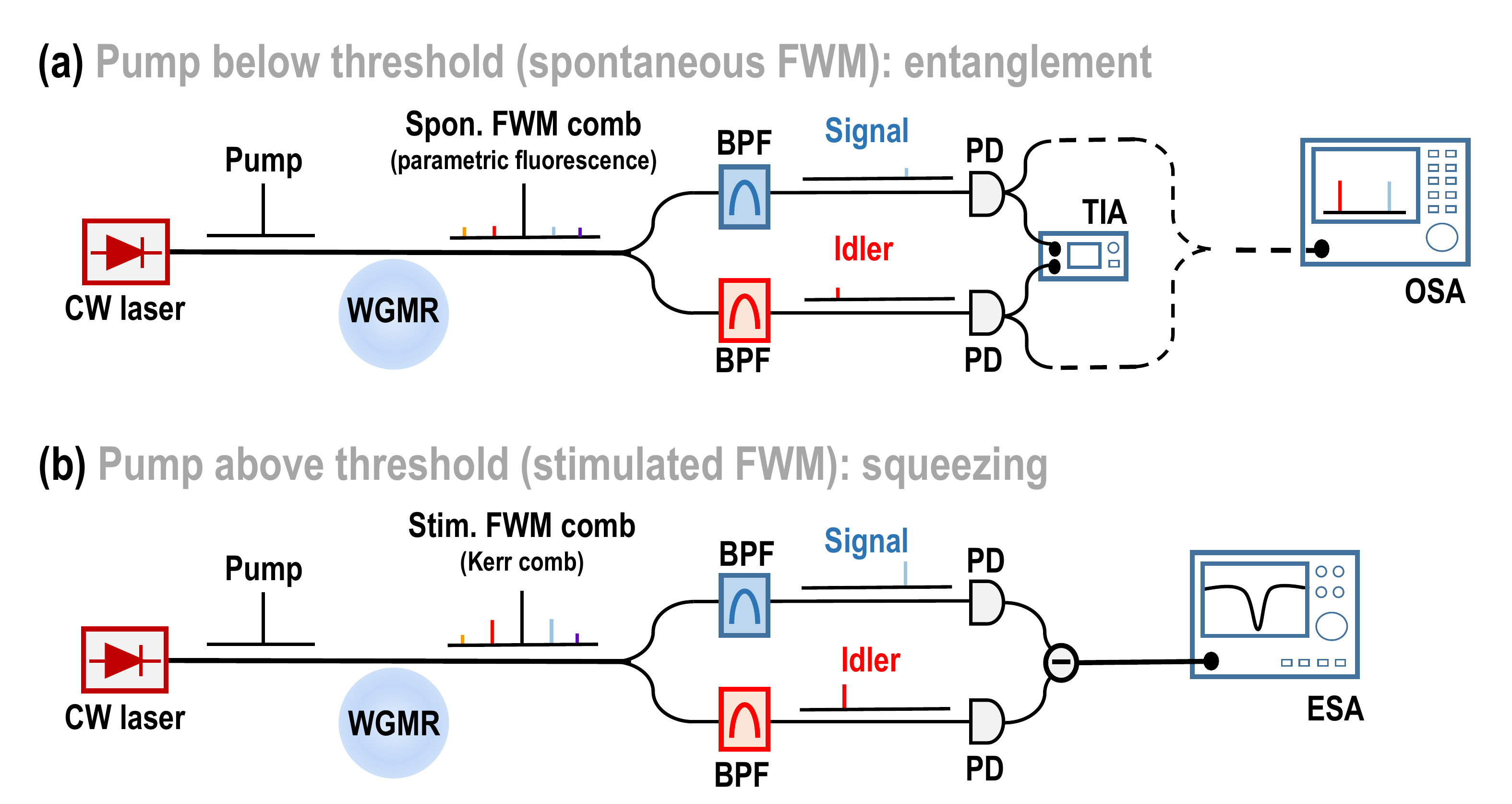}
\end{center}
\caption[Experimental_setups]
{\label{Experimental_setups} 
(Color online) Schematic representation of the experimental setups used to analyze the quantum states of a Kerr comb.
Practical elements such as polarization controllers, amplifiers, variable attenuators, isolators, etc. 
have been ommitted for the sake of conceptual clarity.
(a)~Pump below threshold. Two symmetric sidemodes from the parametric fluorescence spectrum can be isolated and   
can be processed in the time domain using a time interval analyzer (TIA), or in the spectral domain using an optical spectrum analyzer(OSA). Note that the TIA and the OSA should normally not be used simultaneously. 
(b)~Pump above threshold. Two symmetric sidemodes from the Kerr comb are isolated and the difference between the photodetected signal can be monitored using an electrical spectrum analyzer (ESA). Note that in the ESA the baseband spectrum will be single-side band (the double-side band pictogram is only a convenient visual reminder of the squeezing spectra we are theoretically plotting in this article).  }
\end{figure*}

\section{Quantum model for Kerr optical frequency combs}
\label{Quantummodel}

The construction of quantum models for Kerr combs is required in order to understand the spatio- and spectro-temporal behavior of the system when it is in a dynamical state like one of those displayed in Fig.~\ref{spatiospectroplots}.
The determination of this dynamical behavior at the quantum level can be performed through the canonical quantization of the semi-classical model, or by defining an Hamiltonian operator ruling the relevant interactions in the system.
The first approach has the advantage to be more intuitive, while the second is generally helpful at the time to establish  conservation rules (which are closely related to commutators involving the Hamiltonian).
In the present article we will use both formalisms, which will be introduced in this section to derive the temporal behavior of the Kerr comb.

\subsection{Canonical quantization}
\label{canonical_quantization}

The canonical quantization permits to derive the quantum counterpart of a semi-classical model, and in our case
it consists in three steps~\cite{Gardiner,Grynberg_Aspect_Fabre_book}: 
(\textit{i}) replace all the fields ${\cal A}_l (t)$ and their complex conjugates ${\cal A}_l^* (t)$ by annihilation and creation operators  $\hat{{\mathsf a}}_l (t)$ and $\hat{{\mathsf a}}_l^\dagger (t)$, respectively~\cite{notational_convention};
(\textit{ii}) introduce vacuum fluctuation operators for every loss mechanism (intrinsic or extrinsic) in the optical system;
(\textit{iii}) introduce vacuum fluctuation operators at both the in- and out-coupling ports.

The creation and annihilation operators obey the following boson commutation rules
\begin{align}
& [\hat{{\mathsf a}}_l^{\,\,\,}, \hat{{\mathsf a}}_{l'}^{\dagger} ] = \delta_{l,l'}  \\
& [\hat{{\mathsf a}}_l, \hat{{\mathsf a}}_{l'} ] = [\hat{{\mathsf a}}_l^{\dagger}, \hat{{\mathsf a}}_{l'}^{\dagger} ] =0 \, .
\label{comutation_rules_crea_anihil}
\end{align}
The semi-classical photon number $|{\cal A}_l|^2= {\cal A}_l^*{\cal A}_l $, which was a measure of the intra-cavity optical energy for each mode, is now represented by its quantum counterpart, which is the photon number operator
\begin{align}
\hat{{\mathsf n}}_{l} = \hat{{\mathsf a}}_{l}^{\dagger} \hat{{\mathsf a}}_l \, .
\label{photon_number}
\end{align}
It is useful to recall that the ordering of the operators $\hat{{\mathsf a}}_{l}^{\dagger}$ and $ \hat{{\mathsf a}}_l$ can not be arbitrarily swapped, as these two operators do not commute. We adopt here the so-called normal ordering which consists in placing the creation operators on the left and the annihilation operators on the right. 

The vacuum fluctuations associated with losses and coupling can be explicitly introduced in each mode 
using the vacuum operators
$\hat{{\mathsf V}}_{{\rm i},l}$ for the intrinsic losses, 
$\hat{{\mathsf V}}_{{\rm t},l}$ for the coupling losses in the through port, and
$\hat{{\mathsf V}}_{{\rm d},l}$ for the coupling losses in the drop port. 
These free-field operators have zero-mean value and obey the commutation rules 
\begin{align}
&[\hat{{\mathsf V}}_{{\rm s},l} (t), \hat{{\mathsf V}}_{{\rm s'},l'}^{\dagger}(t') ] =\delta_{{\rm s},{\rm s'}} \, \delta_{l,l'}  \, \delta (t-t') \label{vacuum}  \, ,
\end{align}
where s, s' = t (through), i (intrinsic), or d (drop). The vacuum fluctuations, which are necessary to avoid a violation of the Heisenberg uncertainty principle, have following correlation properties
\begin{align}
& \left< \hat{{\mathsf V}}_{{\rm s},l} (t) \hat{{\mathsf V}}_{{\rm s'},l'}^\dagger  (t') \right>
= \delta_{{\rm s},{\rm s'}} \, \delta_{l,l'} \, \delta (t - t') \label{time_corr_vacuum1} \\
& \left< \hat{{\mathsf V}}_{{\rm s},l}^\dagger (t) \hat{{\mathsf V}}_{{\rm s'},l'} (t') \right>
=0 \, ,\label{time_corr_vacuum2} \\            
& \left< \hat{{\mathsf V}}_{{\rm s},l} (t) \hat{{\mathsf V}}_{{\rm s'},l'} (t') \right>
= \left< \hat{{\mathsf V}}_{{\rm s},l}^\dagger (t) \hat{{\mathsf V}}_{{\rm s'},l'}^\dagger (t') \right> 
=0 \, . \label{time_corr_vacuum3}            
\end{align}
The pumping field is now defined as a coherent state 
\begin{eqnarray}
\hat{{\mathsf A}}_{\rm in} =  {{A}}_{\rm in} + \hat{{\mathsf V}}_{{\rm t},0} \, ,
\label{def_hatAin}
\end{eqnarray}
which is the sum a semi-classical contribution ${{A}}_{\rm in}$ (this is a shorthand for ${{A}}_{\rm in} \hat{\mathbbm{1}}$, 
where $\hat{\mathbbm{1}}$ is the identity operator), and a vacuum fluctuation that will be inserted in the through port.
Its commutation rules is therefore
\begin{align}
[\hat{{\mathsf A}}_{\rm in}(t), \hat{{\mathsf A}}_{\rm in}^{\dagger}(t') ] =  
[\hat{{\mathsf V}}_{{\rm t},0}(t), \hat{{\mathsf V}}_{{\rm t},0}^{\dagger}(t') ] = \delta (t-t') \, ,
\label{commutation_rules_pump}
\end{align}
and it then, has the same quantum-noise properties as a vacuum fluctuation. \\

The canonical quantization can be now be performed by transforming the semi-classical 
Eqs.~(\ref{modal_model_new}), (\ref{Aout_addthrough}) and (\ref{Aout_adddrop})
into their quantum counterparts. 

Let us first introduce the following notation for the sake of conciseness:
\begin{eqnarray}
2\kappa_{{\rm i}} & \equiv & \Delta \omega_{\rm int} \\
2\kappa_{{\rm d}} & \equiv & \Delta \omega_{\rm ext,d} \\
2\kappa_{{\rm t}} & \equiv & \Delta \omega_{\rm ext,t} \\
2\kappa & \equiv & \Delta \omega_{\rm tot} \, .
\label{kappa_notation}
\end{eqnarray}
For the add-through configuration, the quantum model explicitly reads
\begin{align}
 \dot{\hat{{\mathsf a}}}_l =& -\kappa \, {\hat{{\mathsf a}}}_l 
                              +i \left[ \sigma - \frac{1}{2} \zeta_2 l^2 \right] \, {\hat{{\mathsf a}}}_l   
                               + \delta(l) \, \sqrt{2 \kappa_{\rm t}} \, {{A}}_{\rm in} \nonumber \\
  {\,}                      &+ i g_0 \sum_{m,n,p} \delta(m-n+p-l) \, {\hat{{\mathsf a}}}_n^\dagger  {\hat{{\mathsf a}}}_m 
                                                       {\hat{{\mathsf a}}}_p  \nonumber \\
  {\,}                       &  + \sqrt{2 \kappa_{\rm t}} \,  \hat{{\mathsf V}}_{{\rm t},l}                 
                                + \sqrt{2 \kappa_{\rm i}}  \, \hat{{\mathsf V}}_{{\rm i},l}                 
\label{quantum_model_add_through}
\end{align}
with  
\begin{align}
& \kappa = \kappa_{\rm t}+\kappa_{\rm i} \label{quantum_model_add_through_losses} \\
& {\hat{{\mathsf A}}}_{{\rm out},l} = \sqrt{2 \kappa_{\rm t}} \, {\hat{{\mathsf a}}}_l - {{A}}_{\rm in}  \delta(l)  
                                   -\hat{{\mathsf V}}_{{\rm t},l}\, .                
\label{quantum_model_add_through_output}
\end{align}
On the other hand, for the add-drop configuration, the quantum model is 
\begin{align}
 \dot{\hat{{\mathsf a}}}_l =& -\kappa \, {\hat{{\mathsf a}}}_l 
                              +i \left[ \sigma - \frac{1}{2} \zeta_2 l^2 \right] \, {\hat{{\mathsf a}}}_l  
                               + \delta(l) \, \sqrt{2 \kappa_{\rm t}}  \, {{A}}_{\rm in}  \nonumber \\
  {\,}                      &+ i g_0 \sum_{m,n,p} \delta(m-n+p-l) \, {\hat{{\mathsf a}}}_n^\dagger  {\hat{{\mathsf a}}}_m 
                                                       {\hat{{\mathsf a}}}_p  \nonumber \\
  {\,}                       &  + \sqrt{2 \kappa_{\rm t}}  \, \hat{{\mathsf V}}_{{\rm t},l} 
                                + \sqrt{2 \kappa_{\rm i}}  \, \hat{{\mathsf V}}_{{\rm i},l} 
                                + \sqrt{2 \kappa_{\rm d}}  \, \hat{{\mathsf V}}_{{\rm d},l} \, .               
\label{quantum_model_add_drop}
\end{align}
where the losses and the output field operator obey 
\begin{align}
& \kappa = \kappa_{\rm t}+\kappa_{\rm i}+\kappa_{\rm d} \label{quantum_model_add_drop_losses} \\
& {\hat{{\mathsf A}}}_{{\rm out},l} = \sqrt{2 \kappa_{\rm d}}  \, {\hat{{\mathsf a}}}_l - \hat{{\mathsf V}}_{{\rm d},l} \, .               
\label{quantum_model_add_drop_output}
\end{align}
Note that because of the normal ordering, the creation operator in the nonlinear interaction terms is always placed on the left. Also, in the canonical quantization procedure, the pump fields ${{A}}_{\rm in}$ have not been explicitly replaced by
the operator ${\hat{{\mathsf A}}}_{{\rm in}}$, since the related vacuum fluctuation $\sqrt{2 \kappa_{\rm t}}  \, \hat{{\mathsf V}}_{{\rm t},0}$ is already accounted for in the generic term $\sqrt{2 \kappa_{\rm t}}  \, \hat{{\mathsf V}}_{{\rm t},l}$.

\subsection{Hamiltonian formalism}
\label{Hamiltonian_formulation}

The theoretical understanding of the quantum properties of Kerr optical frequency combs 
can also be achieved through an Hamiltonian formalism, and in our case, 
the total Hamiltonian of the system has three contribution.

The first contribution corresponds to the propagation of the fields, following
\begin{align}
{\hat{{\mathsf H}}}_{\rm free} 
& =  \hbar \int_{-\pi}^{+\pi} 
{\hat{{\mathsf a}}}^\dagger \,
\left[ \sigma + \frac{1}{2} \zeta_2  \frac{\partial}{\partial \theta^2 } \right]
\, {\hat{{\mathsf a}}} \, \frac{d \theta}{2 \pi}  \nonumber \\
& = \hbar  \sum_{l} \left[ \sigma - \frac{1}{2} \zeta_2  l^2 \right] 
{\hat{{\mathsf a}}}_l^\dagger {\hat{{\mathsf a}}}_l     \, .    
\label{def_Hfree}
\end{align}
The second contribution originates from the external pump field, and reads
\begin{align}
{\hat{{\mathsf H}}}_{\rm pump} 
= i \hbar \sqrt{2\kappa_{\rm t}} \, {A}_{\rm in} \, \left({\hat{{\mathsf a}}}_0^\dagger -{\hat{\mathsf a}}_0 \right)  \, . 
\label{def_Hext}
\end{align}
The third and last contribution comes from the interactions related to the Kerr nonlinearity:
\begin{align}
{\hat{{\mathsf H}}}_{\rm Kerr} 
& = - \frac{1}{2} \hbar g_0 \int_{-\pi}^{+\pi} 
\left( {\hat{{\mathsf a}}}^\dagger \right)^2
\left( {\hat{{\mathsf a}}}         \right)^2 \, \frac{d \theta}{2 \pi} \label{def_Hint} \\
& =-\frac{1}{2} \hbar g_0 \sum_{m,n,p,q} \delta(m-n+p-q) \, {\hat{{\mathsf a}}}_n^\dagger {\hat{{\mathsf a}}}_q^\dagger 
                                                             {\hat{{\mathsf a}}}_m {\hat{{\mathsf a}}}_p  \, . \nonumber      
\end{align}
For the physical understanding of the quantum phenomena in Kerr media, it is sometimes useful to decompose the interaction Hamiltonian itself into three contributions following
\begin{align}
{\hat{{\mathsf H}}}_{\rm Kerr}={\hat{{\mathsf H}}}_{_{\rm SPM}}+{\hat{{\mathsf H}}}_{_{\rm CPM}}+{\hat{{\mathsf H}}}_{_{\rm FWM}} \, , 
\label{decomp_Hint}
\end{align}
where 
\begin{align}
{\hat{{\mathsf H}}}_{_{\rm SPM}}= -\frac{1}{2} \hbar g_0 \sum_{m}  
                                \, \left( \hat{{\mathsf a}}_m^\dagger \right)^2 \left( \hat{{\mathsf a}}_m \right)^2  
\label{def_HSPM}
\end{align}
is the self-phase modulation (SPM) contribution (a single mode is involved in the interaction), 
\begin{align}
{\hat{{\mathsf H}}}_{_{\rm CPM}}= -2 \hbar g_0 \sum_{m < n}  
                                \, \hat{{\mathsf a}}_m^\dagger \hat{{\mathsf a}}_n^\dagger 
                                   \hat{{\mathsf a}}_m         \hat{{\mathsf a}}_n   
\label{def_HCPM}
\end{align}
is the cross-phase modulation (CPM) contribution (two distinct modes are involved), 
while the four-wave mixing (FWM) term ${\hat{{\mathsf H}}}_{_{\rm FWM}}$ gathers all the remaining monomials of  ${\hat{{\mathsf H}}}_{\rm Kerr}$, which necessarily involve three or four distinct interacting modes.

The total Hamiltonian is therefore 
\begin{align}
{\hat{{\mathsf H}}}_{\rm tot}=  {\hat{{\mathsf H}}}_{\rm free}
                               +{\hat{{\mathsf H}}}_{\rm pump} 
                               +{\hat{{\mathsf H}}}_{\rm Kerr}\, , 
\label{def_Htot}
\end{align}
and it is interesting to note that this Hamiltonian can be very large for Kerr combs.
In earlier studies related to quantum correlations in systems ruled by the LLE, 
the Hamiltonian was always truncated to a maximum of few tens of monomials.
However, in our case, if we consider a comb with $l=-K, \dots, K$ (that is, a comb with $2K+1$ modes), 
then the interaction Hamiltonian ${\hat{{\mathsf H}}}_{\rm Kerr}$ has exactly 
$\frac{1}{3}[2(2K+1)^3 + (2K+1)]$ monomials: this number therefore grows in a cubic polynomial fashion with the 
number of modes, and for a comb with $ \sim 100$~modes, there is already $\sim 10^6$ monomials in 
the Hamiltonian.

The Hamiltonian  ${\hat{{\mathsf H}}}_{\rm tot}$ can now be used to track the temporal dynamics of the quantum Kerr comb, as it permits to obtain an explicit equation for the annihilation operator ${\hat{{\mathsf a}}}_l$ following
\begin{align}
 \dot{\hat{{\mathsf a}}}_l =&  \frac{1}{i \hbar} [{\hat{{\mathsf a}}}_l, {\hat{{\mathsf H}}}_{\rm tot}]   
                             + \sum_\textrm{s} \left[-\kappa_{\rm s}  {\hat{{\mathsf a}}}_l + \sqrt{2 \kappa_{\rm s}}  \, \hat{{\mathsf V}}_{{\rm s},l} \right]   \, .               
\label{Heisenberg_picture}
\end{align}
where the index \textrm{s} runs across the various loss terms corresponding to the configuration under study, that is
\begin{align}
\textrm{s}= 
\left\{
\begin{array}{ll}
{\rm t, i}                    & \rm{for}\,\,\,\rm{add}-\rm{through} \\
{\rm t, i, d}                 & \rm{for}\,\,\,\rm{add}-\rm{drop}
\end{array}
\right. \, , 
\label{def_s}
\end{align}
The term $\kappa = \sum_\textrm{s} \kappa_\textrm{s}$ stands for the total losses [see Eqs.~(\ref{d_omegatot_AT}) 
and~(\ref{d_omegatot_AD})], 
and $\hat{{\mathsf V}}_{{\rm s},l}$ represent the vacuum fluctuations corresponding to these losses.
On the other hand, the output field is
\begin{align}
{\hat{{\mathsf A}}}_{{\rm out},l} =   \sqrt{2 \kappa_{\rm r}} \, {\hat{{\mathsf a}}}_l 
                                    -  {{A}}_{\rm in} \delta_{\rm t,r} \delta(l)  
                                      -\hat{{\mathsf V}}_{{\rm r},l}\, .                
\label{qmodel_output_field}
\end{align}
where the index ${\rm r}$ stands for the output port following
\begin{align}
\textrm{r}= 
\left\{
\begin{array}{ll}
{\rm t}                    & \rm{for}\,\,\,\rm{add}-\rm{through}  \\
{\rm d}                    & \rm{for}\,\,\,\rm{add}-\rm{drop}
\end{array}
\right. \, . 
\label{def_r}
\end{align}
Equation~(\ref{Heisenberg_picture}) is  identical to Eqs.~(\ref{quantum_model_add_through}) and~(\ref{quantum_model_add_drop}), and 
the output field operators defined in Eq.~(\ref{qmodel_output_field}) in the add-through and add-drop configurations obey the same relationships as in Sec.~\ref{canonical_quantization}. 
The commutator $[{\hat{{\mathsf a}}}_l, {\hat{{\mathsf H}}}_{\rm tot}]$ generates exactly 
$3 K^2 + 3 K - l^2 +1$ monomials, and accordingly, Eq.~(\ref{Heisenberg_picture}) includes a large number of terms as well. 
We also note that this formalism is close to the one adopted by Matsko {\em et al.} to investigate the temporal dynamics of Kerr combs in the deterministic regime, that is, when all the vacuum noise terms are uniformly set to zero~\cite{Matsko_Normal_Comb}.

Another approach is to study the following Master Equation~\cite{Lugiato_Castelli}:
\begin{align}
\dot{\hat{\rho}}  = \sum_l \Lambda_l  {\hat{\rho}} - \frac{1}{i \hbar} [{\hat{\rho}}, {\hat{{\mathsf H}}}_{\rm tot}] \, . 
\label{shrodinger_picture}
\end{align}
where ${\hat{\rho}}$ is the density operator for the comb, and $\Lambda_l $ is a Liouvillian explicitly defined as
\begin{align}
\Lambda_l  =  [{\hat{\mathsf a}}_l {\hat{\rho}}, {\hat{\mathsf a}}_l^\dagger] +  [{\hat{\mathsf a}}_l , {\hat{\rho}} {\hat{\mathsf a}}_l^\dagger ] \, . 
\label{Liouvillian}
\end{align}
In this article, we will however only consider the Hamiltonian in the context of Eq.~(\ref{Heisenberg_picture}), which yields a set of equations that are formally identical to those obtained through the canonical quantization in Sec.~\ref{canonical_quantization}.

\subsection{Spatiotemporal formalism}
\label{spatiotemporal_formulation}

The quantum form of the spatio-temporal LLE for Kerr comb generation is
\begin{eqnarray}
\frac{\partial }{\partial t} \, \hat{{\mathsf a}}
   &=&  -(\kappa - i\sigma )\, \hat{{\mathsf a}} + i g_0 \, \hat{{\mathsf a}}^\dagger \hat{{\mathsf a}}^2 
      + i \frac{\zeta_2}{2} \frac{\partial^2 }{\partial \theta^2} \, \hat{{\mathsf a}}
                                          + \sqrt{2\kappa_{\rm t}} \, {A}_{\rm in}  \nonumber \\
   && +  \sum_\textrm{s}  \sqrt{2 \kappa_{\rm s}}  \, \hat{{\mathsf V}}_{{\rm s}}(\theta, t)                                         
\label{eq:quantum_LLE}
\end{eqnarray}
where $\hat{{\mathsf a}}(\theta, t) = \sum_l {\hat{{\mathsf a}}}_l (t) \, e^{i l \theta} $ is the total intra-cavity annihilation operator.
The quantum equation in the case where higher-order dispersion is accounted for 
is straightforwardly obtained by replacing 
$({\zeta_2}/{2}) \partial^2 \hat{{\mathsf a}}/\partial {\theta}^2 $ by
$v_g \sum_{k=2}^{k_{\rm max}} (i\Omega_{_{\rm FSR}})^k ({\beta_k}/{k !}) \partial^k \hat{{\mathsf a}}/\partial {\theta}^k$.
The multimode vacuum fluctuation operators are analogously defined as $\hat{{\mathsf V}}_{{\rm s}} (\theta, t) = \sum_l \hat{{\mathsf V}}_{{\rm s},l}(t)  \, e^{i l \theta}$, and the output field annihilation operator reads ${\hat{{\mathsf A}}}_{{\rm out}}(\theta, t) = \sum_l {\hat{{\mathsf A}}}_{{\rm out},l} (t) \, e^{i l \theta} $. Quantum versions of the LLE for other physical systems have previously been investigated by several researchers in one and two transverse spatial dimensions (see for example 
refs.~\cite{Lugiato_Castelli,Zambrini_PRA,Grynberg_Lugiato,Gatti_Mancini,Hoyuelos_pumpmeter_PRA}).

\section{System under threshold: Spontaneous four-wave mixing}
\label{Spontaneous_FWM}

When the system is pumped under threshold (this is always the case when $P < P_{\rm min}$), only the pumped mode $l=0$ is excited from the semi-classical standpoint, that is, ${\cal A}_0 \neq 0$ and ${\cal A}_l \equiv 0 $ for $l \neq 0$ .
However, from a quantum perspective, there are quantum fluctuations in all modes, which are allowing for the spontaneous photonic interaction $2\hbar \omega_{0} \rightarrow \hbar \omega_{l} + \hbar \omega_{-l}$.
The objective of this section is to determine the power spectra of all the sidemodes and their eventual correlations as a function of pump power, dispersion, detuning and nonlinear gain. 
In the scientifc literature, the topic of quantum dynamics of nonlinear optical systems pumped under threshold has been the focus of several research works, essentially in the context of parametric down conversion~\cite{Lugiato_spatial_PRL,Gatti_quantum_images_PRA,Lugiato_Marzoli,Gatti_Langevin_OPO_PRA,Gatti_Multiphoton_PRA,Zambrini_Polar_PRA,Fabre_J_Phys}.
or for spontaneous FWM~\cite{Garcia_Ferrer_JQE,Brainis_PRA}. A convenient method to determine  consists in establishing the linearized time-domain equation for the quantum fluctuations, and then calculate their Fourier spectra.

\subsection{Quantum Langevin equations}
\label{Spontaneous_FWM_Langevin}

In order to understand the effect of these quantum fluctuations, let us consider that under threshold, the annihilation operator in the various modes of the resonator can be explicitly rewritten as  
\begin{align}
\hat{{\mathsf a}}_l= 
\left\{
\begin{array}{ll}
{\cal A}_0 + \delta \hat{{\mathsf a}}_0         & {\rm{for}} \,\,\, l = 0 \\
\delta \hat{{\mathsf a}}_l                      & {\rm{for}} \,\,\, l \neq 0
\end{array}
\right. \, , 
\label{def_annihilation_ops_SpFWM}
\end{align}
where the operators $\delta \hat{{\mathsf a}}_l$ stand for the quantum fluctuations in a given mode $l \in \{-K, \dots, K \}$.
By inserting Eq.~(\ref{def_annihilation_ops_SpFWM}) into Eq.~(\ref{Heisenberg_picture}), it appears that the quantum dynamics of the system is decomposed under the form of a nonlinear algebraic equation
\begin{eqnarray}
 ( -\kappa +i  \sigma) \, {{\mathcal A}}_0  
+ \sqrt{2 \kappa_{\rm t}} \, {{A}}_{\rm in} 
+ i g_0 | {{\mathcal A}}_0|^2  {{\mathcal A}}_0 =0  \, ,               
\label{steadystate_SpFWM}
\end{eqnarray}
for the central mode $l=0$, while we have the set of $2K$ differential equations
\begin{align}
\delta \dot{\hat{{\mathsf a}}}_l = \mathcal{R}_{l} \, \delta  \hat{{\mathsf a}}_{l} 
                                  +\mathcal{S}_{l} \, \delta \hat{{\mathsf a}}_{-l}^\dagger
                               + \sum_\textrm{s} \sqrt{2 \kappa_{\rm s}} \, {\hat{{\mathsf V}}}_{{\rm s},l} \, ,  
\label{pert_eqs_SpFWM}
\end{align}
for the quantum fluctuations in the sidemodes $\pm l \neq 0$, with 
\begin{eqnarray}
\mathcal{R}_{l} &=& -\left[\kappa -i\left(\sigma -\frac{1}{2} \zeta_2 l^2 \right) \right]
                   + 2i g_0 \, |{{\mathcal A}}_0|^2  \label{Define_Rl_SpFWM} \\
  \mathcal{S}_{l} &=& i g_0 \, {{\mathcal A}}_0^2    \label{Define_Sl_SpFWM}      
\end{eqnarray}
being complex-valued parameters.
Equations~(\ref{pert_eqs_SpFWM}) can be rewritten under the form of $K$ independent sets of $2 \times 2$ quantum-noise driven linear flows, following 
 \begin{eqnarray}
\left[ \begin{array}{l}
	\delta \dot{\hat{{\textsf{a}}}}_{l} \\
    \delta \dot{\hat{{\textsf{a}}}}_{-l}^\dagger
	  \end{array}
\right]
=
 \mathbf{J}_{{{\textsf{a}}},l} \,
\left[ \begin{array}{l}
	\delta {\hat{{\textsf{a}}}}_{ l} \\
    \delta {\hat{{\textsf{a}}}}_{-l}^\dagger
	  \end{array}
\right]
+  \sum_\textrm{s} \sqrt{2 \kappa_{\rm s}} \,
\left[ \begin{array}{l}
	\hat{{\textsf{V}}}_{{\rm s},l}(t) \\
    \hat{{\textsf{V}}}_{{\rm s},-l}^\dagger (t)
	  \end{array}
\right]
\, ,
\label{eq_matrix_pert_SpFWM}
\end{eqnarray}	 
where 
\begin{eqnarray}
\mathbf{J}_{{{\textsf{a}}},l}
=
\left[ \begin{array}{ll}
      \mathcal{R}_l  & \mathcal{S}_l \\
      \mathcal{S}_l^* & \mathcal{R}_l^*
	  \end{array}
\right] 
\label{Jacobian_SpFWM}
\end{eqnarray}	 
is a $2 \times 2$ Jacobian matrix. It is interesting to note that the quantum fluctuations $\delta {\hat{{\textsf{a}}}}_{\pm l}$ are mutually coupled, and are independent from the other modes of order $l' \neq l$.

\subsection{Spontaneous emission spectra}
\label{Spontaneous_FWM_spectra}

In the Fourier domain, we transform the operators as
\begin{align}
{\tilde{{\mathsf X}}} (\omega) = \frac{1}{\sqrt{2\pi}} 
                                \int_{-\infty}^{+\infty} 
                                 {\hat{{\mathsf X}}}(t) \, e^{i \omega t} dt .
\label{def_fourier}
\end{align}
and we find that in the spectral domain, Eq.~(\ref{eq_matrix_pert_SpFWM}) can be rewritten as
\begin{eqnarray}
\left[ \begin{array}{l}
	\delta \tilde{{\textsf{a}}}_{l} (\omega) \\
    \delta \tilde{{\textsf{a}}}_{-l}^\dagger (\omega)
	  \end{array}
\right]
&=& -  [ \mathbf{J}_{{{\textsf{a}}},l} + i \omega  \mathbf{I}_2]^{-1} \nonumber \\
&&  \times \sum_s \sqrt{2 \kappa_{\rm s}}
\left[ \begin{array}{l}
	{\tilde{{\textsf{V}}}}_{{\rm s}, l}(\omega) \\
    {\tilde{{\textsf{V}}}}_{{\rm s},-l}^\dagger(\omega)
	  \end{array}
\right]  \, ,
\label{ops_a_fourier_SpFWM}
\end{eqnarray}	 
where $\mathbf{I}_2$ is the $2 \times 2$ identity matrix. 
Using Eq.~(\ref{qmodel_output_field}), it is easy to find that the output annihilation and creation operators obey
\begin{eqnarray}
\left[ \begin{array}{l}
	\delta \tilde{{\textsf{A}}}_{{\rm out},l} (\omega) \\
    \delta \tilde{{\textsf{A}}}_{{\rm out},-l}^\dagger (\omega)
	  \end{array}
\right]
&=& -  [ \mathbf{J}_{{{\textsf{a}}},l} + i \omega  \mathbf{I}_2]^{-1} \nonumber \\
&&  \times \sum_s \sqrt{4 \kappa_{\rm r} \kappa_{\rm s}}
\left[ \begin{array}{l}
	{\tilde{{\textsf{V}}}}_{{\rm s}, l}(\omega) \\
    {\tilde{{\textsf{V}}}}_{{\rm s},-l}^\dagger(\omega)
	  \end{array}
\right]  \nonumber \\
&& - 
\left[ \begin{array}{l}
	{\tilde{{\textsf{V}}}}_{{\rm r}, l}(\omega) \\
    {\tilde{{\textsf{V}}}}_{{\rm r},-l}^\dagger(\omega)
	  \end{array}
\right] \, .
\label{output_ops_a_fourier_SpFWM}
\end{eqnarray}	 
Using the following correlation properties of the vacuum fluctuations in the Fourier domain 
\begin{align}
& \left< \tilde{{\mathsf V}}_{{\rm s},l} (\omega) \tilde{{\mathsf V}}_{{\rm s'},l'}^\dagger  (\omega') \right>
= \delta_{{\rm s},{\rm s'}} \, \delta_{l,l'} \, \delta (\omega - \omega') \label{fourier_spec_commut_vacuum1} \\
& \left< \tilde{{\mathsf V}}_{{\rm s},l}^\dagger (\omega) \tilde{{\mathsf V}}_{{\rm s'},l'} (\omega') \right>
=0 \, ,\label{fourier_spec_commut_vacuum2} \\            
& \left< \tilde{{\mathsf V}}_{{\rm s},l} (\omega) \tilde{{\mathsf V}}_{{\rm s'},l'} (\omega') \right>
= \left< \tilde{{\mathsf V}}_{{\rm s},l}^\dagger (\omega) \tilde{{\mathsf V}}_{{\rm s'},l'}^\dagger (\omega') \right> 
=0 \, ,\label{fourier_spec_commut_vacuum3}            
\end{align}
together with Eq.~(\ref{output_ops_a_fourier_SpFWM}),
the spectral density of the output photon flux in the sidemodes $\pm l$ can be explicitly calculated as 
\begin{eqnarray}
S_{{\rm sp}, l} (\omega)&=& \left< \delta {\tilde{{\mathsf A}}}_{{\rm out},\pm l}^\dagger (\omega)
                                   \delta {\tilde{{\mathsf A}}}_{{\rm out},\pm l}(\omega) \right> 
                                   \label{fourier_spec_SpFWM} \\
           &=& 4 \rho \kappa^2 \, \frac{g_0^2 |{\mathcal A}_0|^4}{[ \kappa^2 - g_0^2 |{\mathcal A}_0|^4 +  \xi_{l}^2 - \omega^2 ]^2 + 4 \kappa^2 \omega^2 }   \, .              \nonumber
\end{eqnarray}	 
where 
\begin{eqnarray}
\xi_{l}= \Im [\mathcal{R}_l] = \sigma - \frac{1}{2} \zeta_2 l^2 + 2 g_0 |{\mathcal A}_0|^2
\label{def_xi_l}
\end{eqnarray}	 
is  the overall shift induced by laser detuning, group-velocity dispersion and self-phase modulation for a given mode $l$, 
while the parameter $\rho \in \, [0,1[$ is defined as 
\begin{align}
\rho  &= \frac{\kappa_{\rm r} }{\kappa}   \label{def_rho}  \\
      &= 
\left\{
\begin{array}{ll}
\kappa_{\rm t} /(\kappa_{\rm t} +\kappa_{\rm i})                     & \rm{for}\,\,\,\rm{add}-\rm{through}\\
\kappa_{\rm d} /(\kappa_{\rm t} +\kappa_{\rm i}+\kappa_{\rm d} )     & \rm{for}\,\,\,\rm{add}-\rm{drop}
\end{array}
\right. \, . \nonumber
\end{align}
The parameter $\rho$ is the ratio between out-coupling and total losses, and can therefore be interpreted as the ratio between the number of detected photons versus the total number of annihilated photons~\cite{Fabre_J_Phys}. 
The best performance for spontaneous FWM  is achieved for $\rho \rightarrow 1$, which physically corresponds to strong over-coupling in the detection port, that is, to $\kappa_{\rm t} \gg \kappa_{\rm i}$ in the add-through configuration, and 
to $\kappa_{\rm d} \gg \kappa_{\rm t},\kappa_{\rm i}$ in the add-drop configuration. 
Therefore, ultra-low loss resonators are the most perfectly suitable for the purpose of spontaneous FWM, as $\rho$ is anyway maximized when $\kappa_{\rm i} \rightarrow 0$ (or $Q_{\rm int} \rightarrow + \infty$).

\begin{figure}
\begin{center}
\includegraphics[width=7cm]{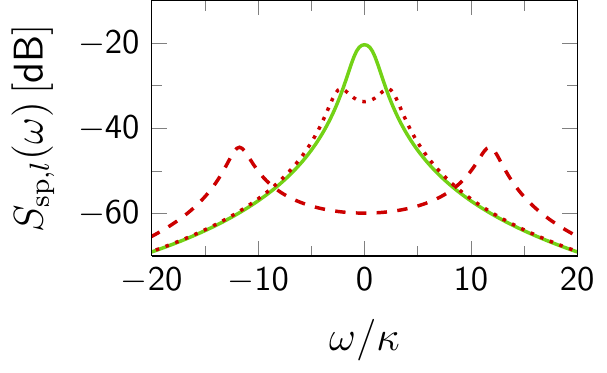}
\end{center}
\caption[Fig_lineshape]
{\label{Fig_lineshape} 
(Color online) Lineshape $S_{{\rm sp},l}(\omega)$ of various sidemodes with different $l$ values when populated by 
spontaneous FWM, as defined in Eq.~(\ref{fourier_spec_SpFWM}).
The parameters are set to $\sigma = \frac{1}{2} \kappa$, $\rho=0.5$, $g_0 |{\mathcal A}_0|^2 = \kappa/10$ and $\zeta_2 = \kappa/100$. Note that these parameters correspond to those of Fig.~\ref{Fig_envelope_sp_spectra}(b).
Continuous green: $l=\pm 1$, the lineshape is single-peaked.
Dotted red: $l=\pm 25$, the lineshape is doubled-peaked.
Dashed red: $l=\pm 50$, the lineshape is still doubled-peaked, and the separation between the peaks is wider.
}
\end{figure}

Equation~(\ref{fourier_spec_SpFWM}) defines the lineshape of the sidemode spectra, when populated by spontaneous FWM. 
Since the spectra can be rewritten as
$S_{{\rm sp}, l} (\omega) =4 \rho \kappa^2 g_0^2 |{\mathcal A}_0|^4/ |{\cal D}_l(\omega)|^2$
with
\begin{eqnarray}
{\cal D}_l(\omega) = [ \kappa^2 - g_0^2 |{\mathcal A}_0|^4 +  \xi_{l}^2 - \omega^2 ] - 2 i \kappa \omega \, ,
\label{denom_SFWM_l}
\end{eqnarray}	   
it is easy to demonstrate that the lineshapes of $S_{{\rm sp}, l} (\omega)$ is either single- or double-peaked, depending on if the bi-quadratic polynomial $|{\cal D}_l(\omega)|^2$ has one or two minima, respectively.
The spectra are thereby found to be single-peaked when 
\begin{eqnarray}
\xi_l^2 \leq \kappa^2 + g_0 |{\mathcal A}_0|^2  \, ,
\label{single_peak_SFWM_l}
\end{eqnarray}	   
and double-peaked otherwise. In other words, single-peaked lineshapes correspond to a small overall detuning $|\xi_l|$, while double-peaked ones indicate large overall detunings. A direct consequence is that the sidemodes are always double-peaked in the asymptotic limit $l \rightarrow \pm \infty$. These two typical lineshape profiles are displayed in
Fig.~\ref{Fig_lineshape}.

\begin{figure}
\begin{center}
\includegraphics[width=8.5cm]{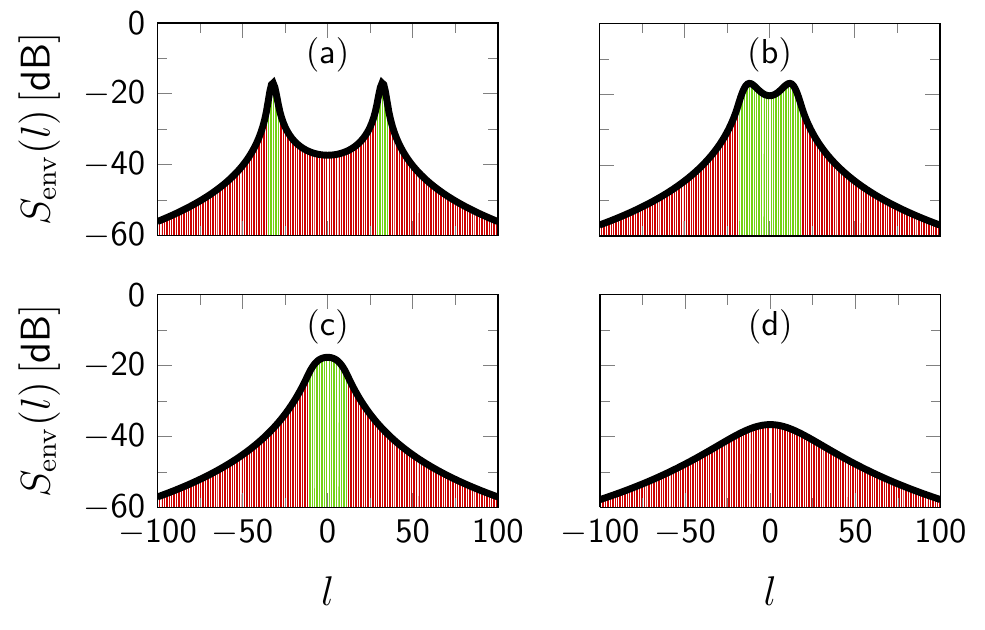}
\end{center}
\caption[Fig_envelope_sp_spectra]
{\label{Fig_envelope_sp_spectra} 
(Color online) Spontaneous FWM (parametric fluorescence) spectra for various values of the laser frequency detuning
$\sigma$. 
The green lines are single-peaked sidemodes, while the red lines are double-peaked sidemodes (see Fig.~\ref{Fig_lineshape}). 
The thick black line is the envelope $S_{{\rm env}}(l)$ of the spectrum as defined in Eq.~(\ref{S_enveloppe}).
The parameters are set to $\rho=0.5$, $g_0 |{\mathcal A}_0|^2 = \kappa/10$ and $\zeta_2 = \kappa/100$.
(a)~$\sigma= 5 \kappa$: the envelope $S_{{\rm env}}(l)$ has two maxima located around $l= \pm 32$ as predicted by Eq.~(\ref{peak_enveloppe}).
(b)~$\sigma= \frac{1}{2}\kappa$: the envelope still has two maxima, located around $l= \pm 12$.
(c)~$\sigma= -\frac{1}{2} \kappa$: the envelope only has one maximum.
(d)~$\sigma= -5 \kappa$: there is only one maximum and all the modes are double-peaked.
}
\end{figure}

From Eq.~(\ref{fourier_spec_SpFWM}), it is possible to define the enveloppe of the spontaneous emission spectrum, which is defined as the continuous line linking the sidemode peaks in the spectral domain.
These maxima are located at the frequency $\omega_{\rm m}=0$ for single-peaked sidemodes (SPS), and at 
$\omega_{\rm m} = \pm [\xi_l^2 - \kappa^2 - g_0 |{\mathcal A}_0|^2 ]^\frac{1}{2}$ for the double-peaked sidemodes (DPS). 
Inserting these frequency values in Eq.~(\ref{fourier_spec_SpFWM}) yields the following envelope
\begin{align}
S_{{\rm env}}(l)  =
\left\{
\begin{array}{ll}
4 \rho  \, \frac{g_0^2 |{\mathcal A}_0|^4 \kappa^2}{[ \kappa^2 - g_0^2 |{\mathcal A}_0|^4 +  \xi_{l}^2]^2 }                     & \rm{for}\,\,\,\rm{SPS} \\
\rho  \, \frac{g_0^2 |{\mathcal A}_0|^4}{ \xi_{l}^2 - g_0^2 |{\mathcal A}_0|^4 }                     
& \rm{for}\,\,\,\rm{DPS}
\end{array}
\right. \, . 
\label{S_enveloppe}
\end{align}
There are therefore two types of enveloppe $S_{{\rm env}}(l)$.
The first kind has two maxima located at the nearest integer approximation of 
\begin{eqnarray}
l \simeq \pm \sqrt{\frac{2}{\zeta_2} (\sigma + 2 g_0 |{\mathcal A}_0|^2)}  
\label{peak_enveloppe}
\end{eqnarray}	   
when $2 ( \sigma + 2 g_0 |{\mathcal A}_0|^2)/\zeta_2 >0$ (this condition can only be fulfilled for single-peaked sidemodes). 
The second kind corresponds the case where $2 ( \sigma + 2 g_0 |{\mathcal A}_0|^2)/\zeta_2 \geq 0$, 
and it yields an envelope that with a single maximum located around the pump frequency ($l=0$). 
The various types of envelopes are displayed in Fig.~\ref{Fig_envelope_sp_spectra}, where is can be seen that when the pumping is resonant ($\sigma < \kappa$), the spectrum configuration is such that there are single-peaked sidemodes around the pump, and double-peakes ones at the edges of teh spectrum. However, for strong detuning, we have either the case where there are no single-peaked sidemodes at all (the envelope only has one maximum), or the one where single- and double-peaked lineshapes alternate as the sidemode order is varies (with single-mode lineshapes located around the two maxima of the envelope).

However, the spectrum  $S_{{\rm sp}, l} (\omega)$ diverges when the denominator function $|{\cal D}_l(\omega)|^2 \rightarrow 0$. In particular, such a divergence is observed when the three following conditions are fulfilled:
\begin{eqnarray}
\omega &=& 0 \label{divergence_S1} \\
g_0 |{\mathcal A}_0|^2 &=& \kappa \implies |{\mathcal A}_0|^2 = |{\mathcal A}_{\rm th}|^2 \label{divergence_S2} \\
\xi_{l} &=&0  \implies \sigma = \frac{1}{2} \zeta_2 l^2 - 2 \kappa  \, . \label{divergence_S3} 
\end{eqnarray}	 
The first condition is an equidistance condition, which indicates that the spontaneous FWM lines are precursors of the stimulated FWM comb that is expected to emerge just above threshold. 
The second equation is the amplitude condition (null gain) which indicates that the FWM is passing from being spontaneous to stimulated. 
The third and last equation is a phase-matching condition. 
However, this unphysical divergence occurs because the linearization procedure fails when the system is pumped close to threshold, since the higher-order contributions are not negligible anymore.
This regime of large quantum fluctuations just below threshold is non-trivial therefore deserves a study of its own.

\subsection{Output photons flux and power of spontaneous emission spectra}
\label{Spontaneous_FWM_flux}

Knowing the spectral power density of the spontaneous FWM spectra, it is possible to calculate analytically the output photon flux $R_{{\rm out},l}$ (or photon production rate, in units of s$^{-1}$) for any mode $l$ using the Parseval theorem following
\begin{eqnarray}
R_{{\rm out},l} = \frac{1}{2 \pi} \int_{-\infty}^{+\infty} S_{{\rm sp}, l} (\omega) \, d \omega \, ,
\label{def_parseval_Rout}
\end{eqnarray}	   
while the output power (in W) for each mode $l$ is simply obtained through 
\begin{eqnarray}
P_{{\rm out},l} = \hbar \omega_{_{\rm L}} \, R_{{\rm out},l}\, ,
\label{def_parseval_Pout}
\end{eqnarray}	   
and the total power emitted in the spontaneous emission spectra is 
\begin{eqnarray}
P_{{\rm out}} =  \hbar \omega_{_{\rm L}}  \sum_{l \neq 0}  R_{{\rm out},l} \, .
\label{def_parseval_P_total}
\end{eqnarray}	   
The explicit calculation of the integral in Eq.~(\ref{def_parseval_Rout}) mathematically leads two different cases, 
depending on if the intra-cavity photon number $|{\mathcal A}_0|^2$ [solution of Eq.~(\ref{steadystate_SpFWM})] is small 
or not with regards to the threshold photon number value $|{\mathcal A}_{\rm th}|^2=\kappa/g_0$.

\subsubsection{Case of weak pumping}
\label{Spontaneous_FWM_flux_weak}

The first case, which is referred to here as the case of weak pumping, mathematically corresponds to 
\begin{eqnarray}
g_0^2 |{\mathcal A}_0|^4 <\kappa^2 + \xi_{l}^2
\label{def_weak_pumping}
\end{eqnarray}	   
and it is particularly important because it physically corresponds to the most widespread experimental configuration.
Effectively, weak pumping permits to avoid parasitic nonlinear (Raman, Brillouin) and thermal effects (such as thermo-optical oscillations, see ref.~\cite{Thermooptic_osc}), thereby allowing for a better control of the spontaneous emission process. 
In this case, the output flux of spontaneously emitted photons is explicitly defined as
\begin{eqnarray}
R_{{\rm out},l} = \rho \kappa \, \frac{g_0^2 |{\mathcal A}_0|^4}{\kappa^2 - g_0^2 |{\mathcal A}_0|^4 +  \xi_{l}^2}  \,
\label{def_parseval_Rout_weak}
\end{eqnarray}	   
However, the nonlinear relationship between $|{\mathcal A}_0|^2$ and $P = \hbar \omega_{_{\rm L}} \, A_{\rm in}^2$ is non-trivial,  as evidenced by the nonlinear equation~(\ref{steadystate_SpFWM}).
This situation impedes a simple quantitative understanding of the interplay between pump power and parametric spontaneous emission. 
This nonlinearity disappears in the asymptotic case of \textit{very} weak pumping ($|{\mathcal A}_0|^2 \ll |{\mathcal A}_{\rm th}|^2 =  \kappa / g_0$), which is the most relevant from a physical standpoint as highlighted above. 
Effectively, when the pump power is extremely small, the intracavity photon number is typically much smaller that the Kerr comb threshold. The nonlinear gain term can therefore be neglected in Eq.~(\ref{steadystate_SpFWM}) and in that case, the intra-cavity photon number in the pumped mode scales with the pump power following 
\begin{eqnarray}
|{\mathcal A}_0|^2 \simeq \frac{2 \kappa_{\rm t}}{\kappa^2 + \sigma^2}\, \frac{P}{\hbar \omega_{_{\rm L}}} \, .
\label{A_to_P_low_pump}
\end{eqnarray}	 
In this very weak pumping regime, the intracavity photon number $|{\mathcal A}_0|^2$ is therefore proportional to the input pump power $P$, and therefore the output photon flux can now be determined as 
\begin{eqnarray}
R_{{\rm out},l} \simeq  \frac{R_{{\rm max}}}{\left[ 1+\left( \frac{\sigma}{\kappa}\right)^2 \right]^2  
                      \left[ 1+\frac{1}{\kappa^2} \left( \sigma - \frac{1}{2} \zeta_2 l^2 \right)^2 \right] } \, ,
\label{def_parseval_Rout_very_weak}
\end{eqnarray}	   
where
\begin{eqnarray}
R_{{\rm max}} &=&  4 \rho  \, \frac{g_0^2 \kappa_{\rm t}^2}{\kappa^5} 
                 \, \left[ \frac{P}{\hbar \omega_{_{\rm L}}} \right]^2 \nonumber \\ 
              &=&  32 \,  \frac{Q_{\rm tot}^6}{Q_{\rm r} Q_{\rm t}^2} \left[ \frac{ \gamma v_g^2 }{2 \pi a}  \right]^2 \frac{P^2}{\omega_{_{\rm L}}^3}     
\label{def_Rmax}
\end{eqnarray}	   
is the maximum photon production rate that can be achieved in a given sidemode. 
As far as orders of magnitude are concerned, if we consider the resonators described in Sec.~\ref{orderofmagnitude},
the maximal photon flux per sidemode is equal $R_{{\rm max}} \sim 10^4$~s$^{-1}$ when the crystalline resonator is pumped with 
$0.1$~mW, or when the integrated ring-resonator is pumped with $1$~mW.

Further simplifications be considered to establish a useful approximation of the sidemode photon flux $R_{{\rm out},l}$, or equivalently,  the sidemode power $P_{{\rm out},l}$. 
For example, in the common case of a ring-resonator of radius $a$ which is resonantly pumped ($\sigma = 0$) and critically  coupled ($\rho = {1}/{2}$) in the add-through configuration  ($\kappa_{\rm t} = \kappa_{\rm i} =\kappa/2$), Eq.~(\ref{def_parseval_Rout_very_weak}) can be simplified and leads to the following formula for the sidemode power
\begin{eqnarray}
P_{{\rm out},l}  & \simeq & \hbar \omega_{_{\rm L}}  \frac{g_0^2}{2 \kappa^3} 
                             \left[ \frac{P}{\hbar \omega_{_{\rm L}}} \right]^2  \nonumber \\
                 & \simeq &  4  \hbar \omega_{_{\rm L}} \,  \left[ \frac{ \gamma v_g^2 }{2 \pi a}  \right]^2  
                             \left[ \frac{Q_{\rm tot}}{\omega_{_{\rm L}}} \right]^3 P^2
\label{def_parseval_Pout_very_weak_simplified}
\end{eqnarray}	   
when dispersion effects are neglected.
It is noteworthy that the above formula exactly corresponds to the one proposed by Azzini~\textit{et al.} 
in ref.~\cite{Azzini_OL}.
It should also be noted that as a general rule of thumb, spontaneous emission is stronger with higher nonlinearity, higher pump power, higher $Q$ factors, and smaller size. 
 
Still in the very weak pumping regime, the total power emitted in the full spontaneous emission spectrum (all the sidemodes) can be calculated using Eq.~(\ref{def_parseval_Rout_very_weak}) and a continuous approximation of the discrete sum of Eq.~(\ref{def_parseval_P_total}), following
\begin{eqnarray}
P_{{\rm out}} & \simeq & \hbar \omega_{_{\rm L}}  \int_{-\infty}^{+\infty}  R_{{\rm out},l} \, dl \\
              & \simeq & \hbar \omega_{_{\rm L}} \, R_{{\rm max}} 
                     \frac{ \pi \sqrt{\frac{\kappa}{|\zeta_2|}}}{
                     \left[ 1+\left( \frac{\sigma}{\kappa}\right)^2 \right]^\frac{9}{4} }
                     \left[1+ \frac{{\rm sgn}(\sigma/\zeta_2)}{\sqrt{1+\left( \frac{\sigma}{\kappa}\right)^2}} \right]^\frac{1}{2} \, , 
                       \nonumber
\label{def_parseval_P_total_integral}
\end{eqnarray}	   
and it appears that spontaneous FWM is globally more effective when $|\zeta_2| \rightarrow 0$ (vanishing dispersion) and $|\sigma| \rightarrow 0$ (resonant pumping). Naturally, in the limit case $\zeta_2 = 0$, other effects such as higher-order dispersion or pump depletion have to be considered in order to prevent the unphysical power divergence.

\subsubsection{Case of strong pumping}
\label{Spontaneous_FWM_flux_strong}

The second case of spontaneous FWM which is refered to as the case of strong pumping corresponds to 
\begin{eqnarray}
g_0^2 |{\mathcal A}_0|^4 > \kappa^2 + \xi_{l}^2 \, .
\label{def_strong_pumping}
\end{eqnarray}	   
This case physically corresponds to the situation  the overall detuning $\xi_{l}$ is very large, so that 
the system remains under threshold even when the pump power (as well as) is very large 
($|{\mathcal A}_0|^2 \sim |{\mathcal A}_{\rm th}|^2$).
The photon flux in a sidemode $l$ is given in that case by
\begin{eqnarray}
R_{{\rm out},l} = \rho \kappa^2 \, \frac{g_0^2 |{\mathcal A}_0|^4}{\sqrt{g_0^2 |{\mathcal A}_0|^4 -\kappa^2} \, 
                [g_0^2 |{\mathcal A}_0|^4 -\kappa^2 -  \xi_{l}^2]}  \, .
\label{def_parseval_Rout_strong}
\end{eqnarray}	   
This case is of strong pumping is scarcely explored experimentally, because as emphasized earlier, the high intracavity power 
triggers many parasitic phenomena.

Note that the limit case $g_0^2 |{\mathcal A}_0|^4 = \kappa^2 + \xi_{l}^2 $ leads to an unphysical divergence that is circumvented by dropping the hypothesis of undepleted pump and pair-wise coupled sidemodes.

\subsection{Quantum correlations and entanglement}
\label{Spontaneous_FWM_correlations}

The correlation of the output annihilation operators can be calculated as
\begin{eqnarray}
{\mathcal C}(\omega)
&=& \left< \delta {\tilde{{\mathsf A}}}_{{\rm out},-l} (\omega)
       \delta {\tilde{{\mathsf A}}}_{{\rm out}, l}(\omega) \right> \label{correl_spec_SpFWM} \\
&=& - \rho  \, \frac{{2 \kappa \mathcal S}_l}{|{\mathcal D}_l(\omega)|^2}  
 \left[{\mathcal D}_l^*(\omega) + 2 \kappa \, ({\mathcal R}_l^* - i \omega) \right]   \, .  \nonumber  
\end{eqnarray}	 
and it appears that it is obviously not null.

It is interesting to note that the dynamical Eqs.~(\ref{pert_eqs_SpFWM}) for the sidemode fields $\pm l$ correspond to a simplified Hamiltonian with the approximation of a strong pump with regards to the sidemodes, that is, $\langle \hat{{\mathsf n}}_{0} \rangle \gg \langle \hat{{\mathsf n}}_{ \pm l} \rangle$. In that case, the interaction between the pump and the sidemodes $\pm l$ is described by the simplified Hamiltonian
\begin{align}
{\hat{{\mathsf H}}}_l 
& = -\frac{1}{2} \hbar g_0 \,  
  \{ ({\cal A}_0^2)^*  \, {\hat{{\mathsf a}}}_l         {\hat{{\mathsf a}}}_{-l} 
+     {\cal A}_0^2     \, {\hat{{\mathsf a}}}_l^\dagger {\hat{{\mathsf a}}}_{-l}^\dagger \}  \nonumber \\ 
& = i \hbar \,
  \{ \zeta^*  \, {\hat{{\mathsf a}}}_l         {\hat{{\mathsf a}}}_{-l} 
-    \zeta    \, {\hat{{\mathsf a}}}_l^\dagger {\hat{{\mathsf a}}}_{-l}^\dagger \}  \nonumber \\ 
\,
\label{def_Hint_SFWM} 
\end{align}
with $\zeta =  -\frac{1}{2} i g_0  {\cal A}_0^2$.
It is well known that the Hamiltonian $ {\hat{{\mathsf H}}}_{l}$ creates entangled photons in pairs 
following~\cite{Grynberg_Aspect_Fabre_book} 
\begin{align}
|\psi_{|l|} (t) \rangle & =  e^{\left[ {\hat{{\mathsf H}}}_{l}/i \hbar \right] t } \, |0,0 \rangle \\
                        & = \frac{1}{\cosh r} \sum_{n=0}^{+ \infty} (-e^{i \varphi})^n \tanh^n r \, |n,n \rangle 
\label{entangled} 
\end{align}
where $\zeta t = r e^{i \varphi}$. 
Hence, when the system is pumped far below threshold, the main characteristics of the spontaneously emitted photons can be estimated analytically.

\section{System above threshold: quantum correlations and squeezing for the photon numbers}
\label{squeezzphotnumbers}

In the frequency comb corresponding to a stationary pattern like rolls of solitons, the photon number in each semi-classical sidemodes is defined as $|{{\cal A}}_{{\rm out}, \pm l}|^2 = N_{{\rm out},\pm L}$, which is proportional to the optical power that can be photo-detected experimentally for each of these two modes.
We have recalled in Sec.~\ref{spatiotemporal_formalism} that for being symmetrical, both sidemodes have the same amplitude,
their photon numbers are equal and the average intensity difference
$\langle N_{{\rm out}, \Delta} \rangle = \langle N_{{\rm out},l} \rangle - \langle N_{{\rm out},- l} \rangle$ is null in the semi-classical limit.
This result indicates that the quantum operator corresponding to this difference in photon numbers could potentially display 
a noticeably non-classical behavior under optimal conditions.

Here, we show that in a stationary Kerr comb (rolls or solitons), the photon number difference $N_{{\rm out},\Delta} = N_{{\rm out},l} - N_{{\rm out},- l}$ which experimentally corresponds to difference of optical powers photo-detected for the modes $+l$ and $-l$ can under certain conditions display squeezing. In the literature, this phenomenon is sometimes referred to as {two-mode squeezing} because two optical modes are involved in the process, at the opposite of traditional notion of squeezing where a single mode is considered. 
We will show that this two-mode squeezing can be observed not only for rolls close to threshold within a three-modes approximation, but also for any type of stationary Kerr comb, regardless of the number of modes involved and the dispersion regime, and even far above threshold. \\

\subsection{General case of combs with arbitrary number of modes}
\label{generalsqueezzphotnumbers}

Let us  consider the modal photon number operators $\hat{{\mathsf n}}_{l}=\hat{{\mathsf a}}_{l}^{\dagger} \hat{{\mathsf a}}_l$ and $\hat{{\mathsf n}}_{-l}=\hat{{\mathsf a}}_{-l}^{\dagger} \hat{{\mathsf a}}_{-l}$, which
correspond to the modes $+l$ and $-l$, respectively.
From the Heisenberg Eq.~(\ref{Heisenberg_picture}), we can determine the time-domain dynamics of these operators as
\begin{align}
\dot{\hat{{\mathsf n}}}_{\pm l} = & \,\,\, \dot{\hat{{\mathsf a}}}_{\pm l}^\dagger \hat{{\mathsf a}}_{\pm l}
                                        + \hat{{\mathsf a}}_{\pm l}^\dagger \dot{\hat{{\mathsf a}}}_{\pm l}  \nonumber \\
                                 = & - 2 \kappa {\hat{{\mathsf n}}}_{\pm l} 
                               + \frac{1}{i \hbar} [{\hat{{\mathsf n}}}_{\pm l}, {\hat{{\mathsf H}}}_{\rm tot}] \nonumber \\   
                               & + \sum_\textrm{s} \sqrt{2 \kappa_{\rm s}} \, 
                               (\hat{{\mathsf V}}_{{\rm s},_{\pm l}}^\dagger \hat{{\mathsf a}}_{\pm l} 
                                + \hat{{\mathsf a}}_{\pm l}^\dagger \hat{{\mathsf V}}_{{\rm s},_{\pm l}})  \, .               
\label{photon_number_dyn}
\end{align}
with {\rm s = t, i} for the add-through configuration, and {\rm s = t, i, d} for the add-drop configuration.
It can be demonstrated that the photon numbers $\hat{{\mathsf n}}_{\pm l}$ do not commute with the Hamiltonian ${\hat{{\mathsf H}}}_{\rm tot}$. 

We can use Eqs.~(\ref{photon_number_dyn}) to show that the operator 
\begin{align}
{\hat{{\mathsf n}}}_{_\Delta} =  {\hat{{\mathsf n}}}_{+l} - {\hat{{\mathsf n}}}_{-l}
\label{def_photon_number_difference_op}
\end{align}
standing for the photon number difference obeys the following time-domain equation:  
\begin{align}
 \dot{\hat{{\mathsf n}}}_{_\Delta} =    - 2 \kappa {\hat{{\mathsf n}}}_{_\Delta} 
                                     + \frac{1}{i \hbar} [{\hat{{\mathsf n}}}_{_\Delta}, {\hat{{\mathsf H}}}_{\rm tot}]
                                     + \sum_\textrm{s} \sqrt{2 \kappa_{\rm s}}  \, {\hat{{\mathsf G}}}_{{\rm s}}  \, ,               
\label{photon_number_difference_dyn}
\end{align}
where
\begin{align}
\hat{{\mathsf G}}_{{\rm s}} = \hat{{\mathsf V}}_{{\rm s},_{+l}}^\dagger \hat{{\mathsf a}}_{+ l} 
                                + \hat{{\mathsf a}}_{+ l}^\dagger \hat{{\mathsf V}}_{{\rm s},_{+ l}}
                             -\hat{{\mathsf V}}_{{\rm s},_{- l}}^\dagger \hat{{\mathsf a}}_{- l} 
                                - \hat{{\mathsf a}}_{- l}^\dagger \hat{{\mathsf V}}_{{\rm s},_{- l}}  \, .             
\label{def_G}
\end{align}
The expectation values related to $\hat{{\mathsf G}}_{{\rm s}}$ are 
\begin{align}
\langle \hat{{\mathsf G}}_{{\rm s}}(t) \rangle & = 0 \\
\langle \hat{{\mathsf G}}_{{\rm s}}(t)  \hat{{\mathsf G}}_{{\rm s'}}(t') \rangle 
          & =  \langle {\hat{{\mathsf n}}}_{_\Sigma} \rangle \, \delta_{{\rm s, s'}} \, \delta(t-t') \, ,
\label{expectation__G}
\end{align}
where
\begin{align}
{\hat{{\mathsf n}}}_{_\Sigma} =  {\hat{{\mathsf n}}}_{+l} + {\hat{{\mathsf n}}}_{-l}
\label{def_photon_number_sum_op}
\end{align}
is the photon number operator for the sum of the sidemodes $\pm l$~\cite{Gatti_Mancini}.

In the general case, Eq.~(\ref{photon_number_difference_dyn}) ruling the dynamics of the photon number difference is highly  nonlinear. However, it degenerates to a linear Langevin equation when ${\hat{{\mathsf n}}}_{_\Delta}$ commutes with ${\hat{{\mathsf H}}}_{\rm tot}$. In particular, this condition is fulfilled when the photon number individually commutes with ${\hat{{\mathsf H}}}_{\rm free}$, ${\hat{{\mathsf H}}}_{\rm pump}$ and 
${\hat{{\mathsf H}}}_{\rm Kerr}$.

It is not difficult to show that ${\hat{{\mathsf n}}}_{_\Delta}$ commutes with both 
${\hat{{\mathsf H}}}_{\rm free}$ and  ${\hat{{\mathsf H}}}_{\rm pump}$.
However, the determination of the commutator $[{\hat{{\mathsf n}}}_{_\Delta}, {\hat{{\mathsf H}}}_{\rm Kerr}]$ 
is much less trivial.
More explicitly, using the relationships 
\begin{align}
[{\hat{{\mathsf n}}}_{\pm l}, {\hat{{\mathsf H}}}_{\rm Kerr}]
= [\hat{{\mathsf a}}_{\pm l}^\dagger, {\hat{{\mathsf H}}}_{\rm Kerr}] \hat{{\mathsf a}}_{\pm l} 
+ \hat{{\mathsf a}}_{\pm l}^\dagger  [\hat{{\mathsf a}}_{\pm l}, {\hat{{\mathsf H}}}_{\rm Kerr}] \, ,
\label{explicit_commutator_photnumber}
\end{align}
we can derive an explicit expression of the commutator
$[{\hat{{\mathsf n}}}_{_\Delta}, {\hat{{\mathsf H}}}_{\rm Kerr}]$, following
\begin{align}
[{\hat{{\mathsf n}}}_{_\Delta}, {\hat{{\mathsf H}}}_{\rm Kerr}]
=& \,\,  [{\hat{{\mathsf n}}}_{+l}, {\hat{{\mathsf H}}}_{\rm Kerr}] - [{\hat{{\mathsf n}}}_{-l}, {\hat{{\mathsf H}}}_{\rm Kerr}]
 \nonumber \\
=& \,\, \hbar g_0 \sum_{n,p,q} \,{\hat{{\mathsf a}}}_n^\dagger {\hat{{\mathsf a}}}_q^\dagger {\hat{{\mathsf a}}}_p 
 \, \{ \delta(l-n+p-q) \, {\hat{{\mathsf a}}}_{l} \nonumber \\
  &\, -  \delta(l+n-p+q) \, {\hat{{\mathsf a}}}_{-l} \} - \textrm{H. c.}
\label{explicit_commutator}
\end{align}
where {\rm H.~c.} stands for the Hermitian conjugate of all preceding terms.
In fact, by setting $p \equiv q$, it can be shown that $[{\hat{{\mathsf n}}}_{_\Delta}, {\hat{{\mathsf H}}}_{_{\rm SPM}}]$ and $[{\hat{{\mathsf n}}}_{_\Delta}, {\hat{{\mathsf H}}}_{_{\rm CPM}}]$ are both null regardless of the size of the comb.
However, $[{\hat{{\mathsf n}}}_{_\Delta}, {\hat{{\mathsf H}}}_{_{\rm FWM}}]$ is not necessarily null.
This implies that the photon number difference is generally not a conserved quantity.
For example, for a  $5$-modes comb (let's consider $l=-2,\dots,+2$ for the sake of simplicity), the photonic interaction $2\hbar \omega_{-1} \rightarrow \hbar \omega_{0} + \hbar \omega_{-2}$ induces a loss of $2$~photons in the mode $l=-1$ (in favor of the modes $l=0$ and $l=-2$), while its symmetric sidemode counterpart $l=1$ remains unaffected. 
Hence, despite the fact that $\langle N_\Delta \rangle = \langle N_{l} \rangle - \langle N_{- l} \rangle$ is expected to be null (in average) in the semi-classical approximation, the value of $N_\Delta$ itself is not necessarily conserved at the photon level. This phenomenology can be witnessed whenever the size of the comb is strictly larger than~$3$.

However, when the size of the comb is equal to~$3$, the photon number difference \emph{does} commute with
the interaction Hamiltonian, and therefore, is conserved. 
This case corresponds to the problem that was originally investigated by Lugiato and Castelli in ref.~\cite{Lugiato_Castelli}. 
From a physical standpoint, the explanation of this feature is that in a $3$-modes Kerr combs, any variation of photon number in one sidemode \textit{must} induce the very same variation in the other sidemode. Therefore, in this case, the photon number difference itself $N_\Delta = N_{l} - N_{- l}$ (and not only its average value) is \textit{strictly} null in deterministic photon picture. As a consequence, in $3$-modes Kerr combs, non-classical light can be generated in twin-beams, as analyzed in the next sub-section.

\subsection{Particular case of combs with $3$~modes \\
            (pump, signal and idler)}
\label{3modessqueezzphotnumbers}

We aim here to derive the output spectra of the photon-number difference in both the add-through and add-drop configurations, when the Kerr comb is constituted with only $3$~modes. Such combs arise for example in the anomalous dispersion regime just after a super-critical Hamiltonian-Hopf bifurcation ($\sigma > -\frac{41}{30} \, \kappa$, see 
refs.~\cite{LL,PRA_Unified,Parra_Rivas}). The system in that case yields the so-called Turing patterns (or rolls) in the time-domain, and primary combs in the frequency-domain (Kerr combs with multiple-FSR spacing). 
Using the normalized LLE of Eq.~(\ref{eq:LLE}), it has been shown in \cite{PRA_Unified} that the threshold pump power is $F_{\rm th}^2 = 1 + (1-\alpha)^2$, which corresponds in watts to
\begin{align}
P_{\rm th} = P_{\rm min} F_{\rm th}^2 = P_{\rm min} \left[1+ \left(1+ \frac{\sigma}{\kappa} \right)^2 \right] \, ,
\label{def_Pth}
\end{align}
where $P_{\rm min}$ is the absolute minimum power needed for comb generation, and was introduced in Eq.~(\ref{minpowersinwattspump}). 
Above that threshold pump power, a stable roll pattern of order $L$ with  
\begin{align}
L \simeq \sqrt{\frac{2}{\beta} (\alpha - 2)} =\sqrt{\frac{2}{\zeta_2} (\sigma + 2 \kappa)}  
\label{def_L_rolls}
\end{align}
emerges in the $\theta$-domain through modulational instability, and it essentially features $3$~modes $l=0,\pm L$ in the frequency domain (see refs.~\cite{YanneNanPRA,IEEE_PJ,PRA_Unified}). 
At the experimental level, the value of $L$ can be as low as $1$ and as high as $\sim 200$~\cite{Lin_OEng,Lin_OExp}.

\begin{figure}
\begin{center}
\includegraphics[width=7cm]{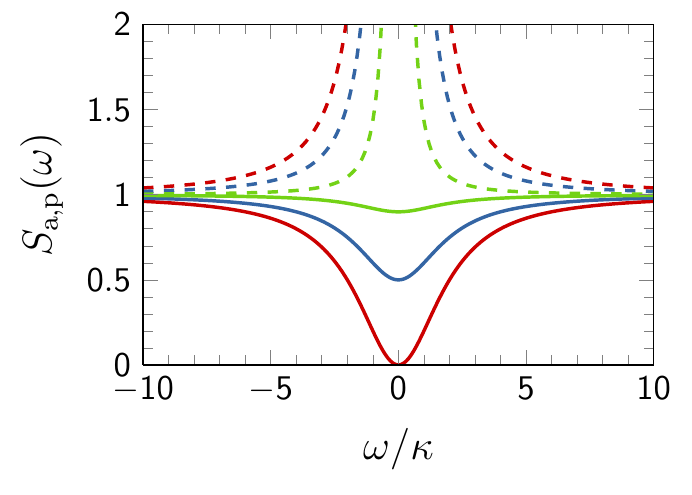}
\end{center}
\caption[Quads_Sa_Sp]
{\label{Quads_Sa_Sp} 
(Color online) Power spectra of pure amplitude and phase quadratures for different values of the squeezing parameter $\rho$. The solid lines correspond at the same time to the photon number difference spectrum of Eq.~(\ref{Power_Fourier_output_difference_number_normalized}), and to the pure amplitude quadrature spectrum of Eq.~(\ref{def_S_a}), since both are identical.
The dashed lines correspond to the phase quadrature spectrum of Eq.~(\ref{def_S_p}).
Green: $\rho = 0.1$; 
Blue:  $\rho = 0.5$; 
Red:   $\rho = 1$.
We have arbitrarily set $\kappa_p \equiv \frac{1}{3} \kappa$. }
\end{figure}

The photon number operators for the output fields $\pm l$ are 
\begin{align}
{\hat{{\mathsf N}}}_{{\rm out}, \pm L} = {\hat{{\mathsf A}}}_{{\rm out}, \pm L}^\dagger
                                         {\hat{{\mathsf A}}}_{{\rm out}, \pm L}
\label{output_number_free_field}
\end{align}
and they can be calculated using Eqs.~(\ref{quantum_model_add_through_output}) and~~(\ref{quantum_model_add_drop_output}).
The difference between these operators is experimentally observable and can explicitly be defined as
\begin{align}
{\hat{{\mathsf N}}}_{{\rm out}, \Delta} & = {\hat{{\mathsf N}}}_{{\rm out}, + L} -{\hat{{\mathsf N}}}_{{\rm out}, - L}
                                            \nonumber  \\
                                        & = 2 \kappa_{\rm r}  {\hat{{\mathsf n}}}_{_\Delta} 
                                            - \sqrt{2 \kappa_{\rm r}} \, {\hat{{\mathsf G}}}_{{\rm r}}
                                            + {\hat{{\mathsf N}}}_{{\rm r}, \Delta}^{\rm vac} \, ,               
\label{output_difference_number_free_field}
\end{align}
where 
\begin{align}
{\hat{{\mathsf N}}}_{{\rm r}, \Delta}^{\rm vac} = 
								\hat{{\mathsf V}}_{{\rm r},_{+L}}^\dagger \hat{{\mathsf V}}_{{\rm r},_{+L}}
						      - \hat{{\mathsf V}}_{{\rm r},_{-L}}^\dagger \hat{{\mathsf V}}_{{\rm r},_{-L}} \, .
\label{def_N_delta_vac}
\end{align}
When we consider the fact that the photon number commutes with the total Hamiltonian following $[{\hat{{\mathsf n}}}_{_\Delta}, {\hat{{\mathsf H}}}_{\rm tot}]= 0$, Eq.~(\ref{photon_number_difference_dyn})  becomes linear and can be translated in the Fourier space according to
\begin{align}
 {\tilde{{\mathsf n}}}_{_\Delta}(\omega) &=   \frac{\sum_\textrm{s} \sqrt{2 \kappa_{\rm s}} \, {\tilde{{\mathsf G}}}_{{\rm s}}(\omega)}{2 \kappa - i \omega}  \, ,                  
\label{Fourier_output_difference_number_free_field}
\end{align}
and from Eq.~(\ref{output_difference_number_free_field}), the Fourier spectrum of the difference in photon numbers is found to be
\begin{align}
{\tilde{{\mathsf N}}}_{{\rm out}, \Delta} (\omega) = & \,\,\, 2 \kappa_{\rm r}  {\tilde{{\mathsf n}}}_{_\Delta} (\omega)
                                            - \sqrt{2 \kappa_{\rm r}} \, {\tilde{{\mathsf G}}}_{{\rm r}} (\omega)
                                            + {\tilde{{\mathsf N}}}_{{\rm r}, \Delta}^{\rm vac} (\omega) \nonumber \\
                                             = & \,\,\, \frac{2 \kappa_{\rm r}}{2 \kappa - i \omega}
                                             \, \sum_\textrm{s} \sqrt{2 \kappa_{\rm s}} \, 
                                             {\tilde{{\mathsf G}}}_{{\rm s}}(\omega)  \nonumber \\
                                            &  - \sqrt{2 \kappa_{\rm r}} \, {\tilde{{\mathsf G}}}_{{\rm r}} (\omega)
                                            + {\tilde{{\mathsf N}}}_{{\rm r}, \Delta}^{\rm vac}  (\omega)          \, ,  
\label{Fourier_output_difference_number_free_field_bis}
\end{align}
so that the power spectrum can be determined as
\begin{align}
\langle |{\tilde{{\mathsf N}}}_{{\rm out}, \Delta} (\omega)|^2 \rangle
= {2 \kappa_{\rm r}} \langle {\hat{{\mathsf n}}}_{_\Sigma} \rangle \,
\frac{\omega^2 + 4 \kappa (\kappa - \kappa_{\rm r})}{\omega^2 + 4 \kappa^2}\, .
\label{Power_Fourier_output_difference_number}
\end{align}
Since the shot noise level is $  {2 \kappa_{\rm r}}\langle {\hat{{\mathsf n}}}_{\Sigma} \rangle$, 
it is convenient to rewrite this spectrum under the following normalized form
\begin{align}
S(\omega) = & \,\,\, \frac{\langle |{\tilde{{\mathsf N}}}_{{\rm out}, \Delta} (\omega)|^2 \rangle}{ {2 \kappa_{\rm r}} \langle {\hat{{\mathsf n}}}_{_\Sigma} \rangle}  \nonumber \\
          = & \,\,\, 1 - \rho \, \frac{4 \kappa^2}{\omega^2 + 4 \kappa^2}
\label{Power_Fourier_output_difference_number_normalized}
\end{align}
where the parameter $\rho \in \, ]0,1]$ is defined in Eq.~(\ref{def_rho}).
The spectrum described by $S(\omega)$ is an inverted Lorentzian which qualitatively displays a dip below the shot noise level close to the zero frequency. It converges to~$1$ at $\omega = \pm \infty$, and to 
$1-\rho$ at $\omega =0$. The parameter $\rho$ therefore represents a direct indicator of the squeezing efficiency, as 
$\rho \rightarrow 1$ leads to quasi-perfect squeezing at zero frequency, while $\rho \rightarrow 0$ leads to no squeezing at all frequencies. 
The case of perfect squeezing would theoretically correspond to an ideal cavity with null intrinsic losses in the add-through configuration, since $\rho =1$ for $\kappa_{\rm i} =0$~\cite{Lugiato_Castelli,Gatti_Mancini}.
Some squeezing spectra of the photon number difference with different values of $\rho$ are displayed in 
Fig.~\ref{Quads_Sa_Sp}, where they have been plotted as solid lines. 

In Kerr comb generation, efficient squeezing ($\rho \rightarrow 1$) is achieved with strong over-coupling ($\kappa_{\rm t} \gg \kappa_{\rm i}$ in the add-through configuration, and 
to $\kappa_{\rm d} \gg \kappa_{\rm t},\kappa_{\rm i}$ in the add-drop configuration). 
Hence, exactly as for spontaneous FWM, ultra-low loss resonators are ideal since they systematically maximize $\rho$ 
because $\kappa_{\rm i} \rightarrow 0$. Around $1550$~nm, the record intrinsic $Q$ factor is $3 \times 10^{11}$ with a CaF$_2$ resonator~\cite{Record_Q}.
Intrinsic quality factors of the order of $10^9$ are routinely obtained with crystalline or amorphous WGM resonators.
Hence, these ultra-low-loss resonators are therefore idoneous candidates for highly efficient squeezing, and the technological solutions for their large-scale fabrication~\cite{NIST_PRX}, and integration in chip-scale devices~\cite{Integrated_WGMR} are already available. 
Finally, it is very important to note that achieving strong over-coupling in the output port (${\kappa_{\rm r}} \rightarrow +\infty$) is important not only to increase the efficiency of the squeezing ($\rho \rightarrow 1$), but also to increase its bandwidth ($\kappa \rightarrow +\infty$). 
However, one should also keep in mind that the pump power $P_{\rm min}$ needed to trigger comb generation will grow as  $\kappa^2$ in this strongly over-coupled regime [see Eq.~(\ref{minpowersinwattspump})], so that an optimal power vs bandwidth balance has to be found depending on the targeted application.   

\section{System above threshold: quantum correlations and squeezing for the amplitude and phase quadratures}
\label{squeezedquad}

For a wide range of parameters (pump power, cavity detuning and dispersion), Kerr combs can be phase-locked and lead to the emergence of stationary spatio-temporal patterns which can be extended (rolls) or localized (solitons).
Hence, beside amplitude correlations, the phase of the optical fields can display strong correlations as well.

These phase correlations at the semi-classical level can lead to phase quadrature squeezing from a quantum perspective.
We hereafter determine the linearized input-output relationship that is needed to track the temporal dynamics of the modal fluctuation operators under the influence of vacuum noise. 
This fluctuation flow will allow us to determine some relevant phase quadratures for rolls, bright and dark solitons.

\begin{figure}
\begin{center}
\includegraphics[width=7cm]{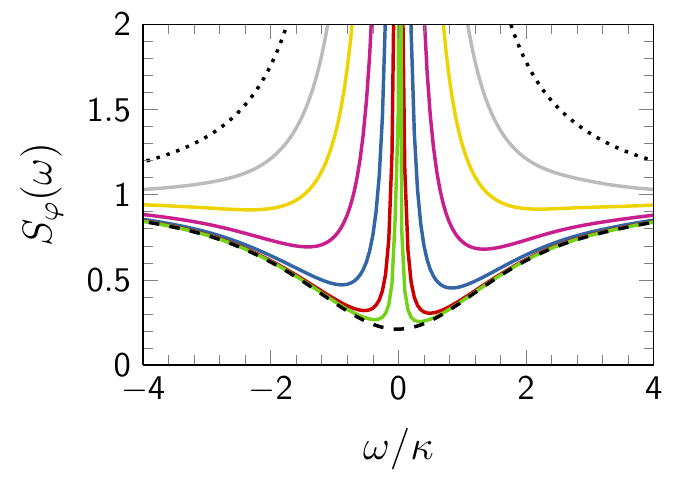}
\end{center}
\caption[Plotting_quads_as_fct_of_psi]
{\label{Plotting_quads_as_fct_of_psi} 
(Color online) Spectra of quadratures with different angles $\varphi$ in a Kerr comb close to threshold, where the $3$-modes approximation is valid. 
The spectra are calculated using  Eq.~(\ref{spectrum_varphi}).
The system is in the add-through configuration with $\rho=0.8$. The threshold power is $P_{\rm th}= 2.06$~mW and the pump power is set to $P= 1.01 \, P_{\rm th}$ and the laser detuning is $\sigma = - \kappa$. 
The comb features sidemodes at $L = \pm 18$.
Pure amplitude quadrature (dashed black) is not exactly achieved for 
$\varphi = \Phi$ as expected, but rather for
$\varphi = \Phi + \delta \Phi \equiv \Phi_{\rm opt}$, where $\delta \Phi$ is an additional offset
angle that is generally small. In this case, we have found $\delta \Phi = 0.022$.
Accordingly, pure phase quadrature (dotted black) is obtained for $\varphi = \Phi_{\rm opt} + \frac{\pi}{2}$.
The other spectra correspond to the following quadrature angles:
$\varphi = \Phi_{\rm opt}+ \pi/100$ (green);
$\varphi = \Phi_{\rm opt}+ \pi/50$ (red);
$\varphi = \Phi_{\rm opt}+ \pi/20$ (blue);
$\varphi = \Phi_{\rm opt}+ \pi/10$ (pink);
$\varphi = \Phi_{\rm opt}+ \pi/6$ (yellow);
$\varphi = \Phi_{\rm opt}+ \pi/4$ (gray).
Note that once we deviate from the pure amplitude spectrum, we here have a divergence at zero-frequency.}
\end{figure}

\subsection{Linearized dynamics of the modal fluctuation operators}
\label{generalcase}

Let us consider a stationary Kerr comb spanning from $l=-K$ to $l=K$ (total of $2K+1$ modes).
The intra-cavity modal fields can be perturbed as   
\begin{eqnarray}
\hat{{\mathsf a}}_{l} = {{\mathcal A}}_{l} +  \delta \hat{{\mathsf a}}_{l}
\label{def_fluctuat}
\end{eqnarray}
where ${{\mathcal A}}_{l}$ are the constant complex-valued numbers representing the semi-classical stationary states, and 
$\delta \hat{{\mathsf a}}_{l}$ are the fluctuation operators.

\begin{figure*}
\begin{center}
\includegraphics[width=12cm]{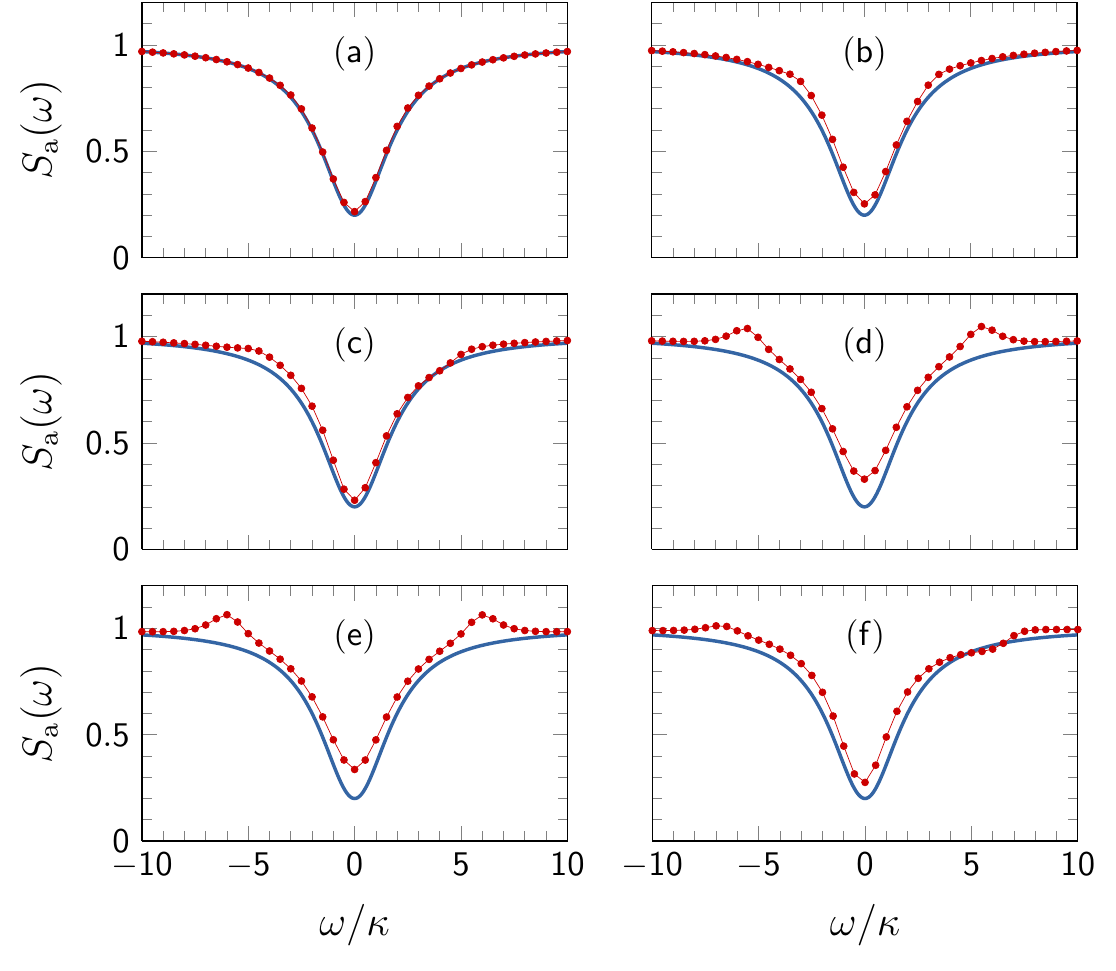}
\end{center}
\caption[Plotting_quads_Sa_fct_of_power]
{\label{Plotting_quads_Sa_fct_of_power} 
(Color online) 
Spectra of amplitude quadratures as the pump power is increased in a Kerr comb originating from a roll pattern.
Except the pump power $P$, the parameters of the system are the same as in Fig.~\ref{Plotting_quads_as_fct_of_psi}.
The modes $\pm L$ of interest are the first (main) sidemodes of the comb. 
The solid blue lines stand for the analytical and ideal amplitude squeezing spectrum provided by the $3$-modes approximation in Eq.~(\ref{def_S_a}). 
The red dots stand for the numerical spectra obtained with Eq.~(\ref{spectrum_varphi}), where all the modes of the comb are accounted for~\cite{comment_symbols}.
The figures display the best squeezing spectra for the pump powers $P$, oscillating modes $\pm L$, and offsets $\delta \Phi$ listed hereafter.
$(a)$:~$P=1.01 \, P_{\rm th}$  ($L=18$ and $\delta \Phi = 0.015$);
$(b)$:~$P=1.1  \, P_{\rm th}$  ($L=19$ and $\delta \Phi = -0.03$);
$(c)$:~$P=1.5  \, P_{\rm th}$  ($L=22$ and $\delta \Phi = 0.1$);
$(d)$:~$P=2.0  \, P_{\rm th}$  ($L=24$ and $\delta \Phi = 0.02$);
$(e)$:~$P=2.5  \, P_{\rm th}$  ($L=25$ and $\delta \Phi = -0.0025$);
$(f)$:~$P=3.0  \, P_{\rm th}$  ($L=26$ and $\delta \Phi = 0.18$).
As the pump is increased, the $3$-modes approximation becomes less and less valid, but very efficient squeezing can still be achieved with the fundamental pair of sidemodes.
}
\end{figure*}

Then by plugging Eq.~(\ref{def_fluctuat}) into Eqs~(\ref{Heisenberg_picture}), it can easily be found that the steady state amplitude of the oscillating modes obey the set of $(2K+1)$ nonlinear algebraic equations 
\begin{eqnarray}
&&  - \left[ \kappa -i \left( \sigma - \frac{1}{2} \zeta_2 l^2 \right) \right]\, {{\mathcal A}}_l  
                               + \delta(l) \, \sqrt{2 \kappa_{\rm t}} \, {{A}}_{\rm in}  \nonumber \\
&& + i g_0 \sum_{m,n,p} \delta(m-n+p-l) \, {{\mathcal A}}_n^*  {{\mathcal A}}_m  {{\mathcal A}}_p  =0  \, ,               
\label{steadystate}
\end{eqnarray}
while the noise driven fluctuations are ruled by the following set of equations: 
\begin{eqnarray}
  \delta \dot{\hat{{\mathsf a}}}_l &=&   - \left[ \kappa  - i \left( \sigma - \frac{1}{2} \zeta_2 l^2 \right) \right]
   \delta \hat{{\mathsf a}}_{l}  + \sum_\textrm{s} \sqrt{2 \kappa_{\rm s}} \, {\hat{{\mathsf V}}}_{{\rm s},l} \nonumber \\
   && + i g_0 \sum_{m,n,p} \delta(m-n+p-l)  \label{pert_eqs} \\
   &&   \times  \{  \delta \hat{{\mathsf a}}_{n}^\dagger  {{\mathcal A}}_{m} {{\mathcal A}}_{p} 
                   + {{\mathcal A}}_{n}^* \delta  \hat{{\mathsf a}}_{m} {{\mathcal A}}_{p}
                   + {{\mathcal A}}_{n}^* {{\mathcal A}}_{m} \delta \hat{{\mathsf a}}_{p}
       \} \nonumber \, .  
\end{eqnarray}
The above fluctuation flow can be synthetically rewritten as 
\begin{align}
\delta \dot{\hat{{\mathsf a}}}_l &= \sum_{p=-K}^K \mathcal{R}_{lp} \, \delta  \hat{{\mathsf a}}_{p} 
                               +\sum_{p = -K}^K \mathcal{S}_{lp} \, \delta \hat{{\mathsf a}}_{p}^\dagger
                               + \sum_\textrm{s} \sqrt{2 \kappa_{\rm s}} \, {\hat{{\mathsf V}}}_{{\rm s},l} \, ,  
\label{pert_eqs_bis_new}
\end{align}
where
\begin{eqnarray}
\mathcal{R}_{lp} &=& -\left[\kappa -i\left(\sigma -\frac{1}{2} \zeta_2 l^2 \right) \right]\, \delta(p-l) \nonumber \\
                  && + 2i g_0 \sum_{m,n} \delta(m-n+p-l) \, {{\mathcal A}}_m {{\mathcal A}}_n^*   \label{Define_R} \\
  \mathcal{S}_{lp} &=& i g_0 \sum_{m,n} \delta(m+n-p-l) \, {{\mathcal A}}_m {{\mathcal A}}_n   \label{Define_S}      
\end{eqnarray}
can be considered as the elements of the $(2K+1)^{\rm th}$-order square matrices  $\mathbf{R}$ and $\mathbf{S}$, and the driving quantum noise term  
$\sum_\textrm{s} \sqrt{2 \kappa_{\rm s}} \, {\hat{{\mathsf V}}}_{{\rm s},l}$
represents the sum of all vacuum fluctuations for a given mode $l$.

If we introduce the $(2K+1)$-dimensional fluctuation and vacuum noise vectors 
\begin{align}
\delta {\hat{\textbf{\textsf{a}}}}(t) = \left[ \begin{array}{c}
                    	          \delta  \hat{{\mathsf a}}_{-K}(t) \\
                    	          \vdots \\
                    	          \delta  \hat{{\mathsf a}}_{+K}(t) 
	                              \end{array}
                                   \right] ; \,\,\,\,\,\,\,\,\,\,
 \hat{\textbf{\textsf{V}}}_{\rm s}(t) = \left[ \begin{array}{c}
                    	            {\hat{{\mathsf V}}}_{{\rm s},-K}(t)  \\
                    	            \vdots \\
                    	            {\hat{{\mathsf V}}}_{{\rm s},+K}(t) 
	                                 \end{array}
                                      \right] \, ,
\label{Define_vect_delta_psi_zeta}
\end{align}
then we can write Eq.~(\ref{pert_eqs_bis_new}) under the form of a quantum-noise-driven linear flow:
 \begin{eqnarray}
\left[ \begin{array}{l}
	\delta \dot{\hat{\textbf{\textsf{a}}}} \\
    \delta \dot{\hat{\textbf{\textsf{a}}}}^\dagger
	  \end{array}
\right]
=
\mathbf{J_{{\textsf{\textbf{a}}}}}  \,
\left[ \begin{array}{l}
	\delta {\hat{\textbf{\textsf{a}}}} \\
    \delta {\hat{\textbf{\textsf{a}}}}^\dagger
	  \end{array}
\right]
+  \sum_\textrm{s} \sqrt{2 \kappa_{\rm s}} \,
\left[ \begin{array}{l}
	\hat{\textbf{\textsf{V}}}_{\rm s}(t) \\
    \hat{\textbf{\textsf{V}}}_{\rm s}^\dagger (t)
	  \end{array}
\right]
\, ,
\label{eq_matrix_pert}
\end{eqnarray}	 
where 
\begin{eqnarray}
\mathbf{J_{{\textsf{\textbf{a}}}}} =
\left[ \begin{array}{ll}
      \mathbf{R}  & \mathbf{S} \\
      \mathbf{S}^* & \mathbf{R}^*
	  \end{array}
\right] 
\label{Jacobian}
\end{eqnarray}	 
is a composite (block matrix) Jacobian of order $2 \times (2K+1)$. 
It should be noted that this Jacobian matrix has to be determined numerically, since its components exclusively depend on the steady state values of the  semi-classical modal fields ${{\mathcal A}}_l$.

\subsection{Dynamics of the quadrature operators}
\label{quadraturedynamics}

Quadratures operators are observables of particular interest for the study of the quantum properties of multimode fields.
They are Hermitian operators that correspond to linear combinations of annihilation and creation operators, and usually,
the relevant linear combinations can be inferred from the conserved quantities in the semi-classical limit.

In the case of Kerr combs, it is known that in the asymptotic limit, the amplitudes of two symmetric modes $-l$ and $+l$ are equal ($|{\cal A}_{+l}| = |{\cal A}_{-l}|$), and the sum of their phases is a constant, following
\begin{align}
\phi_{l} + \phi_{-l} = {\rm Const} = 2 \Phi_l \, .
\label{def_const_phase_sum}
\end{align}
The constant $\Phi_l$ depends on the modes $\pm l$ under consideration, but not on the initial conditions. 
In other words, once a symmetric pair of modes has been chosen, the sum of its steady-state slowly-varying phases is a constant of motion. For that particular pair of mode, the phase reference can be shifted so that $\phi_{l} = \phi_{-l}  \equiv \Phi_l$, leading to the conservation law $\phi_{l} - \phi_{-l}=0$ with the in frame.

From a quantum perspective, the corresponding two-modes quadratures are~\cite{diff_1mode_2modes_quads} 
\begin{align}
\delta \hat{{\mathsf q}}_{{\varphi,l}} = \frac{1}{\sqrt{2}}(\delta \hat{{\mathsf a}}_{{+l}} - \delta \hat{{\mathsf a}}_{{-l}}) \, e^{-i \varphi} + {\rm H.~c.} \, ,
\label{quadratureroperator_3modes}
\end{align}
with $l = -K, \dots, K$.
It is therefore interesting to investigate the dynamics of all the quadratures $\delta \hat{{\mathsf q}}_{{\varphi,l}}$ altogether.
For this purpose, we can build the $K$-dimensional operator
\begin{align}
\delta {\hat{\textbf{\textsf{q}}}}_{{\varphi}} =  \left[ \begin{array}{c}
                    	            \delta \hat{{\mathsf q}}_{{\varphi,1}}   \\
                    	            \vdots \\
                    	            \delta \hat{{\mathsf q}}_{{\varphi,K}}  
	                                 \end{array}
                                      \right] 
\label{Define_qvect}
\end{align}
and from Eq.~(\ref{quadratureroperator_3modes}), it is found that the vectorial quadrature 
$\delta {\hat{\textbf{\textsf{q}}}}_{{\varphi}}$ can be rewritten as 
\begin{align}
\delta \hat{\textbf{\textsf{q}}}_{{\varphi}} = \delta \hat{\textbf{\textsf{q}}}_{{0}} \cos \varphi 
                                      + \delta \hat{\textbf{\textsf{q}}}_{{\frac{\pi}{2}}} \sin \varphi 
\label{quadratureroperator_3modes_bis}
\end{align}
where $\delta \hat{\textbf{\textsf{q}}}_{{0}}$ and $\delta \hat{\textbf{\textsf{q}}}_{{\frac{\pi}{2}}}$ are the amplitude and phase vectorial quadratures, respectively. 

The dynamics of these quadratures can be obtained from Eq.~(\ref{quadratureroperator_3modes}) as 
\begin{align}
\delta \dot{\hat{{\mathsf q}}}_{{0,l}} &= \frac{1}{\sqrt{2}}(\delta \dot{\hat{{\mathsf a}}}_{{+l}}
                                                        - \delta \dot{\hat{{\mathsf a}}}_{{-l}}) 
                                                        + {\rm H.~c.}  \\
\delta \dot{\hat{{\mathsf q}}}_{{\frac{\pi}{2},l}} &= -\frac{i}{\sqrt{2}}(\delta \dot{\hat{{\mathsf a}}}_{{+l}}
                                                        - \delta \dot{\hat{{\mathsf a}}}_{{-l}}) + {\rm H.~c.}  \, 
\label{eqts_dyn_q0_qpiover2}
\end{align}
with $l=1,\dots,K$.
Using Eqs.~(\ref{eq_matrix_pert}), it can be demonstrated that the amplitude and phase vectorial quadrature operators obey the closed form Langevin equation:
\begin{eqnarray}
\left[ \begin{array}{l}
	\delta \dot{\hat{\textbf{\textsf{q}}}}_{{0}} \\
    \delta \dot{\hat{\textbf{\textsf{q}}}}_{{\frac{\pi}{2}}}
	  \end{array}
\right]
=
 \mathbf{J_{{\textsf{\textbf{q}}}}} 
\left[ \begin{array}{l}
	\delta {\hat{\textbf{\textsf{q}}}}_{{0}} \\
    \delta {\hat{\textbf{\textsf{q}}}}_{{\frac{\pi}{2}}}
	  \end{array}
\right] 
 +  \sum_\textrm{s} \sqrt{2 \kappa_{\rm s}} 
\left[ \begin{array}{l}
	{\hat{\textbf{\textsf{W}}}}_{{\rm s},0}(t) \\
    {\hat{\textbf{\textsf{W}}}}_{{{\rm s},\frac{\pi}{2}}}(t)
	  \end{array}
\right]
\label{eq_matrix_pert_quad}    
\end{eqnarray}	 
where 
\begin{eqnarray}
\mathbf{J_{{\textsf{\textbf{q}}}}} =
\left[ \begin{array}{lr}
      \Re (\mathbf{U}_+)  & -\Im (\mathbf{U}_+) \\
      \Im (\mathbf{U}_-) &   \Re (\mathbf{U}_-)
	  \end{array}
\right]  \, 
\label{Jacobian_quad}
\end{eqnarray}	 
is a Jacobian block matrix of order $2K$, the $K$-dimensional matrices $\mathbf{U}_\pm$ are explicitly defined 
through their complex-valued components 
\begin{align}
\mathcal{U}_{\pm,lp}= (\mathcal{R}_{l,p}-\mathcal{R}_{l,-p}) \pm (\mathcal{S}_{l,p}-\mathcal{S}_{l,-p})^* 
\label{def_U}
\end{align}
with $l,p \in \{1,\dots,K\}$, while the $K$-dimensional vacuum noise operators ${\hat{\textbf{\textsf{W}}}}_{{\rm s},{0}}(t)$ and ${\hat{\textbf{\textsf{W}}}}_{{\rm s},{\frac{\pi}{2}}}(t)$ are explicitly defined as
\begin{align}
&	{\hat{\textbf{\textsf{W}}}}_{{\rm s},{0}}(t)   \! =    \! \frac{1}{\sqrt{2}} 
                                       \!  \! \left[ \begin{array}{c}
                    	         \hat{{\mathsf V}}_{{\rm s},{+1}}(t) -  \hat{{\mathsf V}}_{{\rm s},{-1}}(t)   \\
                    	        \vdots \\
                    	         \hat{{\mathsf V}}_{{\rm s},{+K}}(t) -  \hat{{\mathsf V}}_{{\rm s},{-K}}(t)  
	                                 \end{array}
                                      \right] \!   +  \!  {\rm H.~c.}  \label{ext_noise_pert_quad_1} \\
&	{\hat{\textbf{\textsf{W}}}}_{{\rm s},{\frac{\pi}{2}}}(t)   \! =   \! - \frac{i}{\sqrt{2}} 
                                      \!  \!  \left[ \begin{array}{c}
                    	         \hat{{\mathsf V}}_{{\rm s},{+1}}(t) -  \hat{{\mathsf V}}_{{\rm s},{-1}}(t)   \\
                    	        \vdots \\
                    	         \hat{{\mathsf V}}_{{\rm s},{+K}}(t) -  \hat{{\mathsf V}}_{{\rm s},{-K}}(t)  
	                                 \end{array}
                                      \right]  \!  +   \! {\rm H.~c.} \label{ext_noise_pert_quad_2}  
\end{align}	 
It is interesting to note that the Jacobian matrix $\mathbf{J_{{\textsf{\textbf{q}}}}}$ is real-valued, as it is built with the real and imaginary parts of the complex-valued matrices $\mathbf{U}_\pm$. This Jacobian is generally \textit{not} diagonal, meaning that in a Kerr comb, all these quadratures are coupled.

\begin{figure}
\begin{center}
\includegraphics[width=7cm]{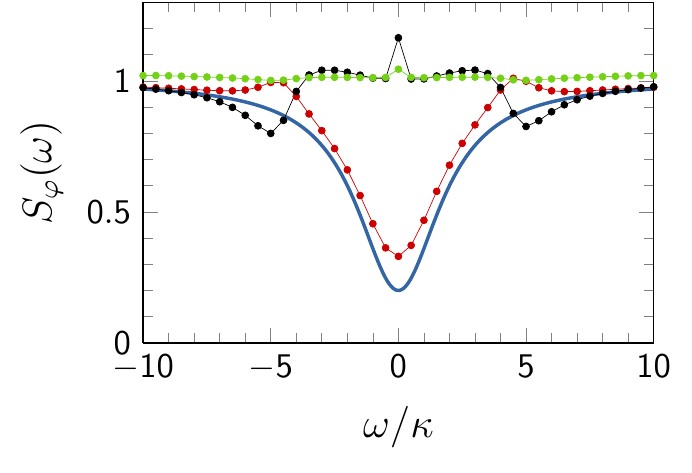}
\end{center}
\caption[Plotting_quads_as_fct_of_order]
{\label{Plotting_quads_as_fct_of_order} 
(Color online) 
Spectra of amplitude quadratures for different mode of order $\pm kL$ in a primary comb of a roll pattern.
The parameters of the system are the same as in Fig.~\ref{Plotting_quads_as_fct_of_psi}(c) [with $P=1.5  \, P_{\rm th}$ and $L=22$], except the detuning $\delta \Phi = 0.03$ that has been applied to all mode quadratures.
The solid blue line is the ideal amplitude squeezing spectrum obtained from Eq.~(\ref{def_S_a}). 
The dots stand for the numerical spectra obtained with Eq.~(\ref{spectrum_varphi})~\cite{comment_symbols}.
Red: modes $\pm L$;
Black: modes $\pm 2L$;
Green: modes $\pm 3L$.}
\end{figure}

Equations~(\ref{qmodel_output_field}) and~(\ref{Define_qvect}) allow to determine the output vectorial quadratures following 
\begin{eqnarray}
\left[ \begin{array}{l}
	\delta {\hat{\textbf{\textsf{Q}}}}_{{\rm out},{0}}(t) \\
    \delta {\hat{\textbf{\textsf{Q}}}}_{{\rm out},{\frac{\pi}{2}}}(t)
	  \end{array}
\right]
=
\sqrt{2 \kappa_{\rm r}} 
\left[ \begin{array}{l}
	\delta {\hat{\textbf{\textsf{q}}}}_{{0}}(t) \\
    \delta {\hat{\textbf{\textsf{q}}}}_{{\frac{\pi}{2}}}(t)
	  \end{array}
\right] 
- 
\left[ \begin{array}{l}
	{\hat{\textbf{\textsf{W}}}}_{{\rm r},{0}}(t) \\
    {\hat{\textbf{\textsf{W}}}}_{{\rm r},{\frac{\pi}{2}}}(t)
	  \end{array}
\right] .
\label{output_quad}
\end{eqnarray}	 
The time-domain dynamics of the generic quadrature $\delta {\hat{{\mathbf q}}}_{{\varphi}}$, as well as its output counterpart $\delta {\hat{{\mathbf Q}}}_{{\rm out},{\varphi}}$ are determined by combining the above equation with Eq.~(\ref{quadratureroperator_3modes_bis}).

\subsection{Correlations and squeezing spectra}
\label{Corrsqueezspec}

The Fourier spectra of the output signals can be determined using the output correlation matrix.
After translating Eqs.~(\ref{eq_matrix_pert_quad}) and ~(\ref{output_quad}) in the Fourier domain, the Fourier spectrum of the output vectorial quadrature is found to be equal to 
\begin{eqnarray}
\left[ \begin{array}{l}
	\delta \tilde{\textbf{\textsf{Q}}}_{{\rm out},{0}}(\omega) \\
    \delta \tilde{\textbf{\textsf{Q}}}_{{\rm out},{\frac{\pi}{2}}}(\omega)
	  \end{array}
\right]
&=&
- \sqrt{2 \kappa_{\rm r}} \, [ \mathbf{J}_\textbf{\textsf{q}} + i \omega  \mathbf{I}]^{-1} \nonumber \\
&&  \times \sum_s \sqrt{2 \kappa_{\rm s}}
\left[ \begin{array}{l}
	{\tilde{\textbf{\textsf{W}}}}_{{\rm s},{0}}(\omega) \\
    {\tilde{\textbf{\textsf{W}}}}_{{\rm s},{\frac{\pi}{2}}}(\omega)
	  \end{array}
\right] \nonumber \\
&& - 
\left[ \begin{array}{l}
	{\tilde{\textbf{\textsf{W}}}}_{{\rm r},{0}}(\omega) \\
    {\tilde{\textbf{\textsf{W}}}}_{{\rm r},{\frac{\pi}{2}}}(\omega)
	  \end{array}
\right] \, .
\label{output_quad_fourier}
\end{eqnarray}	 
We can use the above equation to determine the $2K$-dimensional output correlation matrix, following 
\begin{eqnarray}
\mathbf{C}^{{\rm out}} (\omega) \!\! &=& \!\!
\int_{-\infty}^{+\infty} \! d \omega' \!
\left< 
\left[ \begin{array}{l}
	\delta {\tilde{\textbf{\textsf{Q}}}}_{{\rm out},{0}}(\omega) \\
    \delta {\tilde{\textbf{\textsf{Q}}}}_{{\rm out},{\frac{\pi}{2}}}(\omega)
	  \end{array}
\right]\!\!
\left[ \begin{array}{l}
	\delta {\tilde{\textbf{\textsf{Q}}}}_{{\rm out},{0}}(\omega') \\
    \delta {\tilde{\textbf{\textsf{Q}}}}_{{\rm out},{\frac{\pi}{2}}}(\omega')
	  \end{array}
\right]^{\rm T}
\right> \nonumber \\ 
&=& 
\,\,\,\,\,\,  \{ 2 \kappa \rho[ \mathbf{J}_\textbf{\textsf{q}}+i\omega  \mathbf{I}]^{-1} + \mathbf{I} \} \, \mathbf{C}^{{\rm in}}(\omega)  \nonumber \\
&& \times  \, \{ 2 \kappa \rho[ \mathbf{J}_\textbf{\textsf{q}} - i \omega  \mathbf{I}]^{-1} + \mathbf{I} \}^{\rm T}  \nonumber \\
&& +  \, {4\kappa^2 \rho (1- \rho)} \, \{ [\mathbf{J}_\textbf{\textsf{q}}+i\omega  \mathbf{I}]^{-1} \} \, \mathbf{C}^{{\rm in}}(\omega) \nonumber \\
&&\,\,\,\,\,\,\,\,\,\,\,\,\,\,\,\,\,\,\,\, \,\,\,\,\,\,\,\,\,\,\, \times  \, \{ [ \mathbf{J}_\textbf{\textsf{q}} - i \omega  \mathbf{I}]^{-1}\}^{\rm T} \, ,
\label{output_corr_matrix}
\end{eqnarray}	 
where $\rho$ is the squeezing parameter defined in Eq.~(\ref{def_rho}), and
$\mathbf{C}^{{\rm in}} (\omega)$ is the $2K$-dimensional input correlation matrix
\begin{eqnarray}
\mathbf{C}^{{\rm in}} (\omega) \!\! &=& \!\! 
\int_{-\infty}^{+\infty}\! d \omega' \!
\left< 
\left[ \begin{array}{l}
	{\tilde{\textbf{\textsf{W}}}}_{{\rm s},{0}}(\omega) \\
    {\tilde{\textbf{\textsf{W}}}}_{{\rm s},{\frac{\pi}{2}}}(\omega)
	  \end{array}
\right] \!\!
\left[ \begin{array}{l}
	{\tilde{\textbf{\textsf{W}}}}_{{\rm s},{0}}(\omega') \\
    {\tilde{\textbf{\textsf{W}}}}_{{\rm s},{\frac{\pi}{2}}}(\omega')
	  \end{array}
\right]^{\rm T}
\right> \nonumber \\ 
 \!\! &=& \!\! 
\left[ \begin{array}{rr}
      \mathbf{I}      & i\mathbf{I} \\
      -i\mathbf{I}    &  \mathbf{I}
	  \end{array}
\right]  
\label{input_corr_matrix}
\end{eqnarray}	 
where ${{\mathbf{I}}}$ is the $K$-dimensional identity matrix.
It is interesting to note that this input correlation matrix is found to be frequency independent.

\begin{figure}
\begin{center}
\includegraphics[width=7cm]{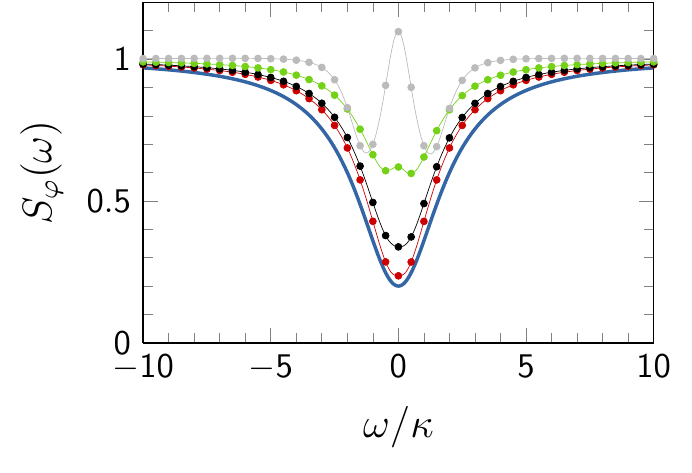}
\end{center}
\caption[Plotting_quads_as_fct_of_order_bright_solitons]
{\label{Plotting_quads_as_fct_of_order_bright_solitons} 
(Color online) 
Spectra of amplitude quadratures for different mode of orders $\pm l$ of the comb from a isolated bright soliton.
The power is set to $P= 4$~mW, and $\sigma = -2 \kappa$.
The same offset  $\delta \Phi = \frac{\pi}{2} - 0.04$ has been applied to all mode quadratures.
The solid blue line is the ideal amplitude squeezing spectrum obtained from Eq.~(\ref{def_S_a}). 
The dots stand for the numerical spectra obtained with Eq.~(\ref{spectrum_varphi})~\cite{comment_symbols}.
Red:   $l \pm 1$;
Black: $l \pm 5$;
Green: $l \pm 10$.
Gray:  $l \pm 20$.}
\end{figure}

For each sidemode pair $\pm l$, the quadrature spectra are explicitly defined as 
\begin{eqnarray}	 
S_{\varphi,l} (\omega) &=&
     \mathcal{C}^{{\rm out}}_{{\rm 11};(l,l)} \cos^2 \varphi
+    \mathcal{C}^{{\rm out}}_{{\rm 22};(l,l)} \sin^2 \varphi \label{spectrum_varphi} \\
&& +[\mathcal{C}^{{\rm out}}_{{\rm 12};(l,l)} 
   + \mathcal{C}^{{\rm out}}_{{\rm 21};(l,l)} ] \cos \varphi \sin \varphi \nonumber
\end{eqnarray}	 
where the coefficients are complex-valued coefficients $\mathcal{C}^{{\rm out}}_{{ab};(l,l)}$ with $a,b \in \{1,2\}$ are diagonal elements of the $K$-dimensional matrices $\mathbf{C}^{{\rm out}}_{{ab}}(\omega)$ that are used to write  $\mathbf{C}^{{\rm out}} (\omega)$ in Eq.~(\ref{output_corr_matrix}) under the form of the block matrix
\begin{eqnarray}
\mathbf{C}^{{\rm out}} (\omega) = 
\left[ \begin{array}{cc}
      \mathbf{C}^{{\rm out}}_{\rm 11} (\omega)   & \mathbf{C}^{{\rm out}}_{\rm 12} (\omega) \\
      \mathbf{C}^{{\rm out}}_{\rm 21} (\omega)  &  \mathbf{C}^{{\rm out}}_{\rm 22} (\omega)
	  \end{array}
\right] \, .
\label{output_corr_matrix_bis}
\end{eqnarray}	 
The analytical expression provided by Eq.~(\ref{spectrum_varphi}) allows to plot the spectra of any 
quadrature for any pair of sidemodes $\pm l$, regardless of the size of the Kerr comb.

In the next two sections, we investigate in more detail the squeezing phenomena that can take place in Kerr combs originating from roll patterns and from solitons.  
For all our simulations, we will consider a calcium fluoride (CaF$_2$) resonator with main radius $a=2.5$~mm, and pumped around $1550$~nm in the add-through configuration.
The intrinsic and extrinsic $Q$-factors are fixed to $Q_{\rm int} \equiv Q_{\rm i} = 10^9$ and  
$Q_{\rm ext} \equiv Q_{\rm t} = 0.25 \times 10^9$, respectively, 
yielding loaded quality factor $Q_{\rm tot} =  Q_{\rm t}Q_{\rm i} /(Q_{\rm i}+Q_{\rm t}) = 0.2 \times 10^9$,
a full linewidth at half-maximum  $2 \kappa = \omega_{_{\rm L}}/Q_{\rm tot} = 2 \pi \times  0.97$~MHz, 
and a squeezing factor $\rho = Q_{\rm i} /(Q_{\rm i}+Q_{\rm t}) = 0.8$. 
The group velocity index is $n_g = 1.43$, so that the free-spectral range is $\Delta \omega_{_{\rm FSR}}= 2 \pi \times 13.35$~GHz. 
The nonlinear coefficient is set to $\gamma = 0.001$, corresponding to $g_0 = 2 \pi  \times 57.2$~$\mu$Hz.
For simulations in the anomalous dispersion regime (rolls and bright solitons), the overall second-order dispersion is fixed to $\beta_2 =-12.4 \times 10^{-27}$~s$^2$/m, which translates to $\zeta_2= 2 \pi  \times 2.9$~kHz.
In the normal dispersion regime (dark solitons), the dispersion parameters are set values opposite to those of the anomalous case.

\begin{figure}
\begin{center}
\includegraphics[width=7cm]{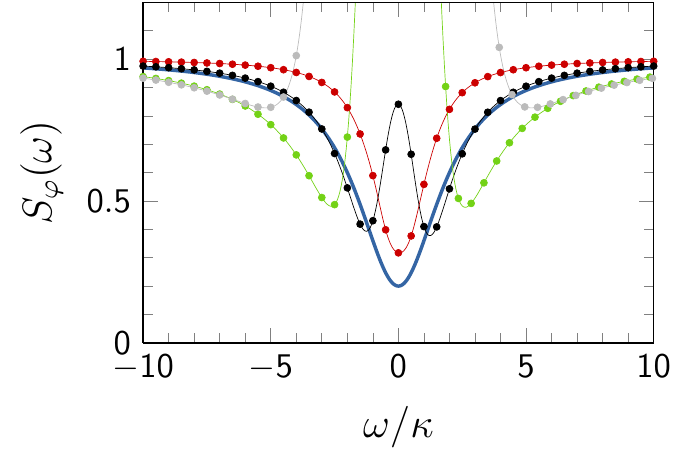}
\end{center}
\caption[Plotting_quads_as_fct_of_order_dark_solitons]
{\label{Plotting_quads_as_fct_of_order_dark_solitons} 
(Color online) 
Spectra of amplitude quadratures for different mode of orders $\pm l$ of the comb from a isolated dark soliton.
The power is set to $P= 5.3$~mW, and $\sigma = -2.5 \kappa$.
The same offset  $\delta \Phi = 0.72$ has been applied to all mode quadratures.
The solid blue line is the ideal amplitude squeezing spectrum obtained from Eq.~(\ref{def_S_a}). 
The dots stand for the numerical spectra obtained with Eq.~(\ref{spectrum_varphi})~\cite{comment_symbols}.
Red:   $l \pm 1$;
Black: $l \pm 5$;
Green: $l \pm 10$.
Gray:  $l \pm 20$.}
\end{figure}

\section{System above threshold: quadrature squeezing in roll patterns and solitons}
\label{quadssqueezingrollssolitons}

Rolls are azimuthal Turing patterns that emerge in the system when the resonator is pumped above a certain critical value.   
In the temporal domain, they are characterized by an integer number $L$ of stationary nodes and anti-nodes of the optical energy in the azimuthal direction of the resonator. In the spectral domain, they yield a comb where only the sidemodes of order $l =\pm kL$ ($k$ being an integer) do oscillate.  These combs, which are sometimes referred to as primary combs, are particularly important because they are the most robust and stable patterns that can be obtained experimentally.
Bright solitons, on the other hand, emerge in the system in the regime of anomalous dispersion, after a sub-critical bifurcation. Finally, dark solitons can be excited in the regime of normal dispersion, and in first approximation, they topologically connect the (hysteretic) upper and lower flat states inside the resonator. These various dynamical states have been investigated extensively in ref.~\cite{PRA_Unified}. Their spatio- and spectro-temporal representation is displayed in Fig.~\ref{spatiospectroplots}. In the forthcoming sub-sections, we will determine the squeezing spectra for the combs corresponding to all these  stationary states. 

\subsection{Quadrature squeezing in roll patterns with $3$ modes}
\label{quadssqueezingrolls_3modes}

In order to understand the spectra of amplitude and phase quadratures, it is important to analyze in detail the case where there are only $3$~modes in the comb.
As explained in Sec.~\ref{3modessqueezzphotnumbers}, such $3$-modes combs emerge in the supercritical case close to threshold, and they feature a multiplicity $L  \simeq \sqrt{(2/\zeta_2)[\sigma + 2 \kappa]}$.
It is useful to recall that regardless of the initial conditions, the two sidemodes $|{\cal A}_{\pm L}|$ have the same amplitude and according to Eq.~(\ref{def_const_phase_sum}), the sum of their phases is a constant, following 
$\phi_{L} + \phi_{-L} = {\rm Const} = 2 \Phi_L $.
Without loss of generality, we can consider in this $3$-modes case that the 
semi-classical solutions ${\cal A}_{\pm L}$ have the same phase $\Phi_L \equiv \Phi$, i.~e., they are considered identical.
It is also noteworthy that close to threshold, the phase $\phi_0$ is a constant that is independent from the sidemodes.

In this $3$-modes configuration, there is a only a single pair of amplitude and phase quadratures, namely 
$\delta \hat{{\mathsf q}}_{_{{0,L}}}$ and $\delta \hat{{\mathsf q}}_{_{{\frac{\pi}{2}},L}}$.
Therefore, the matrices $\mathbf{U}_\pm$ degenerate to scalars following $\mathcal{U}_{\pm} \equiv \mathcal{U}_{\pm,LL}$, 
yielding
\begin{align}
\mathcal{U}_{+} & = +2 g_0 \{ |{\cal A}_L|^2 \sin \Phi + |{\cal A}_0|^2 \sin (2 \phi_0 - \Phi) \}\, e^{-i\Phi} 
\nonumber \\
\mathcal{U}_{-} & = -2 ig_0 \{ |{\cal A}_L|^2 \cos \Phi + |{\cal A}_0|^2 \cos (2 \phi_0 - \Phi) \} \, e^{-i\Phi} \, .
\label{def_Uscalar_pm}
\end{align}
Accordingly, the Jacobian matrix $\mathbf{J_{{\textsf{q}}}}$ becomes $2$-dimensional. 
Interestingly, it already appears that when the quadratures are rotated by an angle $\Phi$, the value of $\mathcal{U}_{+}$ becomes pure real, while $\mathcal{U}_{-}$ becomes pure imaginary. 
In other words, the quadrature $\delta \hat{{\mathsf q}}_{_{{\Phi,L}}}$ is a pure amplitude quadrature, while 
the quadrature $\delta \hat{{\mathsf q}}_{_{{\Phi+\frac{\pi}{2},L}}}$ corresponds to a pure phase quadrature.

Using Eqs.~(\ref{quadratureroperator_3modes}) and~(\ref{eq_matrix_pert_quad}), it can be shown that the dynamics of these pure quadratures can be explicitly determined as
\begin{align}
\delta \dot{\hat{{\mathsf q}}}_{_{{\Phi,L}}} &= -2 \kappa_{\rm a} \, \delta {\hat{{\mathsf q}}}_{_{{\Phi,L}}}
 +  \sum_\textrm{s} \sqrt{2 \kappa_{\rm s}}  \, {\hat{{\mathsf W}}}_{_{{\rm s},{\Phi}}}(t) 
 \label{eqts_dyn_q0_3modes_combs} \\
\delta \dot{\hat{{\mathsf q}}}_{_{{\Phi+\frac{\pi}{2},L}}} &= 
 -2 \kappa_{\rm p} \, \delta {\hat{{\mathsf q}}}_{_{{\Phi,L}}}
 +  \sum_\textrm{s} \sqrt{2 \kappa_{\rm s}}  \, {\hat{{\mathsf W}}}_{_{{\rm s},{\Phi+\frac{\pi}{2}}}}(t) \label{eqts_dyn_qpiover2_3modes_combs}
\end{align}
where the linear coefficients are
\begin{align}
\kappa_{\rm a} &= - g_0 |{\cal A}_0|^2 \sin (2 \phi_0- 2 \Phi ) \label{def_Kappa_a} \\
\kappa_{\rm p} &=  g_0 \{ |{\cal A}_L|^2  + |{\cal A}_0|^2 \cos (2 \phi_0 -2 \Phi) \} \label{def_Kappa_p}
\end{align}
while the noise driving terms are defined analogously to the quadratures of Eq.~(\ref{quadratureroperator_3modes})
using Eqs.~(\ref{ext_noise_pert_quad_1}) and~(\ref{ext_noise_pert_quad_2}).
The normalized spectra corresponding to the pure amplitude and phase output quadratures can finally be calculated as 
\begin{eqnarray}
S_{\rm a}(\omega)&=& \left< |\delta {\tilde{{\mathsf Q}}}_{{\rm out},{\Phi}}(\omega) |^2 \right>  \nonumber \\
           &=& 1 - \rho \, \frac{4 \kappa_{\rm a}^2}{\omega^2 + 4 \kappa_{\rm a}^2} \label{def_S_a} \\
S_{\rm p}(\omega) &=& \left< |\delta {\tilde{{\mathsf Q}}}_{{\rm out},{\Phi+\frac{\pi}{2}}} (\omega)|^2 \right> \nonumber \\
          &=& 1 + \rho \, \frac{4 \kappa_{\rm a}^2}{\omega^2} 
                \left[ 1+ \frac{4\kappa_{\rm p}^2}{\omega^2 + 4 \kappa_{\rm a}^2}\right]^2  \, .
\label{def_S_p}
\end{eqnarray}
It can be demonstrated that $\kappa_{\rm a} \equiv \kappa$ in a $3$-modes comb, and as a consequence, Eq.~(\ref{def_S_a}) becomes in fact identical to Eq.~(\ref{Power_Fourier_output_difference_number_normalized}).
This is explained by the fact that $\delta \hat{{\mathsf q}}_{_{{\Phi,L}}}$ is a pure amplitude quadrature, which exactly corresponds to the case of photon number squeezing we have studied in Sec.~\ref{3modessqueezzphotnumbers}.
On the other hand, the phase quadrature is characterized by a spectrum  that is diverging at $\omega=0$, and this divergence is a generic signature of phase noise spectra.
It is noteworthy that $S_{\rm a}(\omega)$ is always smaller than $1$ and does not depend on the modal amplitudes, while 
$S_{\rm p}(\omega)$ is always larger than $1$,  and does depend on $|{\cal A}_{0}|$ and  $|{\cal A}_{\pm L}|$. 
Figure~\ref{Quads_Sa_Sp} displays both the amplitude (solid lines) and phase (dashed lines) quadratures for various values of the squeezing parameter $\rho$, when the other parameters are kept constant. 
As explained earlier, better squeezing is ensured when $\rho$ gets closer to $1$, which physically corresponds to strong over-coupling.  

The quadratures fluctuations $\delta \hat{{\mathsf q}}_{{\varphi}}$ have been expressed as a linear combination of 
 $\delta \hat{{\mathsf q}}_{{0}}$ and $\delta \hat{{\mathsf q}}_{{\frac{\pi}{2}}}$ in Eq.~(\ref{quadratureroperator_3modes_bis}). However, after rotation by an angle $\Phi$, they can also be expressed as a function of the pure amplitude and phase quadratures as 
\begin{align}
\delta \hat{{\mathsf q}}_{_{{\varphi,L}}} =  \delta \hat{{\mathsf q}}_{_{{\Phi,L}}}               \cos (\varphi - \Phi) 
                                      + \delta \hat{{\mathsf q}}_{_{{\Phi+\frac{\pi}{2},L}}} \sin (\varphi - \Phi)  \, ,  \label{quadratureroperator_3modes_ter}
\end{align}
which is just another way to express the fact that  we have a pure amplitude quadrature for $\varphi = \Phi$, 
and a pure phase quadrature for  $\varphi = \Phi + \frac{\pi}{2}$.
Therefore, since the quadratures with phases $\varphi \neq \Phi, \Phi +\frac{\pi}{2}$ are mixtures of pure amplitude and phase quadratures, their spectra are expected to have intermediate characteristics.
This phenomenology is displayed in Fig.~\ref{Plotting_quads_as_fct_of_psi}, where the power spectra $S_{\varphi,L} (\omega)$ explicitly defined in Eq.~(\ref{spectrum_varphi}) have been plotted for various values of the quadrature angle $\varphi$.
The resonator is pumped very close to threshold (in excess of $1\%$), and in that case, the $3$ modes approximation is very accurate.

It can be seen in Fig.~\ref{Plotting_quads_as_fct_of_psi} that the pure amplitude quadrature with inverted Lorentzian spectrum predicted by Eq.~(\ref{def_S_a}) does not exactly correspond to the angle  $\varphi = \Phi$ predicted theoretically. 
Instead, amplitude quadrature corresponds to a slightly different angle $\varphi = \Phi + \delta \Phi \equiv \Phi_{\rm opt}$, where the offset angle $\delta \Phi$ is generally found to be small close to threshold. 
When the angle of the quadrature is slightly detuned from the optimal value $\Phi_{\rm opt}$, the spectra maintain the inverted Lorentzian structure (like $S_{\rm a}$) but feature a sharp divergence at zero frequency (like $S_{\rm p}$). 
As the detuning is further increased, the spectra $S_{\varphi}$ loose the inverted Lorentzian structure and start to converge continuously towards the phase quadrature spectra  $S_{\rm p}$, which corresponds to  
$\varphi = \Phi_{\rm opt} + \frac{\pi}{2}$. A similar phenomenology has been analyzed in depth by Gatti and Mancini in ref.~\cite{Gatti_Mancini} in the context of quantum correlations in hexagonal spatial patterns.

It is interesting to emphasize the physical interpretation of strong squeezing ($\rho \rightarrow 1$) in Kerr combs in the context of Kerr combs. In ref.~\cite{Grynberg_Lugiato}, Grynberg and Lugiato did discuss the physical implication of the two-mode amplitude-phase squeezing. In particular, they emphasized that if one succesfully achieves perfect squeezing of the photon number difference $(N_+ - N_-)$, the conjugate variable (which is here the phase difference) becomes undetermined and 
``\emph{as a consequence, the position of the rolls [...] cannot be known}'', thereby impeding a ``\emph{direct}'' observation of the roll pattern (however, indirect detection using correlation techniques might remain possible). The authors where discussing the physical manifestation of two-mode squeezing in the context of the original Lugiato-Lefever experimental system (spatial patterns, free space signals, etc.): since Kerr combs translate the problem to a much more controllable environment (temporal patterns, guided signals, etc.), the phenomenon of two-mode squeezing could here enable to explore the phase-amplitude complementarity to a unprecedented extent.

\subsection{Quadrature squeezing in roll patterns with more than $3$~modes}
\label{quadssqueezingrolls_morethan3modes}

When the resonator is pumped far above threshold, the primary comb grows accordingly and features an increasing number of sidemodes.
As analyzed in Sec.~\ref{generalsqueezzphotnumbers}, squeezing is not guaranteed anymore in the system when there are more than $3$~modes involved. However, the regime of large primary combs (with $5$~modes or more) is interesting for various reasons. For instance, when the system is restricted to $3$~modes close to threshold, the amplitude of the sidemodes is very weak and detection can be problematic. Pumping the system far above threshold yields significantly more powerful signals. Another interesting point is that in the super-critical regime, the higher-order sidemodes ($|l| > L$) do not appear discontinuously: they are in fact always present, even though their amplitude is extremely small close to threshold. 
However, their effect does never completely vanish (for example, they contribute to the offset $\delta \Phi$).
It is therefore pertinent to investigate systematically how the quantum correlations are affected by these higher-order sidemodes in a primary Kerr comb.
 
In Figure~\ref{Plotting_quads_Sa_fct_of_power}, we display the best squeezing spectra for the amplitude quadratures as the pump power is increased from $1.01$ to $3$~times the threshold for comb generation. 	   
It should be recalled that as the pump power is increased, the parametric gain bandwidth is shifted outwards, and this explains why the mode orders $L$ increase with the pump (see refs.~\cite{YanneNanPRL,YanneNanPRA,PRA_Unified}).
When the system is very close to threshold ($P=1.01 \, P_{\rm th}$), the $3$-modes approximation holds and the numerical simulations provide results that are in quasi-perfect agreement with the theoretical prediction. 
As the pump power is increased, it can be seen that there is a deviation from the ideal inverted Lorentzian profile, but excellent squeezing performance is still achieved up to $P=3 \, P_{\rm th}$ where there are more than $15$ oscillating modes. This results therefore show that even in the highly multimode regime corresponding to a resonator pumped far above threshold, very efficient squeezing is still possible in Kerr combs. 
It is interesting to note that the spectra and the offsets $\delta \Phi$ are not invariant, as they depend on initial conditions. This is explained by the fact that the spectra depend on the Jacobian matrices $\mathbf{J_q}$, which are built with the complex-valued modal amplitudes, and which depend themselves on these initial conditions.

We have also investigated  the quantum correlation properties of higher-order modes in the comb originating from a roll pattern (modes of order $l = \pm kL$ with $k = 2,3,\dots$).  
Figure~\ref{Plotting_quads_as_fct_of_order} shows that the first order modes displays very good squeezing as already discussed earlier, but the spectrum of the second order modes still feature some weak squeezing in a frequency band were it seems that there is excess noise in the spectrum of the fundamental pair of sidemodes. Squeezing is numerically found to be 
quasi-inexistant for the third order pair, as well as for the higher orders with $k>3$.

\subsection{Quadrature squeezing in bright and dark solitons}
\label{quadssqueezingsol}

An interesting open point is to determine if squeezing with symmetric pairs of sidemodes is still possible in solitons.
Solitons in WGM resonators do not emerge super-critically -- their amplitude can not be arbitrary small. As a consequence,  they always induce combs with a large numbers of phase-locked modes. 

Figure~\ref{Plotting_quads_as_fct_of_order_bright_solitons} displays the quadrature spectra for some modes of a Kerr comb originating from a bright soliton. It can be seen that there is a certain angle of quadrature for which the closest sidemode pair ($l = \pm 1$) features very good squeezing, of the order the ideal squeezing of the $3$-modes comb.
As the mode order $|l|$ is increased and the offset $\delta \Phi$ is kept constant, the squeezing degrades and eventually disappears beyond  
$|l| \sim 20$. The case of dark solitons is presented in Fig.~\ref{Plotting_quads_as_fct_of_order_dark_solitons}, where it can be seen that as in the bright soliton case, the sidemode pair $l = \pm 1$ displays good squeezing. However, this squeezing degrades much faster as the mode order is increased while keeping the offset phase  $\delta \Phi$ constant.

\section{Conclusion}
\label{Conclusion}

In this article, we have investigated in detail the quantum correlations that are taking place in stationary Kerr combs below and above threshold, when driven by the quantum noise associated with vacuum fluctuations. 

We have shown that either a canonical quantization procedure or an Hamiltonian formalism can be used to  establish the quantum stochastic equations ruling the time-domain dynamics of each mode, and 
particular emphasis has been laid on the two principal architectures that are routinely used for Kerr comb generation, namely the add-through and the add-drop configurations.

For the system under threshold (spontaneous FWM), we have investigated the coupling between the pump photons and the vacuum fluctuation in the sidemodes, which is at the origin of parametric fluorescence spectra. 
We have analytically determined the main characteristics of the spontaneous emission spectra, with particular emphasis on the lineshape of the individual sidemodes and envelope of the full spectra. We have explained the conditions under which the sidemodes and/or the spectra might have one or two extrema. 
We have also provided a detailed calculation allowing to determine accurately the spontaneous noise power emitted per sidemode, as a function of all the relevant parameters of the system.

For the system pumped above threshold (stimulated FWM), we have provided insight in relation with the essential commutation properties between the interaction Hamiltonian and the photon numbers, which allowed to understand the physical mechanisms leading to photon number squeezing in Kerr combs. 
We have then explicitly defined the quantum Langevin equations ruling the fluctuations of annihilation and creation operators for each mode, regardless of the number of modes in the comb. We have shown that this fluctuation flow can be reduced to a flow of lower dimension, that rules the dynamics of both amplitude and phase quadratures for each pair of sidemodes. 
Our analysis has shown that the reduced $3$-modes model, which is valid close to threshold for roll patterns, allows for the the exact determination of the spectra of the amplitude and phase quadratures.
These exact analytical solutions have been found to be very good approximations even far above threshold for roll patterns.
Squeezing in bright and dark solitons has also been analyzed as well, for various pairs of sidemodes. 
The best squeezing spectra have been shown to be relatively close to the inverted Lorentzian profile that is predicted from the reduced $3$-modes model. In stationary Kerr combs driven by quantum-noise, squeezing can therefore be obtained regardless of the spectral extension of the comb, regardless of the dynamical state, and regardless of the dispersion regime.
Our results also indicate that a key parameter is the so-called squeezing factor, which is the ratio between out-coupling and total losses. Regardless of the architecture of the Kerr comb generator (add-through or add-drop), strong over-coupling has been shown to be the always the best configuration for squeezing purposes.

This work could be extended to the case where non classical light is generated through second-harmonic generation~\cite{Erlangen_PRL_2011,Erlangen_Nature_Comm_2013,Erlangen_arxiv_2014}. 
New bulk materials, such as aluminum nitride (AlN), allow for the efficient excitation of both the 
second and third order susceptibility owing to their non-centrosymmetric crystalline structure~\cite{OL_AlN_comb}, and they could be interesting materials for the exploration of a wide variety of quantum optics phenomena at chip-scale.
The platform of centro-symmetric crystals allowing for ultra-high $Q$ actor is rapidly expanding as 
well~\cite{Henriet_OEng,OL_BaF2,OL_SrF2}, allowing to explore other nonlinear phenomena such as Brillouin and Raman scattering at the quantum level~\cite{APL_Brillouin,PRA_Raman}.  

We have assumed in our investigations that the noise was exclusively of quantum origin, and was associated with the fundamental vacuum fluctuations. At the experimental level, other sources of noise arise~\cite{MatskoJOSAB_I,MatskoJOSAB_II} as well, and it is important to account for this technical noise in order to perform  pertinent comparisons between theory and experiments. 
Future work will address this issue, as well deterministic effects such as higher-order dispersion (which deserves particular attention even for crystalline materials, see ref.~\cite{Grudinin_Optica}), polarization degrees of freedom~\cite{Zambrini_PRA,Hoyuelos_pumpmeter_PRA}, or other experimental imperfections such as unbalanced detection. The investigation of the quantum properties of time-dependent solutions such as soliton breathers is an interesting challenge as well, which can deserve much attention  
We expect these investigations to open the way for new applications in the area of guided quantum optics at telecom wavelengths, and to provide an idoneous platform for the investigation of the fundamental properties of light at the quantum level~\cite{Fabre_review,Sanders}, in general, and for optical frequency combs in particular~\cite{PRL_Treps,NP_Treps}.

\section*{Acknowledgements}

The author would like to acknowledge financial support 
    from the European Research Council (ERC) through the projects StG NextPhase and PoC Versyt,
    from the \textit{Centre National d'Etudes Spatiales} (CNES) through the project SHYRO,
    from the \textit{R\'egion de Franche-Comt\'e} through the project CORPS, and
    from the Labex ACTION.     
The author would also like to thank \textit{Peyresq -- Foyer d'Humanisme} in France, 
where a significant part of this work has been completed.


\begin{thebibliography}{99}
\bibitem{Vahala_PRL_2004} T.~J. Kippenberg, S.~M. Spillane, and K.~J. Vahala,
                          \textit{Kerr-Nonlinearity Optical Parametric Oscillation in an Ultrahigh-Q Toroid Microcavity},
                          Phys. Rev. Lett.~\textbf{93}, 083904 (2004).
\bibitem{Maleki_PRL_LowThres} A.~A. Savchenkov, A.~B. Matsko, D.~Strekalov, M.~Mohageg, V.~S. Ilchenko, and L.~Maleki,
                           \textit{Low Threshold Optical Oscillations in a Whispering Gallery Mode CaF$_2$ Resonator},
                            {Phys. Rev. Lett.}~\textbf{93}, 243905 (2004).
\bibitem{DelhayeKipp} P.~Del'Haye, A.~Schliesser, A.~Arcizet, R.~Holzwarth, and T.~J. Kippenberg,  
                     \textit{Optical frequency comb generation from a monolithic microresonator},
                    {Nature}~{\bf 450}, 1214 (2007).
\bibitem{Lipson_NatPhot} J.~S. Levy, A.~Gondarenko, M.~A. Foster, A.~C. Turner-Foster, A.~L. Gaeta, and M.~Lipson,
                         \textit{CMOS-compatible multiple-wavelength oscillator
                          for on-chip optical interconnects}, 
                          Nature Photonics~\textbf{4}, 37 (2010).
\bibitem{Review_Kerr_combs_Science} T. J. Kippenberg, R. Holzwarth, and S. A. Diddams,
                                    \textit{Microresonator-Based Optical Frequency Combs},
                                    {Science}~{\bf 322}, 555 (2011).
\bibitem{Nature_Ferdous} F. Ferdous, H. Miao, D. E. Leaird,	K. Srinivasan, J. Wang,
                          L. Chen,	L. T. Varghese	and A. M. Weiner,
                          \textit{Spectral line-by-line pulse shaping of on-chip
                          microresonator frequency combs},
                          {Nature Photonics}~\textbf{5}, 770 (2011). 
\bibitem{PRL_Vahala_2012}  J. Li, H. Lee, T. Chen, and K. J. Vahala,
                           \textit{Low-Pump-Power, Low-Phase-Noise, and Microwave 
                            to Millimeter-Wave Repetition Rate Operation in Microcombs},
                           {Phys. Rev. Lett.}~\textbf{109}, 233901 (2012).
\bibitem{PRL_NIST} P.~Del'Haye, S.~B. Papp, and S.~A. Diddams, 
                   \textit{Hybrid Electro-Optically Modulated Microcombs},
                    {Phys. Rev. Lett.} \textbf{109}, 263901 (2012). 
\bibitem{Nat_Gaeta_rev} D. J. Moss, R. Morandotti, A. L. Gaeta, and M. Lipson,
                        \textit{New CMOS-compatible platforms based on
                        silicon nitride and Hydex for nonlinear optics},
                        {Nature Photonics}~\textbf{7}, 597 (2013).                        
\bibitem{PfeifleNatPhot} J. Pfeifle \textit{et al.}, 
                          \textit{Coherent terabit communications with
                          microresonator Kerr frequency combs},
                          {Nature Phot.}~\textbf{8}, 375 (2014).
\bibitem{NIST_Optica} S.~B. Papp, K.~Beha, P.~Del’Haye, F.~Quinlan, H.~Lee, K.~J. Vahala, and S.~A. Diddams,
                      \textit{Microresonator frequency comb optical clock}, 
                      Optica~\textbf{1}, 10 (2014).
\bibitem{Self_injection_NIST} P. Del'Haye, K. Beha, S. B. Papp, and S. A. Diddams,
                              \textit{Self-Injection Locking and Phase-Locked States 
                              in Microresonator-Based Optical Frequency Combs},
                             Phys. Rev. Lett.~112, 043905 (2014).
\bibitem{Kipp_Soliton} T.~Herr, V.~Brasch, J.~D.~Jost, C.~Y.~Wang, N.~M.~Kondratiev, M.~L.~Gorodetsky, 
                       and T.~J.~Kippenberg,
                       \textit{Temporal solitons in optical microresonators}
                       {Nature Photon.}~\textbf{8}, 145 (2014).
\bibitem{PRL_WDM} J.~Pfeifle, A.~Coillet, R.~Henriet, K.~Saleh, P.~Schindler, C.~Weimann,
                  W.~Freude, I.~V.~Balakireva, L.~Larger, C.~Koos, and Y.~K.~Chembo,
                       \textit{Optimally Coherent Kerr Combs Generated with Crystalline Whispering Gallery
                       Mode Resonators for Ultrahigh Capacity Fiber Communications}
                       {Phys. Rev. Lett.}~\textbf{114}, 093902 (2015).
\bibitem{Nanophotonics} Y. K. Chembo,
		\textit{Kerr optical frequency combs: theory, applications and perspectives},
		Nanophotonics, accepted for publication (2015).
\bibitem{YanneNanPRL} Y. K. Chembo, D. V. Strekalov, and N. Yu,  
                      \textit{Spectrum and Dynamics of Optical Frequency Combs Generated
                      with Monolithic Whispering Gallery Mode Resonators},
                      {Phys. Rev. Lett.}~{\bf 104}, 103902 (2010).
\bibitem{YanneNanPRA} Y. K. Chembo and N. Yu,   
                       \textit{Modal expansion approach to optical-frequency-comb generation with
                       monolithic whispering-gallery-mode resonators},
                      {Phys. Rev. A}~{\bf 82}, 033801 (2010).
\bibitem{YanneNanOL} Y. K. Chembo and N. Yu,
                      \textit{On the generation of octave-spanning optical
                      frequency combs using monolithic whispering-gallery-mode microresonators},
                      {Opt. Lett.}~\textbf{35}, 2696 (2010).
\bibitem{Matsko_OL_2} A.~B. Matsko, A.~A. Savchenkov, W.~Liang, V.~S. Ilchenko, D.~Seidel, and L.~Maleki,
                     \textit{Mode-locked Kerr frequency combs},
                     {Opt. Lett.}~\textbf{36}, 2845 (2011).
\bibitem{PRA_Yanne-Curtis} Y.~K. Chembo and C.~R. Menyuk, 
                         \textit{Spatiotemporal Lugiato-Lefever formalism for Kerr-comb generation 
                         in whispering-gallery-mode resonators},
                         {Phys. Rev. A}~\textbf{87}, 053852 (2013).
\bibitem{Coen} S.~Coen, H.~G. Randle, T.~Sylvestre, and M.~Erkintalo,
               \textit{Modeling of octave-spanning Kerr frequency combs  
               using a generalized mean-field Lugiato-Lefever model},
               {Opt. Lett.}~\textbf{38}, 37 (2013). 
\bibitem{IEEE_PJ} A. Coillet, I. Balakireva, R. Henriet, K. Saleh,
                  L. Larger, J. M. Dudley, C. R. Menyuk, and Y. K. Chembo,
					\textit{Azimuthal Turing Patterns, Bright and Dark
					Cavity Solitons in Kerr Combs generated with
					Whispering-Gallery Mode Resonators},
                  \emph{IEEE Photonics Journal} \textbf{5}, 6100409 (2013).
\bibitem{OL_phaselocking} A. Coillet and Y. K. Chembo, 
                          \textit{On the robustness of phase locking in Kerr optical frequency combs},
                          {Opt. Lett.} \textbf{39}, 1529 (2014).
\bibitem{Chaos_paper} A. Coillet and Y. K. Chembo, 
                      \textit{Routes to spatiotemporal chaos in Kerr optical frequency combs},
                      {Chaos}~\textbf{24}, 013313 (2014).
\bibitem{LL} L.~A. Lugiato and R.~Lefever, 
            \textit{Spatial Dissipative Structures in Passive Optical Systems},
            {Phys. Rev. Lett.}~\textbf{58}, 2209 (1987).
\bibitem{Sharping_OE} J. E. Sharping, K. F. Lee, M. A. Foster, A. C. Turner, B. S. Schmidt, M. Lipson, A. L. Gaeta
                      and Prem Kumar,
                  \textit{Generation of correlated photons in nanoscale silicon waveguides},
                  Optics Express~\textbf{14}, 12388 (2006).  
\bibitem{Clemmen_OE} S. Clemmen, K. Phan Huy, W. Bogaerts, R. G. Baets, Ph. Emplit, and S. Massar,
                    \textit{Continuous wave photon pair generation in silicon-on-insulator waveguides and ring resonators},
                  Optics Express~\textbf{17}, 16558 (2009).  
\bibitem{Helt_OL} L. G. Helt, Z. Yang, M. Liscidini, and J. E. Sipe,
                  \textit{Spontaneous four-wave mixing in microring resonators},
                  Opt. Lett.~\textbf{35}, 3006 (2010).  
\bibitem{Chen_OE} J. Chen, Z. H. Levine, J. Fan, and A. L. Migdall,
                    \textit{Frequency-bin entangled comb of photon pairs from a Silicon-on-Insulator micro-resonator},
                  Optics Express~\textbf{19}, 1470 (2011).  
\bibitem{Azzini_OE} S. Azzini, D. Grassani, M. J. Strain, M. Sorel, L. G. Helt, J. E. Sipe, M. Liscidini, 
                    M. Galli, and D. Bajoni, 
                  \textit{Ultra-low power generation of twin photons in a compact silicon ring resonator},
                  Optics Express~\textbf{20}, 23100 (2012).  
\bibitem{Helt_scale_JOSAB} L. G. Helt, M. Liscidini, and J. E. Sipe, 
                  \textit{How does it scale? Comparing quantum and classical nonlinear optical processes 
                          in integrated devices},
                  J. Opt. Soc. Am.~\textbf{29}, 2199 (2012).  
\bibitem{Azzini_OL} S. Azzini, D. Grassani, M. Galli, L. C. Andreani, M. Sorel, 
                    M. J. Strain, L. G. Helt, J. E. Sipe, M. Liscidini, and D. Bajoni,
                    \textit{From classical four-wave mixing to parametric fluorescence in silicon microring resonators},
                    Optics Express~\textbf{37}, 3807 (2012).  
\bibitem{Camacho_OE} R. M. Camacho,
                  \textit{Entangled photon generation using four-wave mixing in azimuthally symmetric microresonators},
                  Optics Express~\textbf{20}, 21977 (2012).  
\bibitem{Takesue_SciRep} N. Matsuda, H. Le Jeannic, H. Fukuda, T. Tsuchizawa, W. J. Munro,
                         K. Shimizu, K. Yamada, Y. Tokura and H. Takesue,
                         \textit{A monolithically integrated polarization entangled photon pair source on a silicon chip},
                         Sci. Rep.~\textbf{2}, 817 (2012).  
\bibitem{Reimer_OE} C. Reimer, L. Caspani, M. Clerici, M. Ferrera, M. Kues, M. Peccianti, A. Pasquazi, 
                    L. Razzari, B. E. Little, S. T. Chu, D. J. Moss, and R. Morandotti,
                    \textit{Integrated frequency comb source of heralded single photons},
                  Optics Express~\textbf{22}, 6535 (2014).  
\bibitem{Vernon_Arxiv} Z. Vernon and J.E. Sipe, 
                  \textit{Spontaneous four-wave mixing in lossy microring resonators},
                  arXiv:1502.05900 [quant-ph] (2015).
\bibitem{Engin_OE} E. Engin, D. Bonneau, C. M. Natarajan, A. S. Clark, M. G. Tanner, R. H. Hadfield, Sanders N. Dorenbos, 
                   V. Zwiller, K. Ohira, N. Suzuki, H. Yoshida, N. Iizuka, M. Ezaki, J. L. O'Brien, and M. G. Thompson,
                    \textit{Photon pair generation in a silicon micro-ring resonator with reverse bias enhancement},
                  Optics Express~\textbf{21}, 27826 (2013).  
\bibitem{Grassani_Optica} D. Grassani, S. Azzini, M. Liscidini, M. Galli, M. J. Strain, M. Sorel, J. E. Sipe, and D. Bajoni,
                    \textit{Micrometer-scale integrated silicon source of time-energy entangled photons},
                    Optica~\textbf{2}, 88 (2015).                     
\bibitem{Qubit_entanglement} 
			J. W. Silverstone, R. Santagati,	D. Bonneau, M. J. Strain, M. Sorel, J. L. O'Brien and 
			M. G. Thompson
			\textit{Qubit entanglement between ring-resonator photon-pair sources on a silicon chip},
			Nature Commun.~\textbf{6}, 7948 (2015).
\bibitem{Wakabayashi} R. Wakabayashi, M. Fujiwara, K.-I Yoshino, Y. Nambu, M. Sasaki, T. Aoki,
			\textit{Time-bin entangled photon pair generation from Si micro-ring resonator},
			arXiv:1501.05687 [quant-ph] (2015).
\bibitem{Agrawal} G. P. Agrawal, 
                  \textit{Nonlinear Fiber Optics}, Fifth Edition,
                  Academic Press (2013). 
\bibitem{Lugiato_Castelli} L. A. Lugiato and F. Castelli, 
                           \textit{Quantum Noise Reduction in a Spatial Dissipative Structure},
                            {Phys. Rev. Lett.}~\textbf{68}, 3284 (1992).
\bibitem{Zambrini_PRA} R. Zambrini, M. Hoyuelos, A. Gatti, P. Colet, L. Lugiato, and M. San Miguel, 
                       \textit{Quantum fluctuations in a continuous vectorial Kerr medium model},
                        Phys. Rev. A~\textbf{62}, 063801 (2000).
\bibitem{Grynberg_Lugiato} G. Grynberg and L. Lugiato, 
                           \textit{Quantum properties of hexagonal patterns},
                           Opt. Commun.~\textbf{101}, 69 (1993).
\bibitem{Gatti_Mancini} A.~Gatti and S.~Mancini,
                        \textit{Spatial Correlations in Hexagons Generated via Kerr Nonlinearity},        
                        {\rm Phys. Rev. A} {\bf 65}, 013816 (2001).
\bibitem{Arxiv_squeezing_Cornell} A. Dutt, K. Luke, S. Manipatruni, A.~L. Gaeta, P. Nussenzveig, and M. Lipson,
                                 \textit{On-Chip Optical Squeezing},
                                  Phys. Rev. Applied~\textbf{3}, 044005 (2015).  
\bibitem{Nature_Selectable_freq}  A. A. Savchenkov,	A. B. Matsko, W. Liang,	V. S. Ilchenko,	
                                    D. Seidel	and  L. Maleki, 
                                    \textit{Kerr combs with selectable central frequency},
                                    {Nature Photonics}~\textbf{5}, 293 (2011).
\bibitem{Nature_Mid_IR} C. Y. Wang, T. Herr, P. Del'Haye, A. Schliesser, J. Hofer,	R. Holzwarth,	
                        T. W. Haensch, N. Picqu\'e, and T. J. Kippenberg, 
                        \textit{Mid-infrared optical frequency combs at 2.5~$\mu$m
                         based on crystalline microresonators},
                         {Nature Communications}~\textbf{4}, 134 (2013).
\bibitem{OL_QKD_Bloch} M.~Bloch, S.~W. McLaughlin, J.-M. Merolla, and F. Patois,
                       \textit{Frequency-coded quantum key distribution},
                       Opt. Lett.~\textbf{32}, 301 (2007).
\bibitem{PRA_Merolla_2010} L. Olislager, J. Cussey, A. T. Nguyen, P. Emplit, S. Massar, J.-M. Merolla, and K. Phan Huy,
                           \textit{Frequency-bin entangled photons},
                           Phys. Rev. A~\textbf{82}, 013804 (2010).
\bibitem{PRA_Merolla_2014} L.~Olislager, E.~Woodhead, K. Phan Huy, J.-M. Merolla, P. Emplit, and S. Massar,
                           \textit{Creating and manipulating entangled optical qubits in the frequency domain},
                           Phys. Rev. A~\textbf{89}, 052323 (2014).
\bibitem{comment_phase_operators} It is important to note that the definition of well-behaved phase operators 
                                  is theoretically non-trivial.
\bibitem{SelTop_Matsko_I}  A. B. Matsko and V. S. Ilchenko,
                           \textit{Optical resonators with whispering gallery modes I: Basics},
                           {IEEE J. Sel. Top. Quantum Electron.}~\textbf{12}, 3 (2006).
\bibitem{SelTop_Matsko_II}  V. S. Ilchenko and A. B. Matsko,
                           \textit{Optical Resonators With Whispering-Gallery
                            Modes—Part II: Applications},
                           {IEEE J. Sel. Top. Quantum Electron.}~\textbf{12}, 15 (2006).
\bibitem{Feron_Review} A. Chiasera, Y. Dumeige, P. F\'eron, M. Ferrari, Y. Jestin, G. Nunzi Conti,
                       S. Pelli, S. Soria, and G. C. Righini,
                        \textit{Spherical whispering-gallery-mode microresonators}, 
                        {Laser Photon. Rev.}~\textbf{51}, 457 (2010).
\bibitem{Jove} A. Coillet, R. Henriet, K. P. Huy, M. Jacquot, L. Furfaro, I. Balakireva, L. Larger, and Y. K. Chembo,
               \textit{Microwave Photonics Systems Based on Whispering-gallery-mode Resonators}, 
               {J. Vis. Exp.}~\textbf{78}, e50423 (2013).
\bibitem{PRA_Unified} C.~Godey, I.~V.~Balakireva, A.~Coillet, and Y.~K.~Chembo, 
                       \textit{Stability analysis of the spatiotemporal Lugiato-Lefever model
                       for Kerr optical frequency combs in the anomalous and normal dispersion regimes},
                       {Phys. Rev. A}~\textbf{89}, 063814 (2014).
\bibitem{comment_AT} This nomenclature is not standard, 
                     but it has been adopted here for being particularly intuitive.
\bibitem{JSTQE_AurYanne} A. Coillet, R. Henriet, P. Salzenstein, K. Phan Huy, L. Larger, and Y. K. Chembo,
                          \textit{Time-domain Dynamics and Stability Analysis of Optoelectronic Oscillators
                           based on Whispering-Gallery Mode Resonators},
                           {IEEE J. Sel. Top. Quantum Electron.}~\textbf{19}, 6000112 (2013).
\bibitem{Feron_CRD} Y. Dumeige, S. Trebaol, L. Ghisa T. K. N. Nguyen H. Tavernier, and P. F\'eron,
                          \textit{Determination of coupling regime of high-$Q$ 
                           resonators and optical gain of highly selective amplifiers},
                           {J. Opt. Soc. Am. B}~\textbf{25}, 2073 (2008).
\bibitem{platicons} 
V. E. Lobanov, G. Lihachev, T. J. Kippenberg, and M. L. Gorodetsky,
\textit{Frequency combs and platicons in optical microresonators with normal GVD},
Optics Express~\textbf{23}, 7713 (2015).                           
\bibitem{Bachor_Ralph_book} H.-A. Bachor and T. C. Ralph,
                            \textit{A Guide to Experiments in Quantum Optics}, 
                            Wiley-VCH (2004).
\bibitem{Braginski_PLA} V.~B.~Bragisnky, M.~L.~Gorodetsky and V.~S.~Illchenko, 
                  \textit{Quality factor and optical properties of optical whispering gallery modes},
                  Phys. Lett. A~\textbf{137}, 393 (1989).
\bibitem{Comb_2mW} W. Liang, A. A. Savchenkov, A. B. Matsko, V. S. Ilchenko, D. Seidel, and L. Maleki, 
                          \textit{Generation of near-infrared frequency combs from
                           a MgF$_2$ whispering gallery mode resonator},
                          {Opt. Lett.} \textbf{36}, 2290 (2011).
\bibitem{Gardiner} C. W. Gardiner, 
				  \textit{Quantum noise},
                  Springer-Verlag (1991).
\bibitem{Grynberg_Aspect_Fabre_book} G. Grynberg, A. Aspect, and C. Fabre,
                                        \textit{Introduction to Quantum Optics.
                                         From the Semi-classical Approach to Quantized Light},
                                        Cambridge University Press (2010).
\bibitem{notational_convention} As a notational convention, sans serif fonts are reserved to operators. 
                                All operators are in caps, except pure cavity fields. 
                                Calligraphic fonts are reserved for semi-classical complex-valued variables,
                                and bold fonts stand for matrices and vectors of scalars or operators. 
                                The same terminology applies to photon numbers
\bibitem{Matsko_Normal_Comb} A. B. Matsko, A. A. Savchenkov, and L. Maleki,
                          \textit{Normal group-velocity dispersion Kerr frequency comb},
                          {Opt. Lett.}~\textbf{37}, 43 (2012).
\bibitem{Hoyuelos_pumpmeter_PRA} M. Hoyuelos, A. Sinatra, P. Colet, L. Lugiato, and M. San Miguel, 
                  \textit{Spatial pump-meter quantum correlations in a vectorial Kerr-medium model},
                  Phys. Rev. A~\textbf{59}, 1622 (1999). 
\bibitem{Lugiato_spatial_PRL}  L. A. Lugiato and A. Gatti, 
                  \textit{Spatial Structure of a Squeezed Vacuum},
                  Phys. Rev. Lett.~\textbf{70}, 3868 (1993).  
\bibitem{Gatti_quantum_images_PRA} A. Gatti and L. A. Lugiato, 
                  \textit{Quantum images and critical fluctuations in the optical parametric oscillator
                          below threshold},
                  Phys. Rev. A~\textbf{52}, 1675 (1995).  
\bibitem{Lugiato_Marzoli} L. A. Lugiato and I. Marzoli,
                    \textit{Quantum spatial correlations in the optical parametric oscillator with spherical mirrors},
                    Phys. Rev. A~\textbf{52}, 4886 (1995).  
\bibitem{Gatti_Langevin_OPO_PRA}  A. Gatti, H. Wiedemann, L. A. Lugiato, I. Marzoli, G.-L. Oppo, and S. M. Barnett, 
                  \textit{Langevin treatment of quantum fluctuations and optical patterns in optical 
                          parametric oscillators below threshold},
                  Phys. Rev. A~\textbf{56}, 877 (1997).                   
\bibitem{Gatti_Multiphoton_PRA} A. Gatti, R. Zambrini, M. San Miguel, and L. A. Lugiato, 
                  \textit{Multiphoton multimode polarization entanglement in parametric down-conversion},
                  Phys. Rev. A~\textbf{68}, 053807 (2003).
\bibitem{Zambrini_Polar_PRA} R. Zambrini, A. Gatti, L. Lugiato, and M. San Miguel, 
                  \textit{Polarization quantum properties in a type-II optical parametric oscillator below threshold},
                  Phys. Rev. A~\textbf{68}, 063809 (2003).                 
\bibitem{Fabre_J_Phys} C. Fabre, E. Giacobino, A. Heidmann, and S. Reynaud,
                          \textit{Noise characteristics of a non-degenerate Optical Parametric Oscillator 
                           - Application to quantum noise reduction},
                          {J. Phys. France}~\textbf{50}, 1209 (1989). 
\bibitem{Garcia_Ferrer_JQE} F. V. Garcia-Ferrer, C. Navarrete-Benlloch, G. J. de Valcarcel, and E. Roldan, 
                  \textit{Squeezing Via Spontaneous Rotational Symmetry Breaking in a Four-Wave Mixing Cavity},
                  IEEE J. Quantum Electron.~\textbf{45}, 1404 (2009).  
\bibitem{Brainis_PRA}  E. Brainis, 
                  \textit{Four-photon scattering in birefringent fibers},
                  Phys. Rev. A~\textbf{79}, 023840 (2009). 
\bibitem{Thermooptic_osc} S. Diallo, G. Lin and Y. K. Chembo, 
			\textit{Giant thermooptical relaxation oscillations in millimeter-size whispering
			gallery mode disk resonators}, 
			Opt. Lett.~\textbf{40}, 3834 (2015).
\bibitem{Parra_Rivas} P. Parra-Rivas, D. Gomila, M. A. Matias, S. Coen, and L. Gelens, 
                            \textit{Dynamics of localized and patterned structures in the Lugiato-Lefever
                            equation determine the stability and shape of optical frequency combs},
                            Phys. Rev. A~\textbf{89}, 043813 (2014).
\bibitem{Record_Q} A. A. Savchenkov, A. B. Matsko, V. S. Ilchenko, and L. Maleki,
                          \textit{Optical resonators with ten million finesse},
                          {Opt. Express}~\textbf{115}, 6768 (2007). 
\bibitem{NIST_PRX} S. B. Papp, P. Del'Haye, and S. A. Diddams,
                   \textit{Mechanical Control of a Microrod-Resonator Optical Frequency Comb},
                   Phys. Rev. X~\textbf{3}, 031003 (2013).
\bibitem{Integrated_WGMR} L. Maleki, 
                   \textit{The optoelectronic oscillator},
                   Nature Photonics~\textbf{5}, 728 (2011).
\bibitem{diff_1mode_2modes_quads} Note that the single-mode quadrature is a defined with the annihilation 
                                  and creation operators of the same mode $l$, while the two-mode quadrature in 
                                  Eq.~\ref{quadratureroperator_3modes}
                                  is defined using the operators of two different symmetric sidemodes $\pm l$.  
\bibitem{comment_symbols} In Figs.~\ref{Plotting_quads_Sa_fct_of_power}, 
                                   \ref{Plotting_quads_as_fct_of_order},
                                   \ref{Plotting_quads_as_fct_of_order_bright_solitons}, and
                                   \ref{Plotting_quads_as_fct_of_order_dark_solitons}, 
                                   the symbols are obtained numerically and they are linked with a thin line 
                                   for the purpose of eye guidance only. 
                                   The possibility of a divergence to infinity in seemingly continuous lines 
                                   can not be ruled out, 
                                   as it can be understood from Fig.~\ref{Plotting_quads_as_fct_of_psi}. 
\bibitem{Lin_OEng} G. Lin, K. Saleh, R. Henriet, S. Diallo, R. Martinenghi,
			A. Coillet, and Y. K. Chembo,   
			\textit{Wide-range tunability, thermal locking, and mode-crossing effects in Kerr optical
			frequency combs}, 
			Opt. Eng.~\textbf{53}, 122602 (2014).
\bibitem{Lin_OExp} G. Lin and Y. K. Chembo,  
			\textit{On the dispersion management of 
			fluorite whispering-gallery mode resonators for
			Kerr optical frequency comb generation in the telecom and mid-infrared range}, 
			Opt. Express~\textbf{23}, 1594 (2015).
\bibitem{Erlangen_PRL_2011} J. U. F\"urst, D. V. Strekalov, D. Elser, A. Aiello, U. L. Andersen,
                            Ch. Marquardt, and G. Leuchs,
                            \textit{Quantum Light from a Whispering-Gallery-Mode Disk Resonator},
                            Phys. Rev. Lett.~\textbf{106}, 113901 (2011).
\bibitem{Erlangen_Nature_Comm_2013} M. F\"ortsch \textit{et al.},
                                    \textit{A versatile source of single photons 
                                    for quantum information processing},
                                    Nature Communications~\textbf{4}, 1818 (2013).
\bibitem{Erlangen_arxiv_2014} M. F\"ortsch \textit{et al.},
                         \textit{Highly efficient generation of single-mode photon pairs 
                         using a crystalline whispering gallery mode resonator},
                         arXiv:1404.0593v1 (2014).
\bibitem{OL_AlN_comb} H. Jung, C. Xiong, K. Y. Fong, X. Zhang, and H. X. Tang,
                      \textit{Optical frequency comb generation from aluminum nitride microring resonator},
                       Opt. Lett.~\textbf{38}, 2810 (2013).
\bibitem{Henriet_OEng} R. Henriet, A. Coillet, K. Saleh, L. Larger, and Y. K. Chembo,   
		\textit{Barium  fluoride and lithium fluoride whispering gallery mode resonators for photonics applications}, 
		Opt. Eng.~\textbf{53}, 071821 (2014).
\bibitem{OL_BaF2} G. Lin, S. Diallo, R. Henriet, M. Jacquot, and Y. K. Chembo,
                      \textit{Barium fluoride whispering-gallery-mode disk-resonator
                              with one billion quality-factor},
                       Opt. Lett.~\textbf{39}, 6009 (2014).
\bibitem{OL_SrF2} R. Henriet, G. Lin, A. Coillet, M. Jacquot, L. Furfaro, L. Larger, and Y. K. Chembo,
                      \textit{Kerr optical frequency comb generation in strontium
                              fluoride whispering-gallery mode resonators with billion quality factor},
                       Opt. Lett.~\textbf{40}, 1567 (2015).
\bibitem{APL_Brillouin} G. Lin, S. Diallo, K. Saleh, R. Martinenghi, J.-C. Beugnot, 
                        T. Sylvestre, and Y. K. Chembo,
                      \textit{Cascaded Brillouin lasing in monolithic barium fluoride 
                              whispering gallery mode resonators},
                       Appl. Phys. Lett.~\textbf{105}, 231103 (2014).
\bibitem{PRA_Raman} Y. K. Chembo, I. S. Grudinin and N. Yu,
                      \textit{Spatiotemporal dynamics of Kerr-Raman optical frequency combs},
                       Phys. Rev. A~\textbf{92}, 043818 (2015).
\bibitem{MatskoJOSAB_I} A.~B.~Matsko, A.~A.~Savchenkov, N.~Yu, and L.~Maleki,
						\textit{Whispering-gallery-mode resonators as frequency references. 
						I. Fundamental limitations},
                        {J. Opt. Soc. Am. B}~\textbf{24}, 1324 (2007). 
\bibitem{MatskoJOSAB_II} A.~A.~Savchenkov, A.~B.~Matsko, V.~S.~Ilchenko, N.~Yu, and L.~Maleki,
                         \textit{Whispering-gallery-mode resonators as frequency references. 
                        II. Stabilization},
                        {J. Opt. Soc. Am. B}~\textbf{24}, 2988 (2007). 
\bibitem{Grudinin_Optica} I. S. Grudinin and N. Yu,  
		\textit{Dispersion engineering of crystalline resonators via microstructuring}, 
		Optica~\textbf{2}, 221 (2015).
\bibitem{Fabre_review} C. Fabre,
                       \textit{Squeezed states of light},
                       Physics Reports~\textbf{19}, 215 (1992).
\bibitem{Sanders} B. C. Sanders, 
                  \textit{Review of coherent entangled states},
                  J. Phys. A: Math. Theor.~\textbf{45}, 244002 (2012).
\bibitem{PRL_Treps} O. Pinel, P. Jian, R.~M. De Araujo, J. Feng, B. Chalopin, C. Fabre, and N. Treps, 
                  \textit{Generation and characterization of multimode quantum frequency combs},
                  Phys. Rev. Lett.~\textbf{108}, 083601 (2012).
\bibitem{NP_Treps} J. Roslund, R. M. De Araujo, S. Jiang, C. Fabre, and N. Treps, 
                  \textit{Wavelength-multiplexed quantum networks with ultrafast frequency combs},
                  Nature Photonics~\textbf{8}, 113 (2013).
\end{thebibliography}
\end{document}